\documentclass[10pt,twocolumn,a4paper,unpublished,allowtoday]{quantumarticle}

\pdfoutput=1
\usepackage[utf8]{inputenc}
\usepackage[english]{babel}
\usepackage[T1]{fontenc}

\usepackage{amssymb}
\usepackage{amsmath}
\usepackage{amsbsy}
\usepackage{amsfonts}
\usepackage{mathrsfs}
\usepackage{mathtools}
\usepackage{stmaryrd}
\usepackage{adjustbox}

\usepackage[numbers]{natbib} % compress

\allowdisplaybreaks

\usepackage{graphicx}
\graphicspath{{./images}}

\usepackage{tikz}
\usetikzlibrary{positioning}
\usetikzlibrary{calc}
\usetikzlibrary{decorations.pathmorphing}
\usetikzlibrary{arrows}
\usetikzlibrary{arrows.meta}
\usetikzlibrary{shapes.arrows}
\usetikzlibrary{fadings}
\usetikzlibrary{shadings}
\usepackage{tikz-cd}
\usetikzlibrary{quantikz2}

\usetikzlibrary{external}
\tikzsetexternalprefix{images_tikz/}
\usepackage{hyperref}
\hypersetup{
  colorlinks = true,
  urlcolor = blue,
  linkcolor = blue!50!black,
  citecolor = green!50!black,
  filecolor = red!50!black
}
\usepackage[capitalise,nameinlink]{cleveref}

\renewcommand{\leq}{\leqslant}

\newcommand{\Int}{\mathbb{Z}}
\newcommand{\Real}{\mathbb{R}}
\newcommand{\Comp}{\mathbb{C}}

\DeclareMathOperator{\Tr}{Tr}

\DeclarePairedDelimiter{\abs}{\lvert}{\rvert}
\DeclarePairedDelimiter{\norm}{\lVert}{\rVert}

\DeclarePairedDelimiterX{\innerp}[2]{\langle}{\rangle}{#1, #2}
\DeclarePairedDelimiter{\bra}{\langle}{\rvert}
\DeclarePairedDelimiter{\ket}{\lvert}{\rangle}
\DeclarePairedDelimiterX{\braket}[2]{\langle}{\rangle}{#1 \delimsize\vert #2}
\DeclarePairedDelimiterX{\ketbra}[2]{\lvert}{\rvert}{#1 \delimsize\rangle\!\delimsize\langle #2}
\DeclarePairedDelimiterX{\proj}[1]{\lvert}{\rvert}{#1 \delimsize\rangle\!\delimsize\langle #1}

\DeclarePairedDelimiterX{\innerpD}[2]{(}{)}{#1, #2}
\DeclarePairedDelimiter{\braD}{(}{\rvert}
\DeclarePairedDelimiter{\ketD}{\lvert}{)}
\DeclarePairedDelimiterX{\braketD}[2]{(}{)}{#1 \delimsize\vert #2}
\DeclarePairedDelimiterX{\ketbraD}[2]{\lvert}{\rvert}{#1 \delimsize)\!\delimsize( #2}
\DeclarePairedDelimiterX{\projD}[1]{\lvert}{\rvert}{#1 \delimsize)\!\delimsize( #1}

\newcommand{\Id}{\mathrm{Id}}

\newcommand{\NOT}{\mathrm{NOT}}
\newcommand{\CNOT}{C\NOT}
\newcommand{\CZ}{CZ}
\newcommand{\SWAP}{\mathrm{SWAP}}
\newcommand{\BigO}{\mathcal{O}}
\newcommand{\WH}{W\!H}
\newcommand{\Deph}{\mathcal{D}}

\newcommand{\StabGroup}{\mathcal{S}}
\newcommand{\PauliGroup}{\mathcal{P}}
\newcommand{\CliffordGroup}{\mathcal{C}\!\ell}

\begin{document}

%----------------------------------------------------------------------------------------
%  TITLE AND ABSTRACT
%----------------------------------------------------------------------------------------
\title{Further improvements to stabilizer simulation theory: classical rewriting of CSS-preserving stabilizer circuits, quadratic form expansions of stabilizer operations, and framed hidden variable models}

\author{Vsevolod I. Yashin}
\email{yashin.vi@mi-ras.ru}
\orcid{0000-0003-0309-8536}
\affiliation{Steklov Mathematical Institute of Russian Academy of Sciences, Moscow 119991, Russia}
\affiliation{Russian Quantum Center, Skolkovo, Moscow 143025, Russia}

\author{Vladimir V. Yatsulevich}
% \email{iatsulevich.vv@phystech.edu}
\affiliation{Moscow Institute of Physics and Technology, Dolgoprudny 141700, Russia}

\author{Aleksey K. Fedorov}
% \email{akf@rqc.ru}
\orcid{0000-0002-4722-3418}
\affiliation{Russian Quantum Center, Skolkovo, Moscow 143025, Russia}
\affiliation{National University of Science and Technology ``MISIS'', Moscow 119049, Russia}

\author{Evgeniy O. Kiktenko}
% \email{e.kiktenko@rqc.ru}
\orcid{0000-0001-5760-441X}
\affiliation{Russian Quantum Center, Skolkovo, Moscow 143025, Russia}
\affiliation{Steklov Mathematical Institute of Russian Academy of Sciences, Moscow 119991, Russia}
\affiliation{National University of Science and Technology ``MISIS'', Moscow 119049, Russia}

% Watch for this option!!!
\date{November 7, 2025}

\begin{abstract}
  Simulation of stabilizer circuits is a well-studied problem in quantum information processing, with a number of highly optimized algorithms available. Yet, we argue that further improvements can arise from the theoretical structure of stabilizer operations themselves. We focus on the subclass of stabilizer circuits composed of Calderbank-Shor-Steane (CSS)-preserving stabilizer operations, which naturally appear in fault-tolerant computations over CSS stabilizer codes. Using elementary circuit transformation techniques, we show that such circuits can be exactly rewritten as classical probabilistic circuits that reproduce measurement statistics. This rewriting introduces no computational overhead, in contrast to the general case of stabilizer circuits. To clarify the origin of this simplification, we introduce the standard quadratic form representation of general stabilizer operations (Clifford channels). It provides an efficient way to describe compositions of stabilizer operations and thus to simulate stabilizer circuits. CSS-preserving operations correspond to purely linear forms, which under a Walsh-Hadamard-Fourier transform yield a noncontextual hidden variable model, providing an alternative proof of the introduced rewriting. Finally, we develop a theory of reference frames for multiqubit systems, where frames are encoded by quadratic forms. This allows us to express stabilizer operations as probabilistic maps for proper reference frames. Non-CSS-preserving stabilizer circuits require dynamical modifications of reference frames, embodying a contextuality resource that leads to the computational overhead. This framework provides a new perspective on simulating stabilizer and near-stabilizer circuits within dynamically evolving quasiprobability models.
\end{abstract}

\maketitle

%----------------------------------------------------------------------------------------
%  MAIN TEXT
%----------------------------------------------------------------------------------------
\section{Introduction} \label{sec:introduction}

Most of relevant quantum error-correcting codes on qubits belong to the family of stabilizer codes, meaning they are constructed from stabilizer operations and described within the stabilizer formalism \cite{Gottesman_1997, Nielsen_2010}. The celebrated Gottesman–Knill theorem states that stabilizer circuits can be efficiently simulated on a classical computer, which enables tractable modelling of error propagation and syndrome decoding. A variety of approaches exist for simulating multiqubit stabilizer circuits: stabilizer tableau methods \cite{Gottesman_1998, Aaronson_2004, Gidney_2021}, graph-state representations \cite{Anders_2006, Rijlaarsdam_2020, Hu_2022, Pang_2025}, ZX-calculus \cite{Backens_2014,Kissinger_2022,Booth_2022,Booth_2024,Comfort_2023,Booth_2025}, quadratic form expansions \cite{Dehaene_2003, Van_den_Nest_2010, Bravyi_2016, de_Beaudrap_2022, Amy_2023}, Wigner function and other quasiprobability representations \cite{Gross_2006, Veitch_2012, Raussendorf_2020, Park_2024, Pashayan_2015, Kulikov_2024}, and hidden variable models \cite{Raussendorf_2017, Zurel_2020}.

Typically, for a circuit on $n$ qubits of size $L$, to store a stabilizer state in a computer it requires $\BigO(n^2)$ memory. A local Clifford gate can be applied in $\BigO(n)$ time, while a measurement update takes $\BigO(n^2)$. Hence, drawing a single measurement sample (bit string of length $k$) from an adaptive stabilizer circuit (with classical control) costs $\BigO(nL + kn^2)$. In contrast, for a non-adaptive stabilizer circuit one can first compile the entire circuit into a compact representation -- such as a stabilizer tableau of rank $r$ -- in $\BigO(nL)$ time, and then perform weak simulation (sampling $k$ bits) in $\BigO(k r)$ time or strong simulation (computing probabilities of specified outcomes) in $\BigO(kr^2)$ time.

From the complexity-theoretic standpoint, the simulation of stabilizer circuits can be efficiently reduced to simulating classical circuits; non-adaptive stabilizer circuits, in particular, correspond to affine Boolean circuits \cite{Aaronson_2004, Buhman_2006}. The key observation is that sign-free stabilizer tableau updates are affine, while handling phase signs requires only logarithmic additional space (for huge circuits this overhead may become significant in practice).

Refs.~\cite{Johansson_2017, Johansson_2019} proposed a direct rewriting of an $n$-qubit circuit into a $2n$-bit classical circuit without explicit sign processing. Although the resulting classical model does not reproduce all quantum statistics exactly, it captures certain quantum-like behaviour and is closely related to Spekkens' toy model \cite{Spekkens_2007}.

Many practically useful stabilizer codes belong to the Calderbank–Shor–Steane (CSS) family \cite{Calderbank_1996, Steane_1996}, such codes extensively employ CSS-preserving stabilizer operations \cite{Delfosse_2015, Alexander_2023}. It is convenient that CSS codes allow independent treatment of $Z$- and $X$-type errors and admit transversal implementations of logical $\CNOT$ gates.
In this work, we demonstrate that for a CSS-preserving stabilizer circuit -- that is, a circuit composed solely of CSS-preserving operations -- the sign-handling computations become trivial, as the rewriting into a classical circuit is exact. The correctness of this rewriting follows from elementary circuit-transformation techniques. Sampling from the resulting classical circuit requires only $\BigO(L)$ time. Moreover, if the circuit is additionally affine -- as in the non-adaptive case -- it can be compiled in $\BigO(nL)$ time into a Boolean matrix, enabling weak simulation in $\BigO(kn)$ and strong simulation in $\BigO(kn^2)$. For general (non–CSS-preserving) circuits, however, the rewriting in not always correct.

To gain deeper insight into the structure of stabilizer circuits, it is instructive to dive into the mathematical foundations of stabilizer formalism. Recent works \cite{Yashin_2025, Yashin_2025_2, Kliuchnikov_2023, Heimendahl_2022} have introduced and characterized the quantum channels generated by stabilizer circuits -- hereafter referred to as Clifford channels. Abstractly, such channels map stabilizer states to stabilizer states and have a natural analogy with Gaussian bosonic channels \cite{Weedbrook_2012, Bu_2022, Bu_2023, Bu_2023_2, Yashin_2025}. A stabilizer circuit can thus be viewed as a composition of elementary Clifford channels.

The theory of Clifford channels relies on the algebraic structure of Pauli groups, which can be regarded as group extensions over abelian groups \cite{Ceccherini-Silberstein_2022}. We argue that the natural mathematical framework for qubit stabilizer groups involves $\mathbb{Z}_4$-valued quadratic functions on Boolean vector spaces. Within this setting, any Clifford channel admits a quadratic form expansion, for which we introduce a particularly convenient standard representation. Such representations are compact, easily updated under composition, and lead to efficient linear-algebraic description of stabilizer-circuit simulation.

We find that CSS-preserving Clifford channels correspond to cases where the quadratic term vanishes, leaving only linear part and thereby explaining their simplified structure. Applying the Walsh–Hadamard–Fourier transform over the Pauli group (as in discrete Wigner function theory \cite{Gross_2006, Delfosse_2015}) turns CSS-preserving channels into Markov kernels implementing probabilistic affine maps on phase space, which can be directly read off from the standard quadratic form expansion. This provides an alternative correctness proof of the classical rewriting: the Walsh–Hadamard-Fourier transform defines a non-contextual hidden variable model for CSS-preserving circuits.

Finally, we extend this perspective beyond the CSS case by introducing a simple framework for reference frames on multiqubit systems, inspired by \cite{Park_2024}. In our formulation, reference frames are encoded by quadratic forms. Every stabilizer state becomes CSS in an appropriate reference frame, and every Clifford channel is CSS-preserving with respect to suitable input and output reference frames. Thus, any stabilizer circuit admits a classical model when viewed through the correct sequence of reference frames in time. This yields a contextual hidden variable model, where the context is precisely the choice of a reference frame. In this picture, stabilizer circuit simulation amounts to the execution of a classical circuit accompanied by reference frame processing. This approach may also enhance simulation of near-stabilizer circuits by leveraging techniques from quasiprobability models and sum-over-Clifford decompositions \cite{Bravyi_2019}.

\subsection{Main contributions} \label{subsec:main_results}

Our work advances the theory of stabilizer simulation in several complementary directions. Below we summarize the main findings.

\begin{itemize}
  \item We present a set of rewriting rules that transform any $n$-qubit stabilizer circuit $\mathsf{QC}$ into a $2n$-bit classical circuit $\mathsf{CC}$ (\cref{subsec:rewriting_rules}). For CSS-preserving circuits, the output statistics of $\mathsf{QC}$ are exactly reproduced by $\mathsf{CC}$. We provide two independent proofs of this correspondence (\cref{subsec:circuit_proof,subsubsec:symbols_of_CSS}).

  \item We enhance stabilizer formalism by introducing quadratic form expansions for describing (not necessarily maximal) stabilizer groups, mixed stabilizer states, and arbitrary non-adaptive stabilizer operations (\cref{subsec:quadratic_form_expansions}). We propose a standard quadratic form expansion that elegantly characterizes Clifford channels and is particularly convenient for stabilizer simulation (\cref{subsec:quadratic_form_simulation}).

  \item We develop a theory of reference frames encoded by quadratic forms, applicable for multiqubit systems (\cref{subsec:reference_frames}). This framework enables the construction of a contextual hidden variable model that incorporates non-CSS-preserving stabilizer operations, where reference frames serve as contexts. In this model, simulating an arbitrary stabilizer circuit involves rewriting stabilizer circuit $\mathsf{QC}$ to classical circuit $\mathsf{CC}$ and performing additional computations over reference frames (\cref{subsubsec:frames_simulation}). The same construction naturally extends to near-stabilizer circuits (\cref{subsec:magic_simulation}).
\end{itemize}

\begin{figure*}[ht]
  \centering
  \resizebox{\linewidth}{!}{
  \begin{tikzpicture}
    \tikzset{
      node distance=1.8cm,
      base/.style = {
        rectangle, rounded corners, minimum width=6cm, minimum height=3.0cm,
        draw=black, thick, fill=black!5
      }
    }

    % left and right parts
    \node (Q) {};
    \node (C) [right =7.7cm of Q] {};

    % box of stabilizer circuit
    \node (Q-Circuit) [base, above of=Q] {};
      \node[anchor=north west] at (Q-Circuit.north west) {stabilizer circuit $\mathsf{QC}$};
      \node at (Q-Circuit)[yshift=-0.5em] {
        \begin{quantikz}[wire types = {q,q}, row sep={1.2cm,between origins}, column sep={0.9cm,between origins}, align equals at=1.5]
          \lstick{$\ket{+}$}&\ctrl{1} &\ctrl{1} &\ctrl{1} &\meterD{X} &\setwiretype{c} \\
          \lstick{$\ket{0}$}&\targ{}  &\ctrl{0} &\targ{}  &\ground{}
        \end{quantikz}
      };

    % box of classical circuit
    \node (C-Circuit) [base, above of=C] {};
      \node[anchor=north west] at (C-Circuit.north west) {classical circuit $\mathsf{CC}$};
      \node at (C-Circuit)[yshift=-0.5em] {
        \begin{quantikz}[wire types={cz,cx,cz,cx}, row sep={0.35cm,between origins}, column sep={0.55cm,between origins}, align equals at=2.5]
          \lstick{$z$} &\ctrl{0}\wire[d][2]{c} &                      &[0.2cm]\ctrl{0}\wire[d][3]{c} &                      &[0.2cm]\ctrl{0}\wire[d][2]{c} &                      &\ground{} &\setwiretype{n} \\
          \lstick{$0$} &                       &\targ{}\wire[d][2]{c} &                              &\targ{}\wire[d][1]{c} &                              &\targ{}\wire[d][2]{c} &          &\setwiretype{c}    \\[0.6cm]
          \lstick{$0$} &\targ{}                &                      &                              &\ctrl{0}              &\targ{}                       &                      &\ground{} &\setwiretype{n} \\
          \lstick{$x$} &                       &\ctrl{0}              &\targ{}                       &                      &                              &\ctrl{0}              &\ground{} &\setwiretype{n}
        \end{quantikz}
      };

    % arrow "rewriting"
    \draw[->,thick, decorate,decoration={zigzag, pre length=3mm, post length=3mm}]
      ($(Q-Circuit.east) + (2mm,-1mm)$)
      -- node[above] {rewriting}
      ($(C-Circuit.west) + (-2mm,-1mm)$);

      % box of left classical computer
    \node (Q-Control) [rectangle, inner sep=1.2em, below of=Q] {\includegraphics[width=2.0cm]{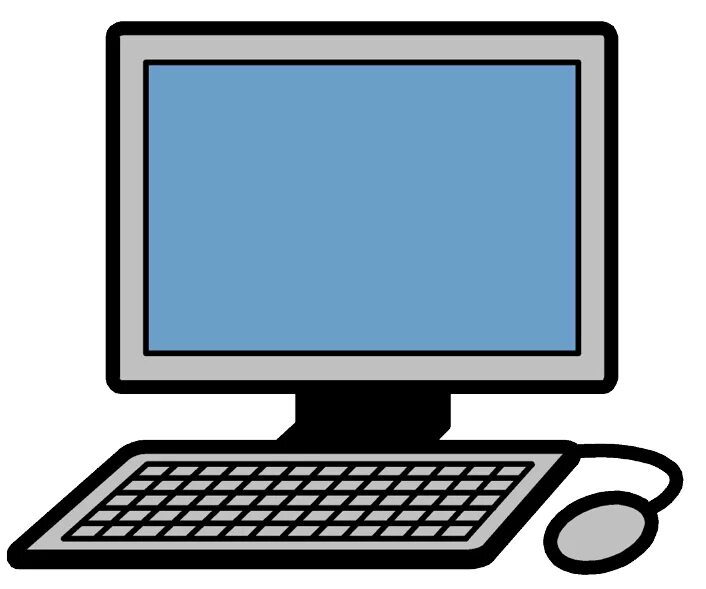}};
      \node[anchor=south] at (Q-Control.south) {classical controller};
    \draw[<-,thick] ($(Q-Control.north) + (-1mm,-3mm)$) -- ($(Q-Circuit.south) + (-1mm,-2mm)$);
    \draw[->,thick] ($(Q-Control.north) + (1mm,-3mm)$) -- node[right]{communication} ($(Q-Circuit.south) + (+1mm,-2mm)$);

    % box of right classical computer
    \node (C-Control) [rectangle, inner sep=1.2em, below of=C] {\includegraphics[width=2.0cm]{fig_controller}};
      \node[anchor=south] at (C-Control.south) {classical controller};
      \node[anchor=west] at ($(C-Control.east) + (-6mm,+1mm)$)(C-Control.east) {+};
      \node[anchor=west,align=center] at ($(C-Control.east) + (-3mm,+1mm)$)(C-Control.east) {reference\\ frames\\ processing};
    \draw[<-,thick] ($(C-Control.north) + (-1mm,-3mm)$) -- ($(C-Circuit.south) + (-1mm,-2mm)$);
    \draw[->,thick] ($(C-Control.north) + (1mm,-3mm)$) -- node[right]{communication} ($(C-Circuit.south) + (+1mm,-2mm)$);
  \end{tikzpicture}
  }
  \caption{
    \emph{Schematic representation of the main setting.}
    A quantum stabilizer circuit $\mathsf{QC}$ interacts with a classical controller through rounds of communication. Each round consists of sending measurement outcomes from the quantum circuit to the classical computer, which processes this information and sends back control parameters for conditional gates; in general, there may be multiple (or no) communication rounds. We rewrite $\mathsf{QC}$ to a classical circuit $\mathsf{CC}$. If the stabilizer circuit $\mathsf{QC}$ is CSS-preserving, then $\mathsf{CC}$ exactly reproduces its output statistics, and the interaction with the controller remains unchanged after the rewriting. If $\mathsf{QC}$ includes non-CSS-preserving operations, the controller must additionally compute information about the reference frames to ensure correctness. These reference frame updates are determined by the quadratic form expansions of the non-CSS-preserving gates. The example circuits used in the figure are taken from \cref{subsec:non-CSS_incorrectness} and further discussed in \cref{subsubsec:frames_as_contexts}.
  }
  \label{fig:main_idea}
\end{figure*}

A schematic illustration of the main setting is shown in \cref{fig:main_idea}.

\subsection{Key previous works} \label{subsec:previous_results}

This work builds upon several earlier studies that provide both theoretical and methodological foundations for our results. Below we briefly outline the most relevant contributions.
\begin{enumerate}
  \item Ref.~\cite{Delfosse_2015} introduced CSS-preserving stabilizer operations and developed the theory of rebit Wigner functions.
  \item Ref.~\cite{Johansson_2019} proposed a set of rewriting rules for transforming quantum circuits into classical ones, referred to as \emph{Quantum Simulation Logic}.
  \item The abstract theory of stabilizer operations (Clifford channels) was formulated in Ref.~\cite{Yashin_2025}.
  \item The idea of employing reference frames to construct Wigner functions for arbitrary stabilizer operations was suggested in Ref.~\cite{Park_2024}.
\end{enumerate}

Additional related works are cited throughout the paper where appropriate.

\subsection{Structure} \label{subsec:structure}

We conclude the Introduction by outlining the structure of the paper. Each Section begins with a short introductory paragraph summarizing its main purpose.

In \cref{sec:rewriting}, we provide preliminaries on stabilizer circuits (\cref{subsec:preliminaries_circuits}) and introduce a set of rewriting rules that reduce CSS-preserving stabilizer operations to classical gates (\cref{subsec:rewriting_rules}). We prove the correctness of this reduction (\cref{subsec:circuit_proof}) and discuss its failure for non-CSS-preserving circuits (\cref{subsec:non-CSS_incorrectness}), as well as describe how to simulate the resulting classical circuits (\cref{subsec:classical_simulation}). \hyperref[appendix:surface_code]{Appendix~A} presents an explicit example of a classical circuit corresponding to a CSS-preserving stabilizer circuit implementing a single-cell defect movement in the surface code, while \hyperref[appendix:simulator]{Appendix~B} reports about the performance of a simple circuit-rewriting simulator.

\Cref{sec:stabilizer} begins with the review of stabilizer formalism and Clifford channels (\cref{subsec:preliminaries_stabilizer}). We then introduce $\mathbb{Z}_4$-valued quadratic forms and show that any stabilizer operation can be represented as a quadratic form expansion (\cref{subsec:quadratic_form_expansions}). We propose a standard form for such expansions and study how these forms evolve during computation. In \hyperref[appendix:trace-decreasing_channels]{Appendix~C}, we extend the analysis to trace-decreasing Clifford channels and show how to compose their standard quadratic form expansions. We redefine the class of CSS-preserving stabilizer operations, showing that they correspond precisely to operations with purely linear term in their quadratic form representation (\cref{subsec:CSS_operations}). By storing each circuit element in the standard quadratic form expansion, we obtain a new simulation procedure that iteratively constructs and simplifies the overall standard quadratic form expansion of the circuit (\cref{subsec:quadratic_form_simulation}).

In \cref{sec:hidden_variables}, we first discuss the noncontextual hidden variable interpretation of CSS-preserving stabilizer circuits (\cref{subsec:CSS_hidden_variables}). We then introduce a reference frame formalism encoded by quadratic forms (\cref{subsec:reference_frames}), which allows us to define contextual hidden variable models for general stabilizer circuits and to study their simulation (\cref{subsec:contextual_hidden_variables}). Finally, we apply the resulting quasiprobability formalism to the simulation of stabilizer circuits with magic states (\cref{subsec:magic_simulation}).

\Cref{sec:conclusion} summarizes the results and discusses possible directions for future research.

\section{Rewriting CSS-preserving stabilizer circuits to classical circuits} \label{sec:rewriting}

In this Section, we discuss how CSS-preserving stabilizer circuits can be reduced to classical circuits. In \cref{subsec:preliminaries_circuits}, we discuss basic classes of quantum and classical circuits and how they are related. In \cref{subsec:rewriting_rules}, we propose a set of rewriting rules transforming stabilizer operations to classical gates, and in \cref{subsec:circuit_proof} we prove that rewriting the whole CSS-preserving stabilizer circuit results in a classical circuit with the same outputs. We employ only elementary circuit-transformation techniques in this proof. In \cref{subsec:non-CSS_incorrectness}, we give an example of the rewriting of non-CSS-preserving circuit which gives incorrect output. In \cref{subsec:classical_simulation}, we discuss methods for simulating classical circuits.

\subsection{Preliminaries on stabilizer circuits} \label{subsec:preliminaries_circuits}

By \emph{classical circuit} we understand a Boolean circuit made of wires transmitting bits and Boolean operations between them. \emph{Probabilistic circuits} allow inputs of random bits with uniform probability $1/2$. We will denote random bits by lowercase latin letters. If a circuit is made of bit addition operations $\{\NOT, \CNOT\}$, then this circuit is called \emph{affine}, its action can be described using Boolean matrices.

We call \emph{stabilizer circuit} a quantum circuit constructed of stabilizer operations: stabilizer states initializations, Clifford unitaries, Pauli measurements, etc. If we additionally allow a classical control via classical probabilistic circuits, we call such circuits \emph{adaptive}. Any stabilizer circuit can be composed of: computational basis state $\ket{0}$ initializations, unitary gates from dictionary $\{H,S,\CNOT\}$, destructive one-qubit measurements $\{\ketbra{0}{0},\ketbra{1}{1}\}$ in $Z$-basis, qubit discarding channels $\rho\mapsto\Tr(\rho)$, and possibly a classical control. Non-destructive measurements can be performed using ancillae. Clifford circuits are known to have simple algebraic structure and are usually described using \emph{stabilizer formalism} \cite{Gottesman_1997, Gottesman_1998, Aaronson_2004}.

\emph{Rebit} (real qubit) is a system of one-qubit states with real coefficients in computational basis $\{\ket{0},\ket{1}\}$. Real quantum operations are the ones that preserve real coefficients in computational basis. Real quantum circuits are known to be computationally universal: any quantum circuit on $n$ qubits may be reduced to a real quantum circuit on $n+1$ rebits \cite{Bernstein_1997, Shi_2003, Rudolph_2002, McKague_2009}. Real Clifford circuits are generated by $\ket{0}$ initialization, real Clifford unitaries $\{H,Z,\CNOT\}$, $Z$-basis measurements, qubit discarding channels, and classical control. Accordingly, real stabilizer circuits are the qubit stabilizer circuits without $S$ gate \cite{Hashagen_2018, Hickey_2018}.

There is an important subclass of \emph{CSS-preserving} rebit stabilizer circuits \cite{Delfosse_2015, Alexander_2023}. These are the circuits without standalone $H$ gate, but with gate $\WH$ (Walsh–Hadamard transform) which acts as $H$ on all rebits together. To be precise, such circuits may be composed of: initialization of $\ket{0}$ and $\ket{+}$, one-rebit gates $X$ and $Z$, two-qubit gate $\CNOT$, Walsh–Hadamard transform $\WH$ on all rebits, $X$- and $Z$-basis measurements, discarding channels, and classical control. (However, in \cref{subsec:CSS_operations} we will discuss why this list of elementary operations can be naturally extended.) The gate $CZ$ is not CSS-preserving. As an important example, CSS-preserving stabilizer operations are those operations that can be fault-tolerantly implemented in surface codes using defect braidings \cite{Raussendorf_2007, Fowler_2012}.

\begin{figure}[h]
  \centering
  \resizebox{0.45\textwidth}{!}{
  \begin{tikzpicture}[font=\sffamily]
    \def\qubitellipse{(0pt,15pt) ellipse (140pt and 70pt)}
    \fill[blue!5!white] \qubitellipse;
    \draw[thick] \qubitellipse;
    \node at (0pt,70pt) {multiqubit operations};

    \def\realellipse{(-30pt,5pt) ellipse (90pt and 50pt)};
    \def\stabilizerellipse{(30pt,5pt) ellipse (90pt and 50pt)};
    \fill[blue!10!white] \realellipse;
    \fill[blue!10!white] \stabilizerellipse;
    \begin{scope}
      \clip \realellipse;
      \fill[blue!15!white] \stabilizerellipse;
    \end{scope}
    \draw \realellipse;
    \draw \stabilizerellipse;
    \node[rotate=40] at (-75pt,20pt) {real operations};
    \node[rotate=-40] at (85pt,20pt) {\begin{tabular}{c}stabilizer\\ operations\end{tabular}};

    \def\CSSellipse{(0pt,0pt) ellipse (50pt and 30pt)};
    \fill[blue!20!white] \CSSellipse;
    \draw \CSSellipse;
    \node at (0pt,0pt) {\begin{tabular}{c}CSS-preserving\\ operations\end{tabular}};
  \end{tikzpicture}
  }
  \caption{
    Venn diagram showing some classes of quantum operations that are relevant to our discussion.
  }
  \label{fig:venn_diagram}
\end{figure}
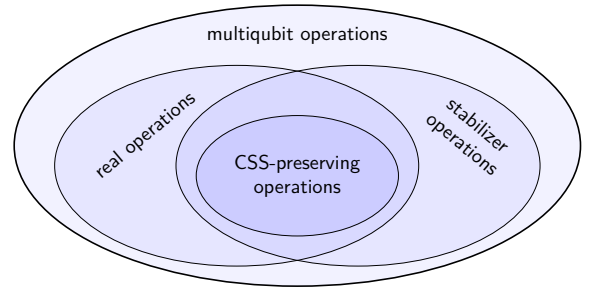

\subsubsection{R-I rebit and state injeciton protocols} \label{subsubsec:state_injection_protocols}

It is possible to reduce a qubit stabilizer circuit $\mathsf{QC}$ to a rebit stabilizer circuit $\mathsf{RC}$. To do it, one adds an additional qubit (R-I rebit) initialised to a state $\ket{0}$ and discarded in the end. This additional qubit describes real and imaginary parts of coefficients \cite{Rudolph_2002}. Instances of gate $S$ then translate to Clifford two-qubit gates $\CNOT\cdot CZ$ with R-I rebit as target:
\begin{equation}
  \adjustbox{valign=m}{$
  \begin{quantikz}[align equals at=1]
    &\gate{S} &\rstick{qubit} \\
    &\setwiretype{n} &
  \end{quantikz}
  \;\leadsto\;
  \begin{quantikz}[align equals at=1]
    &\ctrl{1} &\ctrl{1} &\rstick{rebit} \\
    &\ctrl{0} &\targ{}  &\rstick{R-I rebit}
  \end{quantikz}
  $}
\end{equation}
The resulting circuit $\mathsf{RC}$ acts on $n+1$ rebits and consists of only real stabilizer elements. If the initial circuit $\mathsf{QC}$ itself consists only of real stabilizer elements, then the additional R-I rebit remains disentangled. So, rebit stabilizer circuits are computationally equivalent to qubit stabilizer circuits. Note, however, that real and complex quantum mechanics are distinguishable from the perspective of network communications \cite{Renou_2021, Sarkar_2025}.

Besides, one can realise any complex quantum computation, having the ability to perform real operations and preparations of imaginary state $\ket{+i} = S\ket{+}$, by the following state injection gadget \cite{Zhou_2000}:
\begin{equation} \label{eq:S_injection}
  \begin{quantikz}[row sep={0.7cm,between origins},column sep={0.8cm,between origins}]
    \lstick{$\ket{+i}$}   &\targ{}   &\meterD{Z} &\ctrl{0}\wire[d][1]{c}\setwiretype{c} &\setwiretype{n} \\
    \lstick{$\ket{\psi}$} &\ctrl{-1} &           &\gate{Z}                              &\rstick{$S\ket{\psi}$}
  \end{quantikz}
\end{equation}
The ability to prepare $\ket{+i}$ can be treated as a resource of imaginarity in quantum mechanics \cite{Hickey_2018}. We will also discuss imaginarity in \cref{subsubsec:non-CSS-ness}.

Likewise, arbitrary rebit stabilizer circuits can be reduced to CSS-preserving circuits with state injection \cite{Delfosse_2015}. More concretely, Hadamard gate $H$ or gate $\CZ$ may be realised by preparing the state $\ket{\CZ}$ (the graph state on two qubits) and using the gate teleportation protocols \cite{Gottesman_1999}:
\begin{gather}
  \ket{\CZ} = \frac{\ket{00}+\ket{01}+\ket{10}-\ket{11}}{2}
  \\
  \begin{quantikz}[row sep={0.7cm, between origins}, column sep={0.9cm,between origins}, align equals at=2]
    \lstick{$\ket{\psi}$} &\targ{}   &\meterD{Z} &\setwiretype{c}                       &\ctrl{0}\wire[d][2]{c}\setwiretype{c} &\setwiretype{n}       \\
    \lstick[2]{$\ket{\CZ}$} &\ctrl{-1} &\meterD{X} &\ctrl{0}\wire[d][1]{c}\setwiretype{c} &\setwiretype{n}                       &                      \\
                          &          &           &\gate{X}                              &\gate{Z}                              &\rstick{$H\ket{\psi}$}
  \end{quantikz}
  \label{eq:H_injection}
  \\
  \begin{quantikz}[row sep={0.7cm, between origins}, column sep={0.6cm,between origins}, align equals at=2.5]
    \lstick[2]{$\ket{\CZ}$}  &\targ{}   &&[0.5cm]\meterD{Z} &\setwiretype{c}                       &\ctrl{0}\wire[d][3]{c}\setwiretype{c} &\setwiretype{n}\\
                             &&\targ{}   &\meterD{Z} &\ctrl{0}\wire[d][1]{c}\setwiretype{c} &\setwiretype{n}                       & \\
    \lstick[2]{$\ket{\psi}$} &\ctrl{-2} &&           &\gate{Z}                              &                                      &\rstick[2]{$\CZ\ket{\psi}$} \\
                             &&\ctrl{-2} &           &                                      &\gate{Z}                              & \\
  \end{quantikz}
  \label{eq:CZ_injection}
\end{gather}
So, availability of individual Hadamard gates $H$ or gates $\CZ$ can be viewed as a resource for (not too significant) computational advantage of arbitrary stabilizer circuits over CSS-preserving circuits. This distinction also manifests itself in properties of stabilizer formalism, we will discuss it in \cref{subsec:CSS_operations,sec:hidden_variables}. For deeper elaboration on this resource see \cref{subsubsec:non-CSS-ness}. Also note that Hadamard gates are important in the resource theory of coherence \cite{Jones_2024}.

\subsection{Rewriting rules} \label{subsec:rewriting_rules}

Suppose we have a CSS-preserving stabilizer rebit circuit $\mathsf{RC}$ on $n$ qubits with size $L$ (i.e., the number of elements in the circuit); all outputs are either discarding channels or classical bits obtained after measurements and classical processing, so there are no quantum states on the output; we also can allow using classical bit inputs. Let us show how to rewrite the circuit $\mathsf{RC}$ to a classical circuit $\mathsf{CC}$ reproducing output statistics.

To each rebit we correspond $2$ classical bits: the first is interpreted to represent the result of $Z$-measurement on this rebit, the second represents the result of $X$-measurement:
\begin{equation}
  \begin{quantikz}[align equals at=1]
    & &\rstick{rebit}
  \end{quantikz}
  \quad\leadsto\quad
  \begin{quantikz}[wire types={cz,cx}, row sep={0.4cm,between origins}, align equals at=1.5]
    & &\rstick{$Z$ bit} \\
    & &\rstick{$X$ bit}
  \end{quantikz}
\end{equation}
For the sake of visual clarity, here and thereafter we draw usual classical bit wires in black, $Z$-bits in blue and $X$-bits in red.

The initial state $\ket{0}$ preparation corresponds to $0$ value on the first bit and a uniformly random value $x$ on the second bit. Other one-rebit stabilizer states are rewritten analogously:
\begin{align}
  \begin{quantikz}
    \lstick{$\ket{0}$} &
  \end{quantikz}
  &\quad\leadsto\quad
  \begin{quantikz}[wire types={cz,cx}, row sep={0.4cm,between origins}]
    \lstick{$0$} & \\
    \lstick{$x$} &
  \end{quantikz}
  \label{eq:prep0-rewriting}
  \\
  \begin{quantikz}
    \lstick{$\ket{1}$} &
  \end{quantikz}
  &\quad\leadsto\quad
  \begin{quantikz}[wire types={cz,cx}, row sep={0.4cm,between origins}]
    \lstick{$1$} & \\
    \lstick{$x$} &
  \end{quantikz}
  \label{eq:prep1-rewriting}
  \\
  \begin{quantikz}
    \lstick{$\ket{+}$} &
  \end{quantikz}
  &\quad\leadsto\quad
  \begin{quantikz}[wire types={cz,cx}, row sep={0.4cm,between origins}]
    \lstick{$z$} & \\
    \lstick{$0$} &
  \end{quantikz}
  \label{eq:prepP-rewriting}
  \\
  \begin{quantikz}
    \lstick{$\ket{-}$} &
  \end{quantikz}
  &\quad\leadsto\quad
  \begin{quantikz}[wire types={cz,cx}, row sep={0.4cm,between origins}]
    \lstick{$z$} & \\
    \lstick{$1$} &
  \end{quantikz}
  \label{eq:prepM-rewriting}
\end{align}
The initialization of a chaotic state $\chi = \frac{1}{2}I$ is expressed as using two random bits $z$ and $x$:
\begin{equation}
  \begin{quantikz}
    \lstick{$\chi$} &
  \end{quantikz}
  \quad\leadsto\quad
  \begin{quantikz}[wire types={cz,cx}, row sep={0.4cm,between origins}]
    \lstick{$z$} & \\
    \lstick{$x$} &
  \end{quantikz}
\end{equation}

The action of $Z$ or $X$ gates changes the result of $X$ or $Z$ values correspondingly:
\begin{gather}
  \begin{quantikz}
    &\gate{Z} &
  \end{quantikz}
  \quad\leadsto\quad
  \begin{quantikz}[wire types={cz,cx}, row sep={0.4cm,between origins}]
    &        & \\
    &\targ{} &
  \end{quantikz}
  \label{eq:Z-rewriting}
  \\
  \begin{quantikz}
    &\gate{X} &
  \end{quantikz}
  \quad\leadsto\quad
  \begin{quantikz}[wire types={cz,cx}, row sep={0.4cm,between origins}]
    &\targ{} & \\
    &        &
  \end{quantikz}
  \label{eq:X-rewriting}
\end{gather}

It is often useful to introduce full dephasing gates in $Z$-basis and $X$-basis:
\begin{equation}
\begin{aligned}
  &\Deph_Z[\rho] = \frac{1}{2}\left(\rho + Z \rho Z\right), \\
  &\Deph_X[\rho] = \frac{1}{2}\left(\rho + X \rho X\right).
\end{aligned}
\end{equation}
These channels are CSS-preserving because can be realized by $\CNOT$-gates and ancillae. Dephasing gates rewrite to the process of spoofing a classical bit by a random bit:
\begin{gather}
  \begin{quantikz}
    &\gate{\Deph_Z} &
  \end{quantikz}
  \quad\leadsto\quad
  \begin{quantikz}[wire types={cz,cx}, row sep={0.4cm,between origins}, column sep={0.5cm,between origins}]
    &[0.1cm]          &[0.3cm]                            &                 \\
    &\ground{} &\lstick{$x$}\setwiretype{n} &\setwiretype{cx}
  \end{quantikz}
  \label{eq:dephZ-rewriting}
  \\
  \begin{quantikz}
    &\gate{\Deph_X} &
  \end{quantikz}
  \quad\leadsto\quad
  \begin{quantikz}[wire types={cz,cx}, row sep={0.4cm,between origins}, column sep={0.5cm,between origins}]
    &[0.1cm]\ground{} &[0.3cm]\lstick{$z$}\setwiretype{n} &\setwiretype{cz} \\
    &          &                            &
  \end{quantikz}
  \label{eq:dephX-rewriting}
\end{gather}
where $z,x$ are uniformly random bits.

The Walsh-Hadamard gate $\WH$, which is an application of Hadamard gates on all available rebits, rewrites to swaps on every pair of bits.
\begin{equation}
  \begin{quantikz}[wire types={q,n,q},row sep={0.6cm,between origins},align equals at=2.2]
    &\gate{H} & \\[-0.3em]
    &\vdots   & \\[+0.3em]
    &\gate{H} &
  \end{quantikz}
  \quad\leadsto\quad
  \begin{quantikz}[wire types={cz,cx,n,cz,cx}, row sep={0.4cm,between origins},align equals at=3.2]
    &\targX{}\wire[d][1]{c} & \\
    &\targX{}               & \\[-0.3em]
    &\vdots                 & \\[+0.3em]
    &\targX{}\wire[d][1]{c} & \\
    &\targX{}               & \\
  \end{quantikz}
  \label{eq:WH-rewriting}
\end{equation}

The $\CNOT$ gate corresponds to the Controlled-NOT operation on $Z$-bits and the same operation interchanged on $X$ bits, the $\SWAP$ gate rewrites to the swap of corresponding pairs of bits:
\begin{align}
  \begin{quantikz}[row sep={1.4cm,between origins}, column sep={0.6cm,between origins}, align equals at=1.5]
    &\ctrl{1} & \\
    &\targ{}  &
  \end{quantikz}
  &\quad\leadsto\quad
  \begin{quantikz}[wire types={cz,cx,cz,cx}, row sep={0.4cm,between origins}, column sep={0.6cm,between origins}, align equals at=2.5]
    &\ctrl{0}\wire[d][2]{c} &                       & \\
    &                       &\targ{}                & \\[0.6cm]
    &\targ{}                &                       & \\
    &                       &\ctrl{0}\wire[u][2]{c} &
  \end{quantikz}
  \label{eq:CNOT-rewriting}
  \\
  \begin{quantikz}[row sep={1.4cm,between origins}, column sep={0.6cm,between origins}, align equals at=1.5]
    &\targX{}\wire[d][1]{q} & \\
    &\targX{}               &
  \end{quantikz}
  &\quad\leadsto\quad
  \begin{quantikz}[wire types={cz,cx,cz,cx}, row sep={0.4cm,between origins}, column sep={0.6cm,between origins}, align equals at=2.5]
    &\targX{}\wire[d][2]{c} &                       & \\
    &                       &\targX{}\wire[d][2]{c} & \\[0.6cm]
    &\targX{}               &                       & \\
    &                       &\targX{}               &
  \end{quantikz}
  \label{eq:SWAP-rewriting}
\end{align}

Rebit measurement corresponds to the forgetting of neighbouring bit:
\begin{gather}
  \begin{quantikz}
    &\meterD{Z} &\setwiretype{c}
  \end{quantikz}
  \quad\leadsto\quad
  \begin{quantikz}[wire types={cz,cx}, row sep={0.4cm,between origins}]
    &          &\setwiretype{c} \\
    &\ground{} &\setwiretype{n}
  \end{quantikz}
  \label{eq:Zmeas-rewriting}
  \\
  \begin{quantikz}
    &\meterD{X} &\setwiretype{c}
  \end{quantikz}
  \quad\leadsto\quad
  \begin{quantikz}[wire types={cz,cx}, row sep={0.4cm,between origins}]
    &\ground{} &\setwiretype{n} \\
    &          &\setwiretype{c}
  \end{quantikz}
  \label{eq:Xmeas-rewriting}
\end{gather}
and the discarding channel corresponds to forgetting both bits:
\begin{equation}
  \begin{quantikz}
    &\ground{}
  \end{quantikz}
  \quad\leadsto\quad
  \begin{quantikz}[wire types={cz,cx}, row sep={0.4cm,between origins}]
    &\ground{} \\
    &\ground{}
  \end{quantikz}
\end{equation}

Finally, if there is some classical part to the rebit circuit, then it remains untouched, and the classical control on rebits corresponds to Boolean operations over classical circuit:
\begin{equation}
  \adjustbox{valign=m}{$
  \begin{quantikz}[wire types={c,q}, align equals at=2]
    &\ctrl{0}\wire[d][1]{c}                                        &\setwiretype{n} \\[0.15cm]
    &\gate[1][1.2cm][1.0cm]{\substack{\text{quantum}\\ \text{gate}}} &
  \end{quantikz}
  \quad\leadsto\quad
  \begin{quantikz}[wire types={c,cz,cx}, row sep={0.4cm,between origins}, align equals at=2.5]
    &\ctrl{0}\wire[d][1]{c}                                             &\setwiretype{n} \\[0.6cm]
    &\gate[2][1.2cm][0.55cm]{\substack{\text{classical}\\ \text{gate}}} &                \\
    &                                                                   &
  \end{quantikz}
  $}
  \label{eq:ctrl-rewriting}
\end{equation}

So, we can rewrite any CSS-preserving circuit $\mathsf{RC}$ to some classical circuit $\mathsf{CC}$ with the same output. It takes $\BigO(1)$ additional space and $\BigO(L)$ time to perform such reduction. If the quantum CSS-preserving circuit $\mathsf{RC}$ had $n$ initial rebits, then the corresponding classical circuit $\mathsf{CC}$ has $2n$ bits, where at least $n$ of them are random. One important property of such rewriting is its locality: rewritten elements stay in their place, the structre of the circuit network is not changed globally. Also note that Pauli errors in $\mathsf{QC}$ are expressed as bit-flip errors in $\mathsf{CC}$, which gives a simple classical model for studying error propagation in quantum error-correcting codes.

We present the rewriting of superdense coding and quantum teleportation circuits on \cref{fig:quantum_teleportation}. As an important example, in \hyperref[appendix:surface_code]{Appendix~A} we apply this reduction method to describe one-cell defect movement in surface code \cite{Fowler_2012}. In \cref{subsec:circuit_proof,subsubsec:CSS_hidden_variables_simulation}, we will argue for the correctness of such rewriting for CSS-preserving circuits, and in \cref{subsec:non-CSS_incorrectness} we show the incorrectness for non-CSS-preserving stabilizer circuits.
\begin{figure*}[t]
  (a)
  \vspace{-0.4cm}
  \begin{equation*}
    \begin{quantikz}[wire types={n,q,q}, row sep={1.4cm, between origins}, column sep={0.8cm,between origins}, align equals at=2.5]
                         &         &                                           &                                           &[1.2cm]  &           &                            \\[-0.8cm]
      \lstick{$\ket{+}$} &\ctrl{1} &\gate{Z}\wire[u][1]["a"{above,pos=1.0}]{c} &\gate{X}\wire[u][1]["b"{above,pos=1.0}]{c} &\ctrl{1} &\meterD{X} &\rstick{$a$}\setwiretype{c} \\
      \lstick{$\ket{0}$} &\targ{}  &                                           &                                           &\targ{}  &\meterD{Z} &\rstick{$b$}\setwiretype{c}
    \end{quantikz}
    \qquad\leadsto\qquad
    \begin{quantikz}[wire types={n,cz,cx,cz,cx}, row sep={0.4cm,between origins}, column sep={0.6cm,between origins}, align equals at=3.5]
                   &                       &        & &                                          &[1.0cm]                &        &          &       \\[0.0cm]
      \lstick{$z$} &\ctrl{0}\wire[d][2]{c} &        & &\targ{}\wire[u][1]["b"{above,pos=1.0}]{c} &\ctrl{0}\wire[d][2]{c} &        &\ground{} &\setwiretype{n}       \\
      \lstick{$0$} &                       &\targ{} &\targ{}\wire[u][2]["a"{above,pos=1.0}]{c} & &                       &\targ{} & &\rstick{$a$}\setwiretype{c} \\[0.6cm]
      \lstick{$0$} &\targ{} &                       &                                          & &\targ{} &                       & &\rstick{$b$}\setwiretype{c} \\
      \lstick{$x$} &        &\ctrl{0}\wire[u][2]{c} &                                          & &        &\ctrl{0}\wire[u][2]{c} &\ground{} &\setwiretype{n}
    \end{quantikz}
  \end{equation*}
  (b)
  \begin{equation*}
    \begin{quantikz}[row sep={1.4cm, between origins}, column sep={0.8cm,between origins}, align equals at=2]
      \lstick{$\ket{\psi}$} &         &[1.6cm]\ctrl{1} &\meterD{X} &\setwiretype{c}                       &\ctrl{0}\wire[d][2]{c} &\setwiretype{n}      \\
      \lstick{$\ket{+}$}    &\ctrl{1} &\targ{}         &\meterD{Z} &\ctrl{0}\wire[d][1]{c}\setwiretype{c} &\setwiretype{n}        &                     \\
      \lstick{$\ket{0}$}    &\targ{}  &                &           &\gate{X}                              &\gate{Z}               &\rstick{$\ket{\psi}$}
    \end{quantikz}
    \leadsto\qquad
    \begin{quantikz}[wire types={cz,cx,cz,cx,cz,cx}, row sep={0.4cm,between origins}, column sep={0.6cm,between origins}, align equals at=3.5]
      \lstick{$a$} &                       &                       &[1.0cm]\ctrl{0}\wire[d][2]{c} &                       &\ground{} &\setwiretype{n}        &                          &                \\
      \lstick{$b$} &                       &                       &                              &\targ{}                &          &\setwiretype{c}                       &\ctrl{0}\wire[d][4]{c}    &\setwiretype{n} \\[0.6cm]
      \lstick{$z$} &\ctrl{0}\wire[d][2]{c} &                       &\targ{}                       &                       &          &\ctrl{0}\wire[d][2]{c}\setwiretype{c} &\setwiretype{n}           &                \\
      \lstick{$0$} &                       &\targ{}                &                              &\ctrl{0}\wire[u][2]{c} &\ground{} &\setwiretype{n}        &                          &                \\[0.6cm]
      \lstick{$0$} &\targ{}                &                       &                              &                       &          &\targ{}                &                          &\rstick{$a$}    \\
      \lstick{$x$} &                       &\ctrl{0}\wire[u][2]{c} &                              &                       &          &                       &\targ{}                   &\rstick{$b$}
    \end{quantikz}
  \end{equation*}
  \caption{Simple examples of rewriting quantum circuits to classical circuits. (a) Reduction of the superdense coding protocol \cite{Bennett_1991} to classical circuit. (b) Reduction of the quantum teleportation circuit \cite{Bennett_1993} to classical circuit.}
  \label{fig:quantum_teleportation}
\end{figure*}
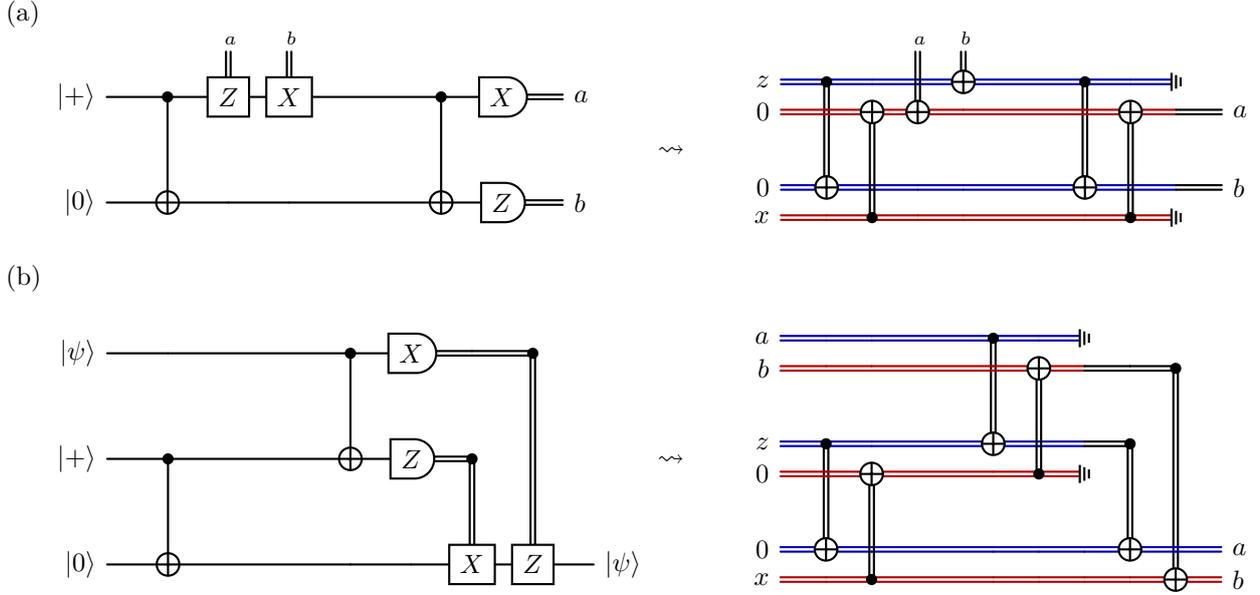

\subsubsection{Rules for non-CSS-preserving gates} \label{subsubsec:non-CSS_rules}

If we were to choose the rewriting rules for Hadamard gate $H$, phase gate $S$ and controlled-$Z$ gate $\CZ$, we would choose the following rules. Hadamard gate $H$ would correspond to interchange of $Z$ and $X$ bits:
\begin{equation}
  \begin{quantikz}
    &\gate{H} &
  \end{quantikz}
  \quad\leadsto\quad
  \begin{quantikz}[wire types={cz,cx}, row sep={0.4cm,between origins}]
    &\targX{}\wire[d][1]{c} & \\
    &\targX{}               &
  \end{quantikz}
  \label{eq:H-rewriting}
\end{equation}
and phase gate $S$ would correspond to adding $X$-bit to $Z$-bit:
\begin{equation}
  \begin{quantikz}
    &\gate{S} &
  \end{quantikz}
  \quad\leadsto\quad
  \begin{quantikz}[wire types={cz,cx}, row sep={0.4cm,between origins}]
    &\ctrl{1} & \\
    &\targ{}  &
  \end{quantikz}
  \label{eq:S-rewriting}
\end{equation}
and $\CZ$ would rewrite to addition of the first $Z$-bit to second $X$-bit and second $Z$-bit to first $X$-bit:
\begin{equation}
  \begin{quantikz}[row sep={1.4cm,between origins}, column sep={0.6cm,between origins}, align equals at=1.5]
    &\ctrl{1} & \\
    &\ctrl{0} &
  \end{quantikz}
  \quad\leadsto\quad
  \begin{quantikz}[wire types={cz,cx,cz,cx}, row sep={0.4cm,between origins}, column sep={0.6cm,between origins}, align equals at=2.5]
    &\ctrl{0}\wire[d][3]{c} &                       & \\
    &                       &\targ{}                & \\[0.6cm]
    &                       &\ctrl{0}\wire[u][1]{c} & \\
    &\targ{}                &                       &
  \end{quantikz}
  \label{eq:CZ-rewriting}
\end{equation}
Rewriting rules \cref{eq:H-rewriting,eq:S-rewriting,eq:WH-rewriting} lead to incorrect outputs for direct rewriting simulator (see \cref{subsec:non-CSS_incorrectness}), but will be used in modified simulator in \cref{subsubsec:frames_simulation}.

Let us note that the rewriting rules of \cref{subsec:rewriting_rules} [including rules \cref{eq:H-rewriting,eq:CZ-rewriting}, but slightly different \cref{eq:S-rewriting}] were first suggested in Refs.~\cite{Johansson_2017, Johansson_2019, Hindlycke_2022} under the name \emph{Quantum Simulation Logic}. The main goal of these works was to compare the query complexity of quantum algorithms in a quantum computer and the correpsonding Quantum Simulation Logic model. However, Quantum Simulation Logic models do not reproduce the measurement outcomes, we give a counterexample in \cref{subsec:non-CSS_incorrectness}, see also Ref.~\cite{Batista_2023}. In the next Section, we show why Quantum Simulation Logic works perfectly for CSS-preserving circuits. Moreover, in \cref{subsubsec:frames_simulation} we enhance a simulator to also give correct results after non-CSS-preserving operations. Also, in addition to Quantum Simulation Logic, there exists a very similar diagrammatic approach to CSS-codes based on ZX-calculus \cite{Kissinger_2022,Comfort_2022}.

\subsection{Circuit transformation based proof of correctness} \label{subsec:circuit_proof}

Let us provide a proof for correctness of the rewriting method described in \cref{subsec:rewriting_rules}. Here we discuss a proof employing only trivial methods and based mainly on the deferred measurent principle \cite{Nielsen_2010}. We will also give a proof relying on advanced developments in stabilizer formalism in \cref{subsubsec:symbols_of_CSS}.

Our aim is to prove that any CSS-preserving circuit $\mathsf{RC}$ having only classical inputs and outputs is equivalent to the rewritten classical circuit $\mathsf{CC}$. Generally, $\mathsf{RC}$ can consist of: preparations of initial states $\ket{0}$ and $\ket{+}$, one-rebit gates $Z$ and $X$, two-rebit gates $\CNOT$, one-rebit $Z$- and $X$-measurements, Walsh-Hadamard transform $\WH$ acting on all rebits, qubit discarding channels, classical computations and control. We will firstly prove the statement for the class of unitary CSS-preserving circuits (\cref{subsubsec:proof_unitary}), then for non-adaptive circuits (\cref{subsubsec:proof_measurements}) and finally include adaptivity (\cref{subsubsec:proof_adaptive}).

\subsubsection{Independence of $Z$- and $X$-parts} \label{subsubsec:ZX_independence}

Let us call a state \emph{pure CSS-stabilizer} if it can be prepared by a unitary CSS-preserving circuit. We will have a deeper discussion on CSS-stabilizer states in \cref{subsec:CSS_operations}, for now one can consult \cite{Delfosse_2015}. One important property of pure CSS-stabilizer states, which we employ in our consideration, is that the outcomes of their $Z$- and $X$-measurements are statistically independent. In other words, preparing a pure CSS stabilizer state $\ket{\psi}$ and measuring some qubits in the $Z$-basis and other qubits in the $X$-basis is equivalent to creating two copies of $\ket{\psi}$ and performing $Z$- and $X$-measurements independently, without altering the measurement statistics:

\begin{equation}
  \begin{quantikz}[row sep={0.8cm,between origins}, align equals at=1.5]
    \lstick[2]{$\ket{\psi}$} &\qwbundle{} &[-0.3cm]\meterD{Z} &\setwiretype{c} \\
                             &\qwbundle{} &\meterD{X} &\setwiretype{c}
  \end{quantikz}
  \quad\approx\quad
  \begin{quantikz}[row sep={0.5cm,between origins}, align equals at=2.5]
    \lstick[2]{$\ket{\psi}$} &\qwbundle{} &[-0.3cm]\meterD{Z} &\setwiretype{c} \\
                             &\qwbundle{} &\ground{}  &\setwiretype{n} \\[0.2cm]
    \lstick[2]{$\ket{\psi}$} &\qwbundle{} &\ground{}  &\setwiretype{n} \\
                             &\qwbundle{} &\meterD{X} &\setwiretype{c} \\
  \end{quantikz}
\end{equation}
(hereinafter, $\approx$ means that the circuits are equivalent: output statistics are the same on all inputs). Let us prove this fact. Every pure CSS-state $\ket{\psi}$ can be represented as $\proj{\psi} = \Pi_Z \Pi_X = \Pi_X \Pi_Z$, where $\Pi_Z$ and $\Pi_X$ are commuting stabilizer projectors on $Z$- and $X$-part correspondingly~\cite{Nielsen_2010}. Consider the probability of a measurement outcome specified by the projector $P = P_Z P_X$, where the projectors $P_Z$ and $P_X$ represent the outcomes of $Z$- and $X$-measurements, respectively. Since $P_Z$ and $P_X$ act non-trivially on distinct qubits, they naturally commute. Using the fact that $\Pi_X$ and $P_X$ (as well as $\Pi_Z$ and $P_Z$) commute, the probability of the given outcome can be expressed in the following way:
\begin{equation}
\begin{aligned}
  \bra{\psi}P\ket{\psi}
  &= \Tr(\Pi_Z\Pi_X\; P_ZP_X) \\
  &= \Tr(\Pi_X\Pi_Z\; P_ZP_X\; \Pi_X\Pi_Z) \\
  &= \Tr(\Pi_X P_Z\Pi_Z \Pi_X P_X \Pi_Z) \\
  &= \bra{\psi}P_Z\ket{\psi}\bra{\psi}P_X\ket{\psi}.
\end{aligned}
\end{equation}
This shows that $Z$- and $X$-outcomes are statistically independent.

\subsubsection{Proof for unitary circuits} \label{subsubsec:proof_unitary}

Now, suppose we have a \emph{unitary} CSS-preserving circuit $\mathsf{RC}$ consisting of the following components: state preparations of $\ket{0}$ and $\ket{+}$ (which are not prohibited from occuring after the initial stage); unitary CSS-preserving gates $X,Z,\CNOT,\WH$; and $Z$- or $X$-basis measurements or discarding, applied to each rebit in the final step. Note that at each point in time prior to the final step, the state of the system is pure CSS-stabilizer.

Let us introduce two identical copies of circuit $\mathsf{RC}$, denoted $\mathsf{RC}_Z$ and $\mathsf{RC}_X$, and refer to them as the $Z$-part and $X$-part, respectively. Let us modify $\mathsf{RC}_Z$ at the final time step -- introduce $Z$-measurements on each rebit and discard the results on wires which do have $Z$-measurements in $\mathsf{RC}$; let us do the same modification on $X$-part analogously.
\begin{equation}
  \begin{array}{c|c|c}
    \mathsf{RC} & \mathsf{RC}_Z & \mathsf{RC}_X \\
    \hline
    \begin{quantikz}[row sep={0.8cm,between origins}, column sep={0.7cm,between origins},align equals at=2]
      &\meterD{Z} &\setwiretype{c} \\
      &\meterD{X} &\setwiretype{c} \\
      &\ground{}  &\setwiretype{n}
    \end{quantikz}
    &
    \begin{quantikz}[row sep={0.8cm,between origins}, column sep={0.7cm,between origins},align equals at=2]
      &\meterD{Z} &\setwiretype{cz}          &[-0.3cm]\setwiretype{c} \\
      &\meterD{Z} &\ground{}\setwiretype{cz} &\setwiretype{n}         \\
      &\meterD{Z} &\ground{}\setwiretype{cz} &\setwiretype{n}
    \end{quantikz}
    &
    \begin{quantikz}[row sep={0.8cm,between origins}, column sep={0.7cm,between origins},align equals at=2]
      &\meterD{X} &\ground{}\setwiretype{cx} &[-0.3cm] \setwiretype{n} \\
      &\meterD{X} &\setwiretype{cx}          &\setwiretype{c}          \\
      &\meterD{X} &\ground{}\setwiretype{cx} &\setwiretype{n}
    \end{quantikz}
  \end{array}
\end{equation}
As explained above, the output of $\mathsf{RC}$ is equivalent to the output of a pair $\mathsf{RC}_Z+\mathsf{RC}_X$. Also, at this stage, the parts $\mathsf{RC}_Z$ and $\mathsf{RC}_X$ are disconnected from each other.
\begin{equation}
  \begin{quantikz}[wire types={c,c},row sep={0.8cm,between origins}]
    \gate[2,style={rounded corners}][1.2cm]{\mathsf{RC}} &\qwbundle{} \\
                                                  &\qwbundle{}
  \end{quantikz}
  \quad\approx\quad
  \begin{quantikz}[wire types={cz,cx},row sep={0.8cm,between origins}]
    \gate[style={rounded corners}]{\mathsf{RC}_Z} &\qwbundle{} &\setwiretype{c}\\
    \gate[style={rounded corners}]{\mathsf{RC}_X} &\qwbundle{} &\setwiretype{c}
  \end{quantikz}
  .
\end{equation}

The idea behind the proof is as follows. We shall commute arrays of $Z$- and $X$- measurements within $\mathsf{RC}_Z$ and $\mathsf{RC}_X$ from the right to the left through all quantum operations, replacing all unitary gates and state preparations with classical operations applied to classical bits. Simultaneously, we shall observe that the resulting classical operations are identical to those obtained from our rewriting rules, with each qubit in $\mathsf{RC}$ represented by a pair of classical bits, one from $\mathsf{RC}_Z$ and the other from $\mathsf{RC}_X$. Notably, $\WH$ operation will connect $\mathsf{RC}_Z$ and $\mathsf{RC}_X$ by the array of swaps.

Let us begin with the single-qubit $Z$ and $X$ gates. If the measurements come across a $Z$ gate, then the results of the $X$-measurements are flipped, analogously for the $X$-gate:
\begin{align}
  \begin{quantikz}[row sep={0.7cm,between origins}, align equals at=1.5]
    &\gate{Z} &\meterD{Z} &\setwiretype{cz} \\
    &\gate{Z} &\meterD{X} &\setwiretype{cx}
  \end{quantikz}
  \quad\approx\quad
  \begin{quantikz}[row sep={0.7cm,between origins}, align equals at=1.5]
    &\meterD{Z} &\setwiretype{cz}        & \\
    &\meterD{X} &\targ{}\setwiretype{cx} &
  \end{quantikz}
  \\
  \begin{quantikz}[row sep={0.7cm,between origins}, align equals at=1.5]
    &\gate{X} &\meterD{Z} &\setwiretype{cz} \\
    &\gate{X} &\meterD{X} &\setwiretype{cx}
  \end{quantikz}
  \quad\approx\quad
  \begin{quantikz}[row sep={0.7cm,between origins}, align equals at=1.5]
    &\meterD{Z} &\targ{}\setwiretype{cz} & \\
    &\meterD{X} &\setwiretype{cx}        &
  \end{quantikz}
\end{align}
This transformation is equivalent to the rewriting rules \cref{eq:Z-rewriting,eq:X-rewriting}.

If there are two pairs of measurements that are connected by a $\CNOT$ gate, then, according to the deferred measurement principle, it holds that
\begin{align}
  \begin{quantikz}[row sep={0.7cm,between origins}, column sep={0.7cm,between origins}, align equals at=1.5]
    &\ctrl{1} &\meterD{Z} &\setwiretype{cz} \\
    &\targ{}  &\meterD{Z} &\setwiretype{cz}
  \end{quantikz}
  \quad\approx\quad
  \begin{quantikz}[row sep={0.7cm,between origins}, column sep={0.7cm,between origins}, align equals at=1.5]
    &\meterD{Z} &\ctrl{0}\wire[d][1]{c}\setwiretype{cz} & \\
    &\meterD{Z} &\targ{}\setwiretype{cz}                &
  \end{quantikz}
  \\
  \begin{quantikz}[row sep={0.7cm,between origins}, column sep={0.7cm,between origins}, align equals at=1.5]
    &\ctrl{1} &\meterD{X} &\setwiretype{cx} \\
    &\targ{}  &\meterD{X} &\setwiretype{cx}
  \end{quantikz}
  \quad\approx\quad
  \begin{quantikz}[row sep={0.7cm,between origins}, column sep={0.7cm,between origins}, align equals at=1.5]
    &\meterD{X} &\targ{}\wire[d][1]{c}\setwiretype{cx} & \\
    &\meterD{X} &\ctrl{0}\setwiretype{cx}              &
  \end{quantikz}
\end{align}
This transformation corresponds to rewriting rule \cref{eq:CNOT-rewriting}.

Commuting $Z$- and $X$-measurements through the Walsh-Hadamard transformation $\WH$ results in the interchange of the measurements: each $Z$-measurement becomes an $X$-measurement and vice versa.
Let us denote $\mathsf{RC}_0$ the part of quantum circuit $\mathsf{RC}$ prior to this $\WH$. Since $\mathsf{RC}_0$ is exactly the same for both $Z$- and $X$-parts, the following identity holds:
\begin{equation}
\begin{gathered}
  \begin{quantikz}[row sep={0.8cm,between origins}, column sep={0.8cm,between origins}, align equals at=1.5]
    \gate[style={rounded corners}]{\mathsf{RC}_0} &\qwbundle{} &[-0.2cm]\gate{\WH} &[0.2cm]\meterD{Z} &[-0.2cm]\setwiretype{cz} \\
    \gate[style={rounded corners}]{\mathsf{RC}_0} &\qwbundle{} &\gate{\WH}         &\meterD{X}        &\setwiretype{cx}
  \end{quantikz}
  \;\approx\;
  \begin{quantikz}[row sep={0.8cm,between origins}, column sep={0.8cm,between origins}, align equals at=1.5]
    \gate[style={rounded corners}]{\mathsf{RC}_0} &\qwbundle{} &[-0.3cm]\meterD{X} &[-0.2cm]\setwiretype{cz} \\
    \gate[style={rounded corners}]{\mathsf{RC}_0} &\qwbundle{} &\meterD{Z}         &\setwiretype{cx}
  \end{quantikz}
  \\
  \quad\approx\quad
  \begin{quantikz}[row sep={0.8cm,between origins}, column sep={0.8cm,between origins}, align equals at=1.5]
    \gate[style={rounded corners}]{\mathsf{RC}_0} &\qwbundle{} &[-0.3cm]\meterD{Z} &\targX{}\wire[d][1]{c}\setwiretype{cz} &[-0.3cm] \\
    \gate[style={rounded corners}]{\mathsf{RC}_0} &\qwbundle{} &\meterD{X}         &\targX{}\setwiretype{cx}               &
  \end{quantikz}
\end{gathered}
\end{equation}
which gives us the rewriting rule \cref{eq:WH-rewriting}.

Ultimately, the backwards propagation of measurements stumbles upon state preparations, which can be handled as follows:
\begin{align}
  \begin{quantikz}[row sep={0.7cm,between origins}, align equals at=1.5]
    \lstick{$\ket{0}$} &\meterD{Z} &\setwiretype{cz} \\
    \lstick{$\ket{0}$} &\meterD{X} &\setwiretype{cx}
  \end{quantikz}
  \quad\approx\quad
  \begin{quantikz}[wire types={cz,cx}, row sep={0.7cm,between origins}, align equals at=1.5]
    \lstick{$0$} & & \\
    \lstick{$x$} & &
  \end{quantikz}
  \\
  \begin{quantikz}[row sep={0.7cm,between origins}, align equals at=1.5]
    \lstick{$\ket{+}$} &\meterD{Z} &\setwiretype{cz} \\
    \lstick{$\ket{+}$} &\meterD{X} &\setwiretype{cx}
  \end{quantikz}
  \quad\approx\quad
  \begin{quantikz}[wire types={cz,cx}, row sep={0.7cm,between origins}, align equals at=1.5]
    \lstick{$z$} & & \\
    \lstick{$0$} & &
  \end{quantikz}
\end{align}
where $x$ and $z$ are uniformly random bits. This corresponds to the rewriting rules \cref{eq:prep0-rewriting,eq:prepP-rewriting}.

As the result, the original $\mathsf{RC}_Z + \mathsf{RC}_X$ pair is transformed to an equivalent fully classical circuit $\mathsf{CC}$, which coincides with the rewritten version of the $\mathsf{RC}$. This finishes our proof for unitary case.

\subsubsection{Proof for circuits with intermediate measurements and classically controlled operations} \label{subsubsec:proof_measurements}

Let us now generalize the proof to include circuits in which $Z$- or $X$-measurements may occur at arbitrary time points, rather than only at the end; and correctness for circuits with clasically controlled elementary operations.

Consider circuit $\mathsf{RC}$ which is unitary CSS-preserving, except that there is a single $Z$-measurement happening somewhere in the middle of the circuit. Such circuit can be presented as
\begin{equation}
  \begin{quantikz}[row sep={0.7cm,between origins},column sep={0.7cm,between origins}, align equals at=1.5]
    \gate[2,style={rounded corners}]{\mathsf{RC}} &\setwiretype{c}            \\
                                                  &\qwbundle{}\setwiretype{c}
  \end{quantikz}
  \quad =\quad
  \begin{quantikz}[row sep={0.7cm,between origins},column sep={0.9cm,between origins}, align equals at=1.5]
    \gate[2,style={rounded corners}]{\mathsf{RC}_1} &[0.2cm]\meterD{Z}  &\setwiretype{c}                                     &                           \\
                                                    &\qwbundle{}        &\gate[style={rounded corners}]{\mathsf{RC}_2^{(w)}} &\qwbundle{}\setwiretype{c}
  \end{quantikz}
\end{equation}
where $\mathsf{RC}_1$ is a unitary CSS-preserving circuit and $\mathsf{RC}_2^{(w)}$ is a unitary CSS-preserving circuit with $w$ occurences of gate $\WH$. Let us defer the $Z$-measurement to the end of the circuit while extending each of $w$ Walsh-Hadamard gates in $\mathsf{RC}_2$ to the ``full'' $\WH$ gates acting on the whole rebit register:
\begin{equation}
  \begin{quantikz}[row sep={0.7cm,between origins},column sep={0.9cm,between origins}, align equals at=1.5]
    \gate[2,style={rounded corners}]{\mathsf{RC}_1} &[0.2cm]     &[-0.2cm]\gate[2,style={rounded corners}]{\mathsf{RC}_3^{(w)}} &[0.4cm]\meterD{Z^{(w)}}        &\setwiretype{c} \\
                                                    &\qwbundle{} &                                                              &\qwbundle{}\setwiretype{c} &
  \end{quantikz}
\end{equation}
where $\mathsf{RC}_3^{(w)}$ contains $w$ Hadamard gate on the first rebit and acts as $\mathsf{RC}_2$ on the remaining rebits, and $Z^{(w)} = H^w Z H^w \in \{X, Z\}$ is the $Z$-($X$-)measurement for even (odd) $w$. We can apply our rewriting rules on the full circuit $\mathsf{RC}_1+\mathsf{RC}_3^{(w)}+Z^{(w)}$, and be sure that they are correct due to the result of the previous subsection.

Next, we observe that the $w$ Hadamard gates on the first rebit in $\mathsf{RC}_3^{(w)}$ correspond, in the classical rewriting $\mathsf{CC}_3^{(w)}$, to $w$ $\SWAP$s between corresponding $X$- and $Z$-part classical bits.
Realization of $Z^{(w)}$ measurement can be rewritten as application of additional $w$ classical $\SWAP$ gates followed by the discarding of the second ($X$-part) bit [in accordance with \cref{eq:Zmeas-rewriting}]. These $2w$ $\SWAP$s cancel each other, and we obtain a circuit that would be obtained by directly applying the rewriting rules to the original $\mathsf{RC}$ circuit:
\begin{equation}
\begin{aligned}
  &
  \begin{quantikz}[wire types={cz,cx,cz,cx},row sep={0.3cm,between origins},column sep={0.5cm,between origins}, align equals at=2.5]
    \gate[4,style={rounded corners}][0.9cm][0.8cm]{\mathsf{CC}_1} &[0.8cm]\gate[4,style={rounded corners}][0.9cm][0.8cm]{\mathsf{CC}_3^{(w)}} &[0.5cm]\gategroup[2,steps=1,style={dashed,rounded corners,yshift=-0.15cm}]{depends on $w$}         &[-0.3cm]\setwiretype{c}        \\
                                                    &                                                                      &\ground{}       &\setwiretype{n} \\[0.8cm]
                                                    &\qwbundle{}                                                           &\setwiretype{n} &                \\
                                                    &\qwbundle{}                                                           &\qwbundle{}\setwiretype{c}     &
  \end{quantikz}
  \\
  &\;\approx\;
  \begin{quantikz}[wire types={cz,cx,cz,cx},row sep={0.3cm,between origins},column sep={0.2cm,between origins}, align equals at=2.5]
    \gate[4,style={rounded corners}][0.9cm][0.8cm]{\mathsf{CC}_1} &[0.3cm]& &[0.2cm]\targX{}\wire[d][1]{c}\gategroup[2,steps=3,style={dashed,rounded corners,yshift=-0.15cm}]{$w$ $\SWAP$s} &&\midstick[2,brackets=none]{$\ldots$} &&\targX{}\wire[d][1]{c} &&[0.4cm] &\targX{}\wire[d][1]{c}\gategroup[2,steps=3,style={dashed,rounded corners,yshift=-0.15cm}]{$w$ $\SWAP$s} &&\midstick[2,brackets=none]{$\ldots$} &&\targX{}\wire[d][1]{c} & &[0.3cm] &[0.2cm]\setwiretype{c} \\
    && &\targX{} && &&\targX{} && &\targX{} && &&\targX{} & &\ground{} &\setwiretype{n} \\[0.8cm]
    &&\qwbundle{} &\gategroup[2,steps=3,style={rounded corners,yshift=0.15cm,fill=white},label style={yshift=-0.8cm}]{$\mathsf{CC}_2^{(w)}$} && && &\setwiretype{n}& & && && & & & \\
    &&\qwbundle{} &  && && &&\qwbundle{}\setwiretype{c} &\setwiretype{n} && && & & &
  \end{quantikz}
  \\
  &\;\approx\;
  \begin{quantikz}[wire types={cz,cx,cz,cx}, row sep={0.3cm,between origins},column sep={0.6cm,between origins}, align equals at=2.5]
    \gate[4,style={rounded corners}][0.9cm][0.8cm]{\mathsf{CC}_1} &[0.4cm]     &[-0.1cm]                                              &[0.4cm]\setwiretype{n} &[-0.4cm] \\
                                                    &\ground{}   &\setwiretype{n}                                       &\setwiretype{n}        &         \\[0.8cm]
                                                    &\qwbundle{} &\gate[2,style={rounded corners}]{\mathsf{CC}_2^{(w)}} &\setwiretype{n}        &         \\
                                                    &\qwbundle{} &                                                      &\qwbundle{}\setwiretype{c}            &
  \end{quantikz}
\end{aligned}
\end{equation}
Any number of interspersed $X$- and $Z$-measurements can be handled similarly: shift each measurement to the end of the circuit, add a Hadamard gate each time it passes whenever it passes $\WH$, then pull it back through the rewritten classical circuit. All ancillary $\SWAP$ gates introduced during rewriting cancel, ensuring that the rewriting is correct.

Next, let us consider the case in which an elementary unitary operation $U$, chosen from the set $\{X, Z, \CNOT, \SWAP, \WH\}$, is applied conditionally based on the value of a classical control bit $b$: we assume that the operation is applied when $b = 1$, and omitted (i.e., replaced by the identity) when $b = 0$. For each fixed value of $b$, the rewriting procedure produces a classical circuit: when $b = 1$, the position of $m$-rebit operation $U$ is occupied by a corresponding classical $2m$-bit operation $U_c$, and when $b = 0$ the operation is simply omitted. Since we have already established the correctness of the rewriting in both cases, we can combine them by introducing a classical conditional control on $U_c$ based on the value of $b$. This leads directly to the rewriting rule \cref{eq:ctrl-rewriting}. If there are many controlled operations, the rewriting also works, since it works for all values of control bits.

\subsubsection{Proof for adaptive circuits} \label{subsubsec:proof_adaptive}

Finally, let us address the case where the circuit involves adaptivity, that is, quantum operations are conditioned on the outcomes of preceding measurements. We show that the rewriting rules remain valid in this setting.

An adaptive quantum computation can be seen as a sequence of communication rounds between a classical controller and a purely quantum device. This interaction can be schematically represented as:
\begin{equation} \label{eq:adaptive_circuit}
  \scalebox{0.8}{
  \begin{quantikz}[wire types={c,c,n},row sep={0.7cm,between origins},column sep={0.7cm,between origins}, align equals at=2]
    \gate[2,style={rounded corners}]{\mathsf{CC}^c_0} &\qwbundle{} &[-0.25cm] &&[-0.25cm]\gate[2,style={rounded corners}]{\mathsf{CC}^c_1} &\qwbundle{} &[-0.25cm] &&[-0.25cm]\ \ldots\ &\qwbundle{}&[-0.25cm] & &[-0.22cm]\gate[2,style={rounded corners}]{\mathsf{CC}^c_L} &\qwbundle{} &[-0.5cm] \\
    &\qwbundle{} &\gate[2,style={rounded corners}]{\mathsf{RC}^q_1} &\qwbundle{} &&\qwbundle{} &\gate[2,style={rounded corners}]{\mathsf{RC}^q_2} &\qwbundle{} &\ \ldots\ &\qwbundle{}&\gate[2,style={rounded corners}]{\mathsf{RC}^q_L} &\qwbundle{} &&\setwiretype{n} & \\
    &&&\qwbundle{}\setwiretype{q} &&&&\qwbundle{} &\ \ldots\ &\qwbundle{} &&\setwiretype{n} &&&
  \end{quantikz}
  }.
\end{equation}
Here, each $\mathsf{RC}^q_i$ consists of CSS-preserving state preparations, (classically controlled) unitary gates, and measurements, while $\mathsf{CC}^c_i$ does arbitrary classical computation acting on the outcomes of prior measurements.

Let us isolate the ``comb'' formed by the sequence of $\mathsf{RC}^q_i$ circuits:
\begin{equation} \label{eq:purely_quantum_part}
  \scalebox{0.8}{
  \begin{quantikz}[wire types={c,n},row sep={0.7cm,between origins},column sep={0.7cm,between origins}, align equals at=1.5]
    &\qwbundle{} &[-0.25cm]\gate[2,style={rounded corners}]{\mathsf{RC}^q_1} &\qwbundle{} &[-0.25cm]\ \quad \ &\qwbundle{} &[-0.25cm]\gate[2,style={rounded corners}]{\mathsf{RC}^q_2} &\qwbundle{} &[-0.25cm]\ \quad \ &\qwbundle{}&[-0.25cm]\gate[2,style={rounded corners}]{\mathsf{RC}^q_L} &\qwbundle{} &[-0.5cm] \\
    &&&\qwbundle{}\setwiretype{q} &&&&\qwbundle{} &\ \ldots\ &\qwbundle{} &&\setwiretype{n} &
  \end{quantikz}
  }
\end{equation}
This fragment can be interpreted as a classically controlled quantum circuit, where the control values originate from the computations performed on $\mathsf{CC}^c_i$. Based on results above, this ``comb'' can be equivalently rewritten to a classical circuit:
\begin{equation} \label{eq:purely_quantum_part_rewritten}
  \scalebox{0.8}{
  \begin{quantikz}[wire types={c,n,n},row sep={0.2cm,between origins},column sep={0.7cm,between origins}, align equals at=1.5]
    &\qwbundle{} &[-0.25cm]\gate[3,style={rounded corners}]{\mathsf{CC}^q_1} &\qwbundle{} &[-0.25cm]\ \quad \ &\qwbundle{} &[-0.25cm]\gate[3,style={rounded corners}]{\mathsf{CC}^q_2} &\qwbundle{} &[-0.25cm]\ \quad \ &\qwbundle{}&[-0.25cm]\gate[3,style={rounded corners}]{\mathsf{CC}^q_L} &\qwbundle{} &[-0.5cm] \\[0.4cm]
    &&&\qwbundle{}\setwiretype{cz} &&&&\qwbundle{} &\ \ldots\ &\qwbundle{} &&\setwiretype{n} & \\
    &&&\qwbundle{}\setwiretype{cx} &&&&\qwbundle{} &\ \ldots\ &\qwbundle{} &&\setwiretype{n} &
  \end{quantikz}
  },
\end{equation}
where each $\mathsf{CC}^q_i$ is a rewriting of $\mathsf{RC}^q_i$.

The key observation is that the causal structure of \cref{eq:purely_quantum_part_rewritten} matches that of \cref{eq:purely_quantum_part}: specifically, the classical outputs of $\mathsf{CC}^q_i$ are computed before the execution of $\mathsf{CC}^q_{i+1}$. This temporal ordering ensures that we can safely substitute the rewritten form \cref{eq:purely_quantum_part_rewritten} back into the original circuit \cref{eq:adaptive_circuit}. As a result, we obtain a fully classical circuit:
\begin{equation}
  \scalebox{0.8}{
  \begin{quantikz}[wire types={c,c,n,n},row sep={0.2cm,between origins},column sep={0.7cm,between origins}, align equals at=2]
    \gate[2,style={rounded corners}]{\mathsf{CC}^c_0} &\qwbundle{} &[-0.25cm] &&[-0.25cm]\gate[2,style={rounded corners}]{\mathsf{CC}^c_1} &\qwbundle{} &[-0.25cm] &&[-0.25cm]\ \ldots\ &\qwbundle{}&[-0.25cm] & &[-0.22cm]\gate[2,style={rounded corners}]{\mathsf{CC}^c_L} &\qwbundle{} &[-0.5cm] \\[0.5cm]
    &\qwbundle{} &\gate[3,style={rounded corners}]{\mathsf{CC}^q_1} &\qwbundle{} &&\qwbundle{} &\gate[3,style={rounded corners}]{\mathsf{CC}^q_2} &\qwbundle{} &\ \ldots\ &\qwbundle{}&\gate[3,style={rounded corners}]{\mathsf{CC}^q_L} &\qwbundle{} &&\setwiretype{n} & \\[0.4cm]
    &&&\qwbundle{}\setwiretype{cz} &&&&\qwbundle{} &\ \ldots\ &\qwbundle{} &&\setwiretype{n} &&& \\
    &&&\qwbundle{}\setwiretype{cx} &&&&\qwbundle{} &\ \ldots\ &\qwbundle{} &&\setwiretype{n} &&& \\
  \end{quantikz}
  }
\end{equation}
which produces output distributions that are indistinguishable from those of the original adaptive quantum circuit~\cref{eq:adaptive_circuit}.

To sum up, using the set of rewriting rules from \cref{subsec:rewriting_rules}, one can reduce any CSS-preserving quantum circuit with classical inputs and outputs to a classical circuit.

\subsection{Incorrectness for non-CSS-preserving stabilizer circuits} \label{subsec:non-CSS_incorrectness}

Let us consider the following example to illustrate the fact that, for a given stabilizer circuit, it is not always the case that its corresponding classical circuit produces the expected output.

We use the non-CSS-preserving stabilizer circuit of the form
\begin{equation}
  \begin{quantikz}[wire types = {q,q}, row sep={0.8cm,between origins}, column sep={0.7cm,between origins}, align equals at=1.5]
    \lstick{$\ket{+}$}&\ctrl{1} &\ctrl{1} &\ctrl{1} &\meterD{X} &\setwiretype{c}\rstick{$1$} \\
    \lstick{$\ket{0}$}&\targ{}  &\ctrl{0} &\targ{}  &\ground{}
  \end{quantikz}.
\end{equation}
This circuit prepares the Bell state $(\ket{00} + \ket{11})/\sqrt{2}$, applies the $CZ$ gate to change it to the state $(|00\rangle - |11\rangle)/\sqrt{2}$, and then disentangles it to $\ket{-}\ket{0}$. The output of this circuit is always $1$.

Applying the rewriting rule \cref{eq:CZ-rewriting}, this circuit is transformed to
\begin{equation}
  \begin{quantikz}[wire types={cz,cx,cz,cx}, row sep={0.35cm,between origins}, column sep={0.6cm,between origins}]
    \lstick{$z$} &\ctrl{0}\wire[d][2]{c} &                      &[0.2cm]\ctrl{0}\wire[d][3]{c} &                      &[0.2cm]\ctrl{0}\wire[d][2]{c} &                      &\ground{} &\setwiretype{n} \\
    \lstick{$0$} &                       &\targ{}\wire[d][2]{c} &                              &\targ{}\wire[d][1]{c} &                              &\targ{}\wire[d][2]{c} &          &\rstick{$0$}\setwiretype{c}    \\[0.6cm]
    \lstick{$0$} &\targ{}                &                      &                              &\ctrl{0}              &\targ{}                       &                      &\ground{} &\setwiretype{n} \\
    \lstick{$x$} &                       &\ctrl{0}              &\targ{}                       &                      &                              &\ctrl{0}              &\ground{} &\setwiretype{n}
  \end{quantikz}
\end{equation}
It can be easily verified that this circuit always outputs $0$, contradicting the result obtained from the quantum circuit. A more detailed understanding of this phenomenon will be provided in \cref{subsubsec:frames_simulation}.

\subsection{Simulation of classical circuits} \label{subsec:classical_simulation}

Let us now review some obvious techniques for simulating a classical circuit $\mathsf{CC}$ obtained after reduction. During this Section we will suppose that $n$ is the number of bits in the circuit, $k$ is the output length, $L$ is the size of the circuit, $m$ is the number of initial uniformly random bits, $r$ is the number of significant random bits (found as the result of the Gaussian elimination, see below), $S$ is the number of samples one wants to obtain in the weak simulation. We will discuss four methods, their expected execution times are compared in \cref{tab:simulation_methods} and \cref{fig:sampling_times}.

\subsubsection{Weak simulation of classical circuits} \label{subsubsec:weak_simulation}

(\textsf{Method~I}) To draw samples from the classical circuit $\mathsf{CC}$, one should generate $m$ random bits and simply act by Boolean gates on the string of all bits. To generate a single sample by this method, one requires $\BigO(n)$ space to store the string and $\BigO(L)$ time. This asymptotics for weak simulation of CSS-preserving stabilizer circuits is better than the usual $\BigO(n L)$ obtained by the use of stabilizer tableau \cite{Aaronson_2004}. Running the strong simulation of arbitrary Boolean circuits is known to be $\mathtt{\# P}$-complete \cite{Van_den_Nest_2010}, which is effectively intractable. Note that it is possible (yet inessential) to store $Z$-, $X$- and measured bits in separate arrays. That makes the operation $\WH$ to run in $\BigO(1)$ by swapping $Z$- and $X$-labels.

\subsubsection{Simulation of affine classical circuits} \label{subsubsec:strong_simulation}

When the CSS-preserving circuit $\mathsf{RC}$ is non-adaptive, the circuit $\mathsf{CC}$ is affine, i.e.\! equivalent to a composition of Boolean additions and negations. This also happens if the circuit $\mathsf{RC}$ has control only over Pauli gates and all classical computation is itself affine \cite{Yashin_2025, Kliuchnikov_2023}. It is possible to efficiently simulate affine Boolean circuits in the strong sense, since they are known to be $\mathtt{\oplus L}$-complete \cite{Aaronson_2004, Damm_1990}.

Let us consider the $m$ random bits as the input of the circuit $\mathsf{CC}$. Let us suppose that this circuit takes random $m$-bit strings on input and outputs $k$-bit strings. Since the circuit is affine, it represents a function $v\mapsto A v \oplus s$, where $A$ is a $k\times m$ Boolean matrix and $s$ is a $k$-bit string (shift). The possible outputs constitute an affine code in the output space $\Int_2^k$ of $k$-bit strings.

(\textsf{Method~II}) To find the matrix $A$ and the shift $s$, one can represent each element in the circuit as an elementary row operation on $(A,s)$: bit $0$ initializations add zero rows, $\NOT$ gates change shift, $\CNOT$ gates add one row to another, discarding channels delete rows. It takes $\BigO(n L)$ time to find the matrix $A$ and the shift $s$. After doing this procedure, one can draw samples in $\BigO(m k)$ time.

(\textsf{Method~III}) Furthermore, one can run a Gaussian elimination on columns of the matrix $A$ to obtain a $k\times r$ \emph{generator matrix} $G$ of the output code ($r$ is a rank of $A$) and a $(k-r)\times k$ \emph{parity-check matrix} $P$
\begin{equation}
\begin{aligned}
  \{ A v \oplus s : v\in\Int_2^m \}
  &= \{ G u \oplus s : u\in\Int_2^r \} \\
  &= \{ y\in\Int_2^k : P y = c\},
\end{aligned}
\end{equation}
where $P G = 0$ and $c = P s$.
The Gaussian elimination takes $\BigO(k m r)$ time, and it allows to draw samples from the circuit in $\BigO(k r)$ time using the function $u \mapsto G u \oplus s$, where $u$ are uniformly random $r$-bit strings. In error correcting codes, if the noise is small, we expect that almost all syndrome measurements are deterministic, in that case $r$ should be significantly smaller than $m$, so doing the Gaussian elimination might be beneficial.

(\textsf{Method~IV}) Also, one can do the Gaussian elimination at each step of the circuit, that would take $\BigO(r^2 L)$ time. In most instances this method should be slow, but it was used in Ref.~\cite{de_Beaudrap_2022} to create phase-sensitive stabilizer simulator.

Using the parity-check matrix $P$, one can run the strong simulation of the circuit: a $k$-bit outcome $y$ has probability $p(y)=2^{-r}$ in case $P(y\oplus s) = 0$ and zero probability $p(y)=0$ otherwise. Therefore, knowing the Parity matrix $P$, the time to strongly simulate a $k$-bit outcome of a non-adaptive CSS-preserving circuit is $\BigO(k(k-r))$.

\begin{figure}[h]
  \centering
  \includegraphics[width=0.5\textwidth]{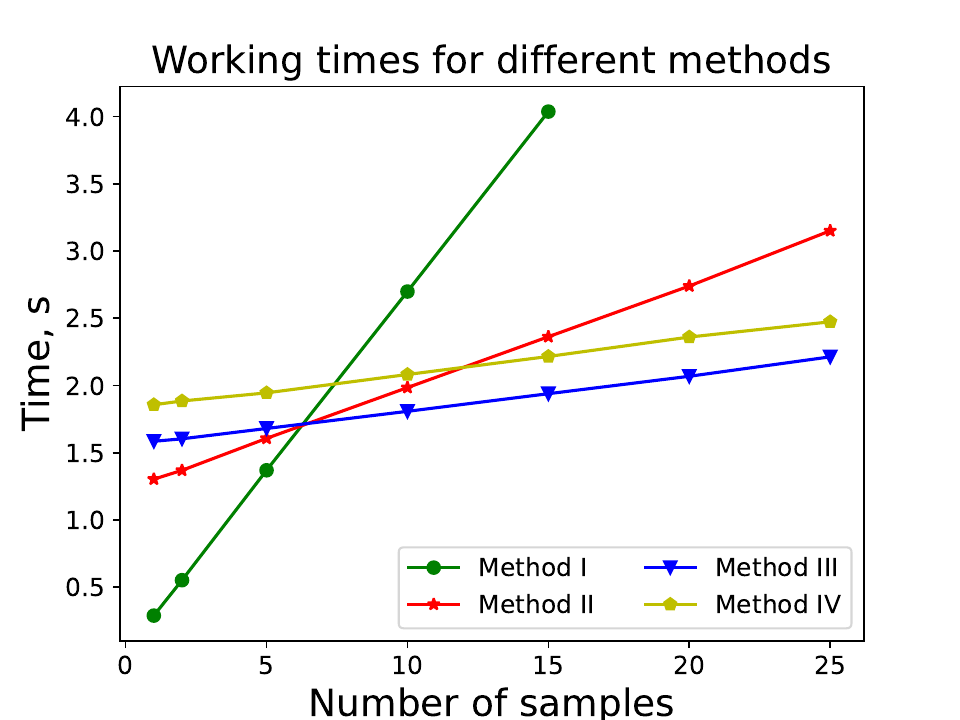}
  \caption{The running times of four described methods depending on the number of drawn samples. The data was obtained by generating some random affine Boolean circuit and simulating it. This plot only illustrates the usual behaviour of the methods and may be inadequate for concrete circuits.}
  \label{fig:sampling_times}
\end{figure}

These methods are compared in \cref{tab:simulation_methods} and \cref{fig:sampling_times}. Also, we developed a simple program to test the correctness for the circuit rewriting, some data on its performance can be seen in \hyperref[appendix:simulator]{Appendix~B}.

\begin{table}[h]
  \centering
  \begin{tabular}{|c|c|c|}
  \hline
  \textsf{Method} & Weak simulation & Strong simulation \\
  \hline
  \hline
  \textsf{I}   & $s(m+L)$              & --- \\
  \textsf{II}  & $n L + S k m$         & --- \\
  \textsf{III} & $n L + k m r + S k r$ & $n L + k m r + k(k-r)$ \\
  \textsf{IV}  & $r^2 L + S k r$       & $r^2 L + k(k-r)$ \\
  \hline
  \end{tabular}
  \caption{The expected execution time (modulo $\BigO$) for four described methods of classical circuits simulation. Here $n$ is the number of bits in the circuit, $k$ is the length of the output, $L$ is the size of the circuit, $m$ is the number of initial random bits, $r$ is the number of significant random bits found as the result of the Gaussian elimination, $S$ is the number of samples.}
  \label{tab:simulation_methods}
\end{table}

\section{Stabilizer formalism and quadratic form expansions} \label{sec:stabilizer}

In this Section, we develop stabilizer formalism to gain deeper understanding of stabilizer circuits simulation. In \cref{subsec:preliminaries_stabilizer}, we give preliminaries on Pauli groups, stabilizer groups, stabilizer states and Clifford channels, and discuss how to treat classical bits inside of stabilizer formalism. In \cref{subsec:quadratic_form_expansions}, we discuss the algebra of $\Int_4$-valued quadratic forms and apply it for describing stabilizer groups. We propose a standard quadratic form representation for arbitrary Clifford channel, including post-selected channels. In \cref{subsec:CSS_operations}, we discuss the special case of CSS-stabilizer theory: we introduce CSS-stabilizer groups and show that CSS-preserving stabilizer operations have only linear part in their quadratic form expansion. Finally, in \cref{subsec:quadratic_form_simulation} we propose a stabilizer simulator based on the theory developed in this Section.

\subsection{Preliminaries on stabilizer formalism} \label{subsec:preliminaries_stabilizer}

Let us begin by discussing the qubit stabilizer formalism originally developed in the works of Gottesman \cite{Gottesman_1997,Gottesman_1998,Aaronson_2004}. More concretely, we will discuss Pauli operators, stabilizer groups, stabilizer states and Clifford channels. Also, we comment on how to include classical bits into consideration. Be aware that in this Section we discuss the formalism for general stabilizer operations, while CSS-preserving stabilizer operations are considered as a special case in \cref{subsec:CSS_operations}.

\subsubsection{Pauli group and stabilizer subgroups} \label{subsubsec:Pauli_group}

Here we discuss the group properties of multiqubit Pauli operators (Pauli group) and stabilizer subgroups inside of this group. For deeper algebraic treatment of the subject it might be beneficial to consult \cite{Heinrich_2021}.

Suppose we examine an $n$-qubit system. Note that we always include trivial zero-qubit system into consideration. To each $i$-th qubit we correspond ``momentum-position'' bit variables $(z_i,x_i)\in\Int_2\!\times\!\Int_2$, and to the whole system we correspond the phase space $\Int_2^{2n}$ made of $2n$-bit strings, each phase space point we denote as $u = (z,x)$. Here and thereafter we treat bit strings $u\in\Int_2^{2n}$ and $z,x\in\Int_2^n$ as Boolean \emph{row-vectors}.

For bit strings $z,x \in \Int_2^n$ we define operators
\begin{equation}
\begin{aligned}
  & Z(z) = Z^{z_1}\otimes\cdots\otimes Z^{z_n}, \\
  & X(x) = X^{x_1}\otimes\cdots\otimes X^{x_n}.
\end{aligned}
\end{equation}
For each phase point $u\in\Int_2^{2n}$ we define a \emph{sign-free Pauli operator} (or \emph{displacement operator})
\begin{equation}
    T(u) = Z(z) X(x).
\end{equation}
Let us denote the set of sign-free Pauli operators $\PauliGroup_+^n$.

Elements of form $i^c T(u)$ for some $c\in\Int_4$ and $u\in\Int_2^{2n}$ are called \emph{Pauli operators}. The set of Pauli operators forms a group called \emph{Pauli group} $\PauliGroup^n = \langle i I, Z, X\rangle$. Pauli operators of form $(-1)^c T(u)$ are real, let us call the set of real Pauli operators the \emph{real Pauli group} $\PauliGroup^n_{\Real} = \langle Z,X \rangle$. Real Pauli group $\PauliGroup^n_{\Real}$ can also be called the \emph{Heisenberg group} over $\Int_2$.

Given two phase points $u$ and $u'$, the product of corresponding Pauli operators $T(u)$ and $T(u')$ corresponds to $T(u\oplus u')$ up to a phase (i.e., satisfies \emph{Weyl-type commutation relations}):
\begin{equation}
    T(u)T(u') = (-1)^{\tau(u,u')}T(u\oplus u'),
\end{equation}
where $\tau$ is the function of sign update
\begin{equation}
  \tau(u,u') = \sum_{i=1}^n x_i z'_i \quad (\mathrm{mod}\; 2).
\end{equation}
Pauli operators $T(u)$ are unitary, real, and either Hermitian or anti-Hermitian, depending on the value $\tau(u,u)$:
\begin{equation}
    T(u)^\dag = (-1)^{\tau(u,u)}T(u).
\end{equation}
Due to the associativity of the product
\begin{equation}
  T(u)\,(\,T(u')\,T(u'')\,) = (\,T(u)\,T(u')\,)\,T(u''),
\end{equation}
the function $\tau$ satisfies $2$-cocycle condition:
\begin{equation}
  \tau(u,u'\oplus u'') \oplus \tau(u',u'') = \tau(u,u') \oplus \tau(u\oplus u',u'').
\end{equation}
(Here and later in the text we write $\oplus$ to indicate summation modulo $2$ and $+$ for summation modulo $4$.) Speaking in group-theoretic terms, real Pauli group $\PauliGroup^n_{\Real}$ is a group extension of translation group $\Int_2^{2n}$ by the group of signs $\Int_2$, the set of sign-free Pauli operators $\PauliGroup^n_+$ is a chosen transversal and $\tau$ is the corresponding $2$-cocycle \cite{Ceccherini-Silberstein_2022}. The complex Pauli group $\PauliGroup^n$ is the extension of $\Int_2^{2n}$ by phases in $\Int_4$.

Commutation relations of Pauli operators are connected with symplectic structure on the phase space:
\begin{equation}
    T(u)T(u') = (-1)^{[u,u']}T(u')T(u),
\end{equation}
where $[u,u'] = \tau(u,u') \oplus \tau(u',u)$ is the $\Int_2$-valued symplectic form on $\Int_2^{2n}$. Pauli operators $T(u)$ constitute an orthonormal basis with respect to Hilbert-Schmidt inner product:
\begin{equation} \label{eq:orthonormal_basis_of_Pauli}
    \frac{1}{2^n} \Tr\left(T(u)^\dag T(u')\right) = \delta_{u,u'},
\end{equation}
where $\delta_{u,u'}$ is the Kronecker symbol. Also, the real linear span of Hermitian operators $T(u)$ [i.e., such that $\tau(u,u) = 0$] is the set of all real symmetric matrices \cite{Delfosse_2015}.

Note that we choose to define phases of Pauli $T(u) = Z(z)X(x)$ as in Ref.~\cite{Delfosse_2015} instead of the more common definition $T(u) = i^{-\scalebox{0.7}{$\sum_i$} z_i x_i}Z(z)X(x)$, used for example in Refs.~\cite{Aaronson_2004, Yashin_2025_2}. Such choice of phases turns out to be important for the theory to work well with respect to our needs, we will discuss it in \cref{subsec:reference_frames}.

A \emph{stabilizer group} $\StabGroup \subseteq \PauliGroup^n$ is an abelian subgroup of Hermitian Pauli operators such that $-I \notin \StabGroup$. Stabilizer group is \emph{real} if every element in it is real; we will discuss a class of CSS-stabilizer groups in \cref{subsec:CSS_operations}. Every stabilizer group is isomorphic to vector space $\Int_2^r$, where $r\leq n$ is a \emph{rank} of $\StabGroup$. One can choose some generating set $\StabGroup = \langle P_1, \dots, P_k \rangle$, where $P_i \in \PauliGroup^n$ are commuting Hermitian Pauli operators. Often it is useful to choose the generating set to be linearly independent, in this case $k=r$. This set can be stored in a computer as a $k\times(2n+1)$ Boolean matrix called \emph{stabilizer tableau}. A substantial part of modern stabilizer circuit simulators are based on the idea of storing the stabilizer tableau and efficiently updating it under stabilizer operations \cite{Gottesman_1997, Gottesman_1998, Aaronson_2004, Gidney_2021, Garner_2025}.

\subsubsection{Stabilizer states and Clifford channels} \label{subsubsec:Clifford_channels}

A $n$-qubit \emph{(mixed) stabilizer state} \cite{Aaronson_2004, Fattal_2004, Audenaert_2005} is a state $\rho$ for which there exists a stabilizer group $\StabGroup \subseteq \PauliGroup^n$ such that
\begin{equation}
  \rho = \frac{1}{2^n} \sum_{P\in\StabGroup} P.
\end{equation}
The stabilizer group $\StabGroup$ is a strong symmetry \cite{Lessa_2025, Buca_2012} of the state $\rho$ in the sense that for all $P\in\StabGroup$ it holds
\begin{equation}
  P \rho = \rho P = \rho.
\end{equation}
If the rank of $\StabGroup$ is maximal $r=n$, then the state is pure $\rho = \proj{\psi}$ and it can be equivalently defined as a unique (up to a phase) pure state $\ket{\psi}$ such that
\begin{equation}
  P \ket{\psi} = \ket{\psi} \quad \text{for all} \; P\in\StabGroup.
\end{equation}
Here are some particular examples of stabilizer states. The computational basis states are stabilizer: $\ket{0}$ has stabilizer group $\langle Z \rangle$ and $\ket{1}$ has stabilizer group $\langle -Z\rangle$. Two-qubit Bell state $\Omega = \proj{\Omega}$, where
\begin{equation}
  \ket{\Omega} = \frac{\ket{00}+\ket{11}}{\sqrt{2}},
\end{equation}
is stabilizer, has stabilizer group $\langle Z\otimes Z, X\otimes X \rangle$ and can be written as
\begin{equation} \label{eq:Bell_state_1}
\begin{aligned}
  \Omega &= \frac{1}{4}\left[I\otimes I + X\otimes X + Z\otimes Z - Y\otimes Y\right] \\
    &= \frac{1}{2^2} \sum_{u\in \Int_2^2} T(u)\otimes T(u).
\end{aligned}
\end{equation}

Unitaries that preserve Pauli group under conjugate action are called \emph{Clifford unitaries}, and the set of such unitaries is called \emph{Clifford group} $\CliffordGroup^n$:
\begin{equation}
  \scalebox{0.97}{$
    \CliffordGroup^n \!= \{ U \text{ unitary} : U P U^\dag \in \PauliGroup^n \text{ for all } P\in \PauliGroup^n \}.
  $}
\end{equation}
Repeating \cref{subsec:preliminaries_circuits}, the full Clifford group $\CliffordGroup^n$ is generated by unitary gates $\langle H,S,\CNOT\rangle$ \cite{Gottesman_1997}, real Clifford unitaries $\CliffordGroup^n_{\Real}$ are generated by $\langle H,Z,\CNOT\rangle$ \cite{Hashagen_2018}, and CSS-preserving Clifford unitary operations $\CliffordGroup^n_{\text{CSS}}$ are generated by $\langle Z,X, \CNOT, \WH \rangle$ \cite{Delfosse_2015}.

Additionally to unitary gates, stabilizer theory takes into account non-invertible operations such as qubit discardings, measurements of Pauli-observables and some forms of noise. The careful theory for general stabilizer operations was developed quite recently in Refs.~\cite{Kliuchnikov_2023, Heimendahl_2022, Yashin_2025}. We argue that the correct mathematical description of stabilizer operations is the notion of \emph{Clifford channels}. A multiqubit quantum channel $\Phi$ is a \emph{Clifford channel} if it satisfies any of the following properties:
\begin{itemize}
  \item $\Phi$ can be realized by a non-adaptive stabilizer circuit, which could be composed of: stabilizer state preparations, Clifford unitary gates, measurements of Pauli observables, qubit discardings. Additionally, it is possible to include classical control over Pauli unitary gates and affine Boolean operations over classical bits: such operations reduce to stabilizer operations under deferred measurement principle. (See also \cref{subsubsec:treatment_of_bits}.)
  \item $\Phi$ maps pure stabilizer states on the input multiqubit system to (mixed) stabilizer states on the output system.
  \item The Choi state of $\Phi$ is a stabilizer state.
  \item $\Phi$ admits a Stinespring dilation consisting of: preparing stabilizer state on environment, acting on the full system with Clifford unitary, and discarding the environment.
  \item The dual channel $\Phi^*$ maps Pauli observables to Pauli observables or zero.
\end{itemize}
These properties were proven to be equivalent in Ref.~\cite{Yashin_2025}. Stabilizer operations with classical non-affine processing were studied in Ref.~\cite{Heimendahl_2022}. Clifford channels can be understood as the multiqubit analogue of bosonic Gaussian channels \cite{Weedbrook_2012}.

\subsubsection{Working with classical bits in stabilizer formalism}  \label{subsubsec:treatment_of_bits}

Stabilizer formalism can handle not only the multiqubit systems, but also the hybrid bits and qubits systems. This opportunity is applicable when dealing with Pauli measurements that tranform quantum information from a qubit to classical bit infromation. Also note that non-adaptive stabilizer circuits on classical bits are exactly affine Boolean circuits \cite{Aaronson_2004, Yashin_2025}.

In this text, we use the following convention. When working with stabilizer operations, we choose to treat classical bits as fully dephased quantum bits. That is, we match a hybrid bits and qubits system with multiqubit system in which some qubits are exposed to fully dephasing noise $\Deph_Z$. For example, the operation of measuring one qubit in $Z$-basis is just equivalent to dephasing this qubit with $\Deph_Z$. Some comments on this convention were also given in Ref.~\cite{Yashin_2025}.

Other possibility is to treat classical bits separately from qubits, as done in Ref.~\cite{Kliuchnikov_2023}. This treatment usually results in case-by-case analysis, which is quite intuitive but complicates the discussion. Yet another possibility is to choose a phase space with $\Int_2$-dimension $n_c + 2n_q$ where $n_c$ is a number of bits and $n_q$ is a number of qubits. On this phase space, one chooses a symplectic form $[\cdot,\cdot]$ and $2$-cocycle $\tau$ to be degenerate on classical bits. We suggest that such meticulous approaches should be realised in program code for better performance.

\subsection{Stabilizer operations as quadratic form expansions} \label{subsec:quadratic_form_expansions}

In this subsection we develop stabilizer theory for Clifford channels further. Firstly, we discuss how $\Int_4$-valued quadratic forms naturally arise in stabilizer theory. Then, we study a variant of Dirac's bra-ket notation for operators and find that it is handy for representing stabilizer states and arbitrary stabilizer operations. Using this notation, we a propose a convenient representation for storing and composing Clifford channels in the classical computer.

\subsubsection{$\Int_4$-valued quadratic forms} \label{subsubsec:quadratic_forms}

We will need to employ some results of the theory of $\Int_4$-valued quadratic forms on Boolean vector spaces \cite{Brown_1972, Schmidt_2009, Bravyi_2016, de_Beaudrap_2022} to understand the structure of stabilizer states and Clifford channels.

By a \emph{linear form} on a Boolean vector space $\Int_2^k$ we mean a function $l : \Int_2^k \to \Int_2$ of form
\begin{equation}
  l(u) = u s = u_1 s_1 \oplus \cdots \oplus u_n s_n,
\end{equation}
where $s\in \Int_2^k$, we treat $u$ as a row-vector and $s$ as a column vector. Linear forms are exactly the functions that satisfy conditions
\begin{equation}
  l(0) = 0, \qquad l(u\oplus u') = l(u)\oplus l(u')
\end{equation}
for all $u,u'\in\Int_2^k$. By a \emph{bilinear form} on $\Int_2^k$ we mean a two-arguments function $b : \Int_2^k\times\Int_2^k\to\Int_2$ that can be represented as
\begin{equation}
  b(u,u') = u B u'^T,
\end{equation}
where $B$ is a $k\times k$ Boolean matrix. Bilinear form is \emph{symmetric} if $b(u,u') = b(u',u)$ for all $u,u'\in\Int_2^k$. Bilinear forms are exactly the functions that are linear when fixing one of the arguments.

The set of Boolean integers $\Int_2 = \{0,1\}$ can be embedded into $\Int_4 = \{0,1,2,3\}$ as the subset of elements (\emph{Teichm{\"u}ller elements}) satisfying the relation
\begin{equation}
  \Int_2 = \{x\in\Int_4 : x^2 = x\}.
\end{equation}
The XOR-summation over such elements can be defined in terms of $\Int_4$ as
\begin{equation} \label{eq:XOR-summation}
  x\oplus x' = (x+x')^2 = x + x' + 2 x x'\quad (\mathrm{mod}\; 4)
\end{equation}
for all $x,x'\in\Int_2$.

By a \emph{$\Int_4$-valued quadratic form on $\Int_2^k$} we will mean a function $q : \Int_2^k \to \Int_4$ that can be expressed as
\begin{equation}
  q(u) = 2 u s + u Q u^T,
\end{equation}
where $s\in \Int_2$, upper index ${}^T$ denotes transposition and $Q$ is a symmetric $k\times k$-matrix of special form: diagonal elements of $Q$ are from $\{0,1,2,3\}$ and taken modulo $4$, while out-of-diagonal elements are from $\{0,1\}$ and taken modulo $2$:
\begin{equation} \label{eq:matrices_of_quadratic_forms}
  Q \in
  \begin{bmatrix}
  \Int_4 & \Int_2 & \cdots & \Int_2 \\
  \Int_2 & \Int_4 & \cdots & \Int_2 \\
  \vdots & \vdots & \ddots & \vdots \\
  \Int_2 & \Int_2 & \cdots & \Int_4
  \end{bmatrix}.
\end{equation}
Let us discuss some properties of such quadratic forms. Quadratic form $q$ takes even values $2 \Int_4$ for all arguments if and only if the diagonal entries of $Q$ are even, in this case the form is called \emph{alternating} \cite{Schmidt_2009} and effectively should be considered as Boolean-valued quadratic form. The class of matrices $Q$ as in \cref{eq:matrices_of_quadratic_forms} is closed under transformations $Q\mapsto V Q V^T$ where $V$ are arbitrary Boolean matrices.

The matrix $Q$ has the capacity to contain the information about the linear part of $q$. That is, if $q$ has a linear Boolean part $q(u) = 2 u s + u Q u^t$ where $s\in\Int_2^k$, then this linear part can be added to the diagonal so that $q(u) = u \tilde{Q} u^T$ where $\tilde{Q} = Q + 2\,\mathrm{diag}(s)$. On the contrary, one can subtract vectors with values in $2\Int_4$ from the diagonal of $Q$ to obtain a linear part $s$. Later in text we choose to distinguish between linear and quadratic parts because the linear part is faster to update during computations and has nice interpretation for CSS-preserving circuits, but one may prefer the convention of storing linear and quadratic parts together as a single matrix \cite{de_Beaudrap_2022}.

Let us define for quadratic form $q$ a \emph{coboundary} $d q : \Int_2^k\times\Int_2^k \to \Int_2$ by the relation:
\begin{equation}
  dq(u,u') = u \mathcal{Q} u'^T \quad (\mathrm{mod}\;2),
\end{equation}
where we calligraphic $\mathcal{Q}$ is a matrix $Q$ in which each element is taken modulo $2$. We will use such convention throughout the text: when taking matrices $Q,M,F$ modulo $2$, we write them as $\mathcal{Q},\mathcal{M},\mathcal{F}$. Coboundary $dq$ is a Boolean symmetric bilinear form. Because of \cref{eq:XOR-summation} and because $Q$ is symmetric, it holds that
\begin{equation} \label{eq:coboundary_definition}
  q(u\oplus u') = q(u) + q(u') + 2\,dq(u,u')
\end{equation}
for all $u,u' \in \Int_2^k$.
A quadratic form $q$ is a linear form if and only if it's coboundary $d q$ is zero.

On the other hand, any function $q:\Int_2^k\to\Int_4$ satisfying $q(0)=0$ and such that
\begin{equation}
  q(u\oplus u')-q(u)-q(u') = 2 b(u,u')
\end{equation}
where $b:\Int_2^k\times\Int_2^k \to \Int_2$ is a Boolean symmetric bilinear form, is a $\Int_4$-valued quadratic form \cite{Schmidt_2009}. Indeed, fix a basis $\{e_i\}_{i=1}^k$ in $\Int_2^k$ and decompose a vector $u\in\Int_2^k$ as $u = \sum_{i=1}^k u_i e_i\, (\mathrm{mod}\, 2)$. Iterating \cref{eq:coboundary_definition} and using $q(u_i e_i) = u_i q(e_i)$, one obtains
\begin{equation}
  q(u) = \sum_i u_i \, q(e_i) + 2 \sum_{i<j} u_i \, b(e_i,e_j) \, u_j,
\end{equation}
which is a manifestly a $\Int_4$-valued quadratic form expression in the chosen basis.

\subsubsection{Quadratic forms and stabilizer groups} \label{subsubsec:quadratic_forms_stabilizer}

Suppose we are given a stabilizer group $\StabGroup\subseteq \PauliGroup^n$ of rank $r$. For each element in the stabilizer group $P\in\StabGroup$ there are unique $c\in\Int_4$ and $u\in\Int_2^{2n}$ such that $P = i^c T(u)$. Uniqueness of phase $c$ follows from $-I\notin \StabGroup$ and uniqueness of $u$ follows from linear independence of $\{T(u)\}$. Let us define $\mathcal{V}$ as the set of such $u$ for varying $P\in\StabGroup$:
\begin{equation}
  \mathcal{V} = \{ u\in \Int_2^{2n} \; : \, i^c T(u)\in\StabGroup \text{ for some } c\in\Int_4\}
\end{equation}
and define $q : \mathcal{V} \to \Int_4$ a function assigning the correct phases, so that
\begin{equation}
  i^{q(u)} T(u) \in \StabGroup \quad\text{ for all } u\in \mathcal{V}.
\end{equation}
Since $\StabGroup$ is commutative, $\mathcal{V}$ is an isotropic vector subspace of $\Int_2^{2n}$, that means the symplectic form $[\cdot,\cdot]$ is zero on $\mathcal{V}$, or equivalently $\tau$ is symmetric on $\mathcal{V}$:
\begin{equation}
  \tau(u,u') = \tau(u',u)\quad \text{ for all } u,u' \in \mathcal{V}.
\end{equation}

The definition of stabilizer group also imposes a number of properties on function $q$:
\begin{itemize}
  \item Hermiticity implies that
  \begin{equation}
    q(u) = \tau(u,u)\quad (\mathrm{mod}\;2),
  \end{equation}
  \item non-degeneracy condition $-I\notin\StabGroup$ implies that
  \begin{equation}
    q(0) = 0,
  \end{equation}
   \item Weyl-type commutation relations result in a $1$-cocycle condition
  \begin{equation}
    q(u\oplus u') = q(u) + q(u') + 2\tau(u,u') \; (\mathrm{mod}\,4)
  \end{equation}
  for all $u,u'\in\mathcal{V}$.
\end{itemize}
That means, the function $q$ is a $\Int_4$-valued quadratic form on $\mathcal{V}$ and the preceding equation rewrites to
\begin{equation} \label{eq:1-cocycle_condition}
  d q = \tau \quad \text{ on }\mathcal{V}\!\times\!\mathcal{V}.
\end{equation}

Let us choose some basis in $\mathcal{V} \cong \Int_2^r$, so that any $u\in \mathcal{V}$ is uniquely represented as $u = t P$ for row-vector $t\in\Int_2^r$ a generating $r\times 2n$ Boolean matrix $P$. Let us set $\tilde{q}(t) = q(t P)$. Function $\tilde{q}$ is a quadratic form, so it can be expressed as
\begin{gather}
  q(t P) = \tilde{q}(t) = 2 t c + t M t^T \quad (\mathrm{mod}\;4), \\
  dq(t P, t' P) = d\tilde{q}(t,t') =  t \mathcal{M} t'^T \quad (\mathrm{mod}\;2).
\end{gather}
where $c\in\Int_2^{r}$ is a column-vector and $M$ is a matrix of form \cref{eq:matrices_of_quadratic_forms}. The stabilizer group $\StabGroup$ is real if and only if $\tilde{q}$ takes only even values $2 \Int_4$, meaning the form is alternating.

Let us show that $q$ can be extended from the domain $\mathcal{V}$ to a quadratic form on the whole $\Int_2^{2n}$. Choose a $2n\times r$ Boolean matrix $J$ which is right inverse to $P$ (i.e., $PJ=I_r$) and set
\begin{equation}
  s = J c,\quad Q = J M J^T
\end{equation}
so that $q(u) = 2 u s + u\,Q\,u^T$ is a $\Int_4$-valued quadratic form on $\Int_2^{2n}$. The condition \cref{eq:1-cocycle_condition} in terms of the matrix $Q$ states that for all $u,u' \in\mathcal{V}$ it holds
\begin{equation} \label{eq:Q_from_tau}
  u \mathcal{Q} u'^T = \tau(u,u') \quad (\mathrm{mod}\, 2),
\end{equation}
but outside of $\mathcal{V}$ the bilinear forms $d q$ and $\tau$ may not be equal. The special case $u=u' \in \mathcal{V}$ means $u \mathcal{Q} u^T = \tau(u,u)$ and corresponds to the Hermiticity of the stabilizer group $\StabGroup$.

Quadratic forms over Boolean vector spaces are known to arise when studying pure stabilizer states \cite{Dehaene_2003, Bravyi_2016, de_Beaudrap_2022, Bu_2022}, but notice that in our case they will be used for describing mixed state dynamics.

\subsubsection{Standard quadratic form expansions of stabilizer operations} \label{subsubsec:standard_form}

Let us introduce a convenient notation for working with Pauli operators. Given a phase point $u\in\Int_2^{2n}$, let us denote the corresponding normalized Pauli operator and linear functional by curly bra-kets:
\begin{equation}
  \ketD{u} = \frac{1}{2^n}T(u), \quad \braD{u} = \Tr\left(T(u)^\dag\, \cdot\,\right).
\end{equation}
Similar notation is often used in works on operator spreading and Krylov complexity \cite{Parker_2019, Ermakov_2025, Lunt_2025, Nandy_2025}. In this notation, the set of ket-vectors $\ketD{u}$ forms an orthonormal basis [see \cref{eq:orthonormal_basis_of_Pauli}]:
\begin{equation} \label{eq:orthonormality_condition}
  \braketD{u}{u'} = \delta_{u,u'}.
\end{equation}
One-qubit computational basis states $\ket{0}$ and $\ket{1}$ are written as
\begin{align}
  &\proj{0} = \ketD{00} + \ketD{10}, \\
  &\proj{1} = \ketD{00} - \ketD{10},
\end{align}
maximally mixed (chaotic) state $\chi$ reads
\begin{equation}
  \chi = \frac{1}{2}I = \ketD{00},
\end{equation}
and two-qubit Bell state $\ket{\Omega}$ is written as (compare with \cref{eq:Bell_state_1}; here we omit tensor product symbol)
\begin{equation}
  \Omega = \sum_{u\in\Int_2^{2n}} \ketD{u}\ketD{u}.
\end{equation}
More generally, suppose $\rho$ is a stabilizer state with stabilizer group $\StabGroup\subseteq\PauliGroup^n$ with set of phase points $\mathcal{V}$ and $q$ a function of signs, then $\rho$ is written as
\begin{equation} \label{eq:state_as_sum_with_q}
  \rho = \sum_{u\in \mathcal{V}} i^{q(u)}\ketD{u}.
\end{equation}
Let us encode $\mathcal{V}$ either by a generating matrix $P$ such that $u\in \mathcal{V}$ if and only if $u = t  P$ for some row-vector $t$, or by a parity-check matrix $H$ such that $u\in \mathcal{V}$ if and only if $u H = 0$, then $\rho$ is represented as
\begin{center}
  \sffamily
  Quadratic form expansions of stabilizer state $\rho$
\end{center}
\begin{equation} \label{eq:state_quadratic_form_expansion}
  \rho = \sum_t i^{q(t P)} \ketD{t P} = \sum_{u: \, u H=0} i^{q(u)}\ketD{u},
\end{equation}
where the quadratic form $q: \mathcal{V}\to \Int_4$ can be written as a sum of linear and quadratic parts:
\begin{equation}
  q(u) = 2 u s + u Q u^T, \quad q(t P) = 2 t c + t M t^T.
\end{equation}
Note that the generator matrix $P$ is exactly a bit representation of a stabilizer tableau \cite{Aaronson_2004}, while signs of the tableau are encoded in $c$ and the diagonal of $M$. So, $H$ can be understood as a parity-check matrix corresponding to a stabilizer tableau. Thus, when working with stabilizer state $\rho$, it suffices to store a triple $(H,s,Q)$ in the computer. In fact, the data $(H,s,Q)$ is overdetermined because $H$ can be degenerate and because we can restore $Q$ from $H$ and $\tau$ by \cref{eq:Q_from_tau}; this abundance makes computations easier. We call the sums as in \cref{eq:state_quadratic_form_expansion} \emph{quadratic form expansions}, following \cite{de_Beaudrap_2022}. Similarly, we will call such sum \emph{linear form expansion} if the quadratic form is in fact linear.

Arbitrary Clifford channels can be represented as some quadratic form expansion. Suppose $\Phi_{A\to B}$ is a Clifford channel from input multiqubit system $A$ to output multiqubit system $B$. (At this place in the text and later we denote systems by lower index latin letters $A,B,C$ for convenience.) The Choi state of $\Phi_{A\to B}$ is
\begin{equation} \label{eq:Choi_state_definition}
  \sigma_{A B} = \Id_{A\to A}\otimes\Phi_{A\to B}[\Omega_{A A}],
\end{equation}
where $\Omega_{A A}$ is a multiqubit Bell state on two copies of input system $A$. The Choi state $\sigma_{A B}$ can be represented as a quadratic form expansion:
\begin{equation} \label{eq:sigma_quadratic_form_expansion}
  \sigma_{A B} = \sum_{t} i^{q(t P)} \ketD{t P_A}\ketD{t P_B},
\end{equation}
where $q$ is a quadratic form and $P = [P_A|P_B]$ is a generator matrix divided into columns over $A$ and $B$, let us choose $P$ to have minimal number of rows. Then, the channel $\Phi_{A\to B}$ is expressed as quadratic form expansion
\begin{equation}
  \Phi_{A\to B} = \sum_{t} i^{q(t P)} \ketbraD{t P_B}{t P_A}.
\end{equation}
That can be checked either by substitution into \cref{eq:Choi_state_definition} or by the inversion formula
\begin{equation}
  \Phi_{A\to B}[\rho_A] = 2^{\abs{A}} \Tr_A[ \sigma_{A B} (\rho_A^T\otimes I_B)],
\end{equation}
where $\abs{A}$ is the number of qubits in the system $A$.

The trace preserving condition of the channel $\Phi_{A\to B}$ is expressed as
\begin{equation}
  \braD{0_B} \Phi_{A\to B} = \braD{0_A}.
\end{equation}
This implies, together with \cref{eq:sigma_quadratic_form_expansion} and $q(0)=0$, that
\begin{equation}
  \sum_{t:\, t\neq 0,\, t P_B = 0} i^{q(tP)} \braD{t P_A} = 0.
\end{equation}
Suppose the sum in left hand side is not trivial and there exists an element $t'$ such that $t'\neq 0$ and $t' P_B = 0$. Then, in order to compensate $i^{q(t'P)} \braD{t' P_A}$ in the sum, there should also exist at least one element $t''$ such that $t''\neq t'$, $t''\neq 0$, $t'' P_A = t' P_A$. Their sum $t''' = t' \oplus t'' \neq 0$ satisfies $t''' P_A = 0$ and $t''' P_B = 0$, which contradicts the linear independence of the rows of $P = [P_A|P_B]$.
Thus, $t P_B=0$ implies $t=0$ and a rank of submatrix $P_B$ equals to the rank of $P$, and the columns of $P_A$ are linearly dependent on the columns of $P_B$. Equivalently, there exists a Boolean $2\abs{B}\times 2\abs{A}$ matrix $V_{BA}$ such that
\begin{equation}
  P_A = P_BV_{BA}.
\end{equation}

Let us find a parity-check matrix $H_B$ corresponding to generator matrix $P_B$. We get a quadratic form expansion of a channel $\Phi_{A \to B}$ in the special representation which we will call \emph{standard}:
\begin{center}
  \sffamily
  Standard quadratic form expansion of \\
  Clifford channel $\Phi$
\end{center}
\begin{equation} \label{eq:standard_form}
  \Phi_{A\to B} = \sum_{u:\, u H_B = 0} (-1)^{u\, s_B} i^{u\, Q_B u^T} \ketbraD{u}{u V_{BA}}.
\end{equation}
Here, $u \in \Int_2^{2\abs{B}}$ is a Boolean row-vector, \emph{parity-check matrix} $H_B$ is a $2\abs{B}\times k$ Boolean parity-check matrix (number of columns $k$ can be arbitrary big for degenerate matrices), \emph{shift vector} $s_B$ is a length $2\abs{B}$ column-vector representing the linear part of a quadratic form, \emph{quadratic form matrix} $Q_B$ is a $2\abs{B}\times 2\abs{B}$ matrix representing the quadratic part [$s_B$ and $Q_B$ describe quadratic form $u_B \mapsto q([u_B|u_BV_{BA}])$ where $q$ is from \cref{eq:sigma_quadratic_form_expansion}], and \emph{transition matrix} $V_{BA}$ is a $2\abs{B}\times 2\abs{A}$ Boolean matrix corresponding to the transformation of the phase space. We will identify a standard quadratic form expansion with it's tuple $(H,s,Q,V)$ (here we omit lower indices for visual clarity). We list tuples of some elementary gates used in stabilizer computations in \cref{tab:standard_forms}.

\begin{table*}[t]
  \centering
  \renewcommand{\arraystretch}{1.7}
  \begin{tabular}{|c | c | c c c c|}
    \hline
    Stabilizer operation & Quadratic form expansion & $H$ & $s$ & $Q$ & $V$ \\
    \hline\hline
      State $\ket{0}$ preparation
      & $\sum_{z} \ketD{z,0}$
      & $\begin{bsmallmatrix} 0 \\ 1 \end{bsmallmatrix}$
      & $\begin{bsmallmatrix} 0 \\ 0 \end{bsmallmatrix}$
      & $\begin{bsmallmatrix} 0 & 0 \\ 0 & 0 \end{bsmallmatrix}$
      & $\begin{bsmallmatrix} - \\ -\end{bsmallmatrix}$
      \\
    \hline
      State $\ket{1}$ preparation
      & $\sum_{z} (-1)^z \ketD{z,0}$
      & $\begin{bsmallmatrix} 0 \\ 1 \end{bsmallmatrix}$
      & $\begin{bsmallmatrix} 1 \\ 0 \end{bsmallmatrix}$
      & $\begin{bsmallmatrix} 0 & 0 \\ 0 & 0 \end{bsmallmatrix}$
      & $\begin{bsmallmatrix} - \\ -\end{bsmallmatrix}$
      \\
    \hline
      State $\ket{+}$ preparation
      & $\sum_{x} \ketD{0,x}$
      & $\begin{bsmallmatrix} 1 \\ 0 \end{bsmallmatrix}$
      & $\begin{bsmallmatrix} 0 \\ 0 \end{bsmallmatrix}$
      & $\begin{bsmallmatrix} 0 & 0 \\ 0 & 0 \end{bsmallmatrix}$
      & $\begin{bsmallmatrix} - \\ -\end{bsmallmatrix}$
      \\
    \hline
      State $\ket{-}$ preparation
      & $\sum_{x} (-1)^x\ketD{0,x}$
      & $\begin{bsmallmatrix} 1 \\ 0 \end{bsmallmatrix}$
      & $\begin{bsmallmatrix} 0 \\ 1 \end{bsmallmatrix}$
      & $\begin{bsmallmatrix} 0 & 0 \\ 0 & 0 \end{bsmallmatrix}$
      & $\begin{bsmallmatrix} - \\ -\end{bsmallmatrix}$
      \\
    \hline
      Chaotic state preparation
      & $\ketD{0,0}$
      & $\begin{bsmallmatrix} 1 & 0\\ 0 & 1 \end{bsmallmatrix}$
      & $\begin{bsmallmatrix} 0 \\ 0 \end{bsmallmatrix}$
      & $\begin{bsmallmatrix} 0 & 0 \\ 0 & 0 \end{bsmallmatrix}$
      & $\begin{bsmallmatrix} - \\ -\end{bsmallmatrix}$
      \\
    \hline
      Identity gate
      & $\sum_{u} \ketbraD{u}{u}$
      & $\begin{bsmallmatrix} - \\ - \end{bsmallmatrix}$
      & $\begin{bsmallmatrix} 0 \\ 0 \end{bsmallmatrix}$
      & $\begin{bsmallmatrix} 0 & 0 \\ 0 & 0 \end{bsmallmatrix}$
      & $\begin{bsmallmatrix} 1 & 0 \\ 0 & 1 \end{bsmallmatrix}$
      \\
    \hline
      Pauli gate $Z$
      & $\sum_{u} (-1)^x \ketbraD{u}{u}$
      & $\begin{bsmallmatrix} - \\ - \end{bsmallmatrix}$
      & $\begin{bsmallmatrix} 0 \\ 1 \end{bsmallmatrix}$
      & $\begin{bsmallmatrix} 0 & 0 \\ 0 & 0 \end{bsmallmatrix}$
      & $\begin{bsmallmatrix} 1 & 0 \\ 0 & 1 \end{bsmallmatrix}$
      \\
    \hline
      Pauli gate $X$
      & $\sum_{u} (-1)^z \ketbraD{u}{u}$
      & $\begin{bsmallmatrix} - \\ - \end{bsmallmatrix}$
      & $\begin{bsmallmatrix} 1 \\ 0 \end{bsmallmatrix}$
      & $\begin{bsmallmatrix} 0 & 0 \\ 0 & 0 \end{bsmallmatrix}$
      & $\begin{bsmallmatrix} 1 & 0 \\ 0 & 1 \end{bsmallmatrix}$
      \\
    \hline
      $Z$-dephasing gate $\Deph_Z$
      & $\sum_{z} \ketbraD{z,0}{z,0}$
      & $\begin{bsmallmatrix} 0 \\ 1 \end{bsmallmatrix}$
      & $\begin{bsmallmatrix} 0 \\ 0 \end{bsmallmatrix}$
      & $\begin{bsmallmatrix} 0 & 0 \\ 0 & 0 \end{bsmallmatrix}$
      & $\begin{bsmallmatrix} 1 & 0 \\ 0 & 0 \end{bsmallmatrix}$
      \\
    \hline
      $X$-dephasing gate $\Deph_X$
      & $\sum_{x} \ketbraD{0,x}{0,x}$
      & $\begin{bsmallmatrix} 1 \\ 0 \end{bsmallmatrix}$
      & $\begin{bsmallmatrix} 0 \\ 0 \end{bsmallmatrix}$
      & $\begin{bsmallmatrix} 0 & 0 \\ 0 & 0 \end{bsmallmatrix}$
      & $\begin{bsmallmatrix} 0 & 0 \\ 0 & 1 \end{bsmallmatrix}$
      \\
    \hline
      Qubit discarding $\Tr$
      & $\braD{0,0}$
      & $\begin{bsmallmatrix} - \end{bsmallmatrix}$
      & $\begin{bsmallmatrix} - \end{bsmallmatrix}$
      & $\begin{bsmallmatrix} - \end{bsmallmatrix}$
      & $\begin{bsmallmatrix} - \\ - \end{bsmallmatrix}$
      \\
    \hline
      Clifford gate $H$
      & $\sum_u (-1)^{z x}\ketbraD{z,x}{x,z}$
      & $\begin{bsmallmatrix} - \\ - \end{bsmallmatrix}$
      & $\begin{bsmallmatrix} 0 \\ 0 \end{bsmallmatrix}$
      & $\begin{bsmallmatrix} 0 & 1 \\ 1 & 0 \end{bsmallmatrix}$
      & $\begin{bsmallmatrix} 0 & 1 \\ 1 & 0 \end{bsmallmatrix}$
      \\
    \hline
      Clifford gate $S$
      & $\sum_u i^{-x}\ketbraD{z,x}{z\oplus x,x}$
      & $\begin{bsmallmatrix} - \\ - \end{bsmallmatrix}$
      & $\begin{bsmallmatrix} 0 \\ 0 \end{bsmallmatrix}$
      & $\begin{bsmallmatrix} 0 & 0 \\ 0 & -1 \end{bsmallmatrix}$
      & $\begin{bsmallmatrix} 1 & 0 \\ 1 & 1 \end{bsmallmatrix}$
      \\
    \hline
      Clifford gate $S^\dag$
      & $\sum_u i^x\ketbraD{z,x}{z\oplus x,x}$
      & $\begin{bsmallmatrix} - \\ - \end{bsmallmatrix}$
      & $\begin{bsmallmatrix} 0 \\ 0 \end{bsmallmatrix}$
      & $\begin{bsmallmatrix} 0 & 0 \\ 0 & 1 \end{bsmallmatrix}$
      & $\begin{bsmallmatrix} 1 & 0 \\ 1 & 1 \end{bsmallmatrix}$
      \\
    \hline
      Clifford gate $\CNOT_{12}$
      & \scalebox{0.9}{$\sum_{u_1,u_2} \ketbraD{z_1,x_1,z_2,x_2}{z_1\oplus z_2,x_1,z_2,x_2\oplus x_1}$}
      & $\begin{bsmallmatrix} - \\ - \\ - \\ - \end{bsmallmatrix}$
      & $\begin{bsmallmatrix} 0 \\ 0 \\ 0 \\ 0 \end{bsmallmatrix}$
      & $\begin{bsmallmatrix} 0&0&0&0 \\ 0&0&0&0 \\ 0&0&0&0 \\ 0&0&0&0 \end{bsmallmatrix}$
      & $\begin{bsmallmatrix} 1&0&0&0\\ 0&1&0&1\\ 1&0&1&0\\ 0&0&0&1\end{bsmallmatrix}$
      \\
    \hline
      Clifford gate $\CZ_{12}$
      & \scalebox{0.8}{$\sum_{u_1,u_2} (-1)^{x_1 x_2}\ketbraD{z_1,x_1,z_2,x_2}{z_1\oplus x_2,x_1,z_2\oplus x_1,x_2}$}
      & $\begin{bsmallmatrix} - \\ - \\ - \\ - \end{bsmallmatrix}$
      & $\begin{bsmallmatrix} 0 \\ 0 \\ 0 \\ 0 \end{bsmallmatrix}$
      & $\begin{bsmallmatrix}0&0&0&0\\ 0&0&0&1\\ 0&0&0&0\\ 0&1&0&0\end{bsmallmatrix}$
      & $\begin{bsmallmatrix} 1&0&0&0\\ 0&1&1&0\\ 0&0&1&0\\ 1&0&0&1\end{bsmallmatrix}$
      \\
    \hline
  \end{tabular}
  \caption{
    The list of common elementary stabilizer operations, their representation as quadratic form expansions, and possible choices of standard quadratic form expansions $(H,s,Q,V)$. Some of matrices have $0$ rows or $0$ columns, we indicate this abscence by the symbol ``$-$''. If the parity-check matrix $H$ has zero columns, then any variable assignment $u$ is a solution to the empty set of equations $u H = 0$.
  }
  \label{tab:standard_forms}
\end{table*}

  Thus, one can store a tuple $(H,s,Q,V)$ in the computer to represent any non-adaptive stabilizer operation, it requires about $\BigO(n^2)$ space or less. There is some freedom and there are some constrictions in definition of standard quadratic form expansion:
\begin{itemize}
  \item The parity-check matrix $H$ and transition matrix $V$ should satisfy relation on symplectic form: for all $u, u'\in\Int_2^{2n}$ such that $u\, H = u' H = 0$ it holds $[u V, u' V] = [u,u']$.
  \item The matrix $H$ has freedom in taking column operations: one can swap columns, add one column to another, also one can attach or delete zero columns to the matrix. Using Gaussian elimination over columns of $H$, one can find it's rank $r$ and take it to reduced column echelon form in time $\BigO(n k r)$.
  \item One can store the shift vector $s$ in the diagonal of $Q$ or redistribute even vectors from the diagonal to the shift vector.
  \item The quadratic form matrix $Q$ should satisfy a $1$-cocycle condition: if $u H = u' H = 0$ then
    \begin{equation}
      u \mathcal{Q} u'^T = \tau(u,u') \oplus \tau(u V, u' V) \quad (\mathrm{mod}\,2),
    \end{equation}
    which also can be written as
    \begin{equation}
      \tau(u V, u' V) = \tau(u,u') \oplus u \mathcal{Q} u'^T \quad (\mathrm{mod}\,2).
    \end{equation}
  \item Shift vector $s$, quadratic form matrix $Q$ and transition matrix $V$ have freedom in updates by column vectors from the image of $H$. The shift vector can be updated as
  \begin{equation}
    s\mapsto s\oplus H w
  \end{equation}
  for some column vector $w$. The quadratic form matrix can be updated as
  \begin{equation}
    Q \mapsto Q + H N^T + N H^T + H M H^T
  \end{equation}
  for some Boolean matrix $N$ with suitable dimensions and symmetric matrix $M$ of form \cref{eq:matrices_of_quadratic_forms}. The transition matrix can be updated as
  \begin{equation}
    V\mapsto V \oplus H W
  \end{equation}
  for some Boolean matrix $W$ with suitable dimensions.
\end{itemize}
These properties show that storing all $(H,s,Q,V)$ in memory is somewhat redundant, because $H$ can be degenerate and because we can restore non-diagonal part of $Q$ from $H$ and $\tau$. This redundancy is not demanding but allows for more efficient computations.

\subsubsection{The updates of standard quadratic form expansions under compositions} \label{subsubsec:standard_form_updates}

Suppose there is a stabilizer state $\rho_A$ described by a tuple $(H_A,s_A,Q_A)$ and a Clifford channels $\Phi_{A\to B}$ described by $(H_B,s_B,Q_B,V_{BA})$. Using orthonormality of bra-kets [see \cref{eq:orthonormality_condition}], one easily obtains a standard quadratic form expansion for the output state $\rho_B = \Phi_{A\to B}[\rho_A]$, it's standard quadratic form expansion $(H_B',s_B',Q_B')$ can be expressed as:
\begin{equation}
\begin{aligned}
  H_B' &= [H_B | V_{BA} H_A], \\
  s_B' &= s_B \oplus V_{BA} s_A, \\
  Q_B' &= Q_B + V_{BA} Q_A V_{BA}^T.
\end{aligned}
\end{equation}
Here, we construct $H_B'$ by appending columns $H_B$ with columns of $V_{AB} H_A$. This usually results in degeneracy of $H_B'$, which can be either ignored or removed by Gaussian elimination.

More generally, suppose there are two Clifford channels $\Phi_{A\to B}$ and $\Phi_{B\to C}$ with corresponding standard quadratic form expansions $(H_B,s_B,Q_B,V_{BA})$ and $(H_C,s_C,Q_C,V_{CB})$. A standard quadratic form expansion for their composition $\Phi_{A\to C} = \Phi_{B\to C}\circ\Phi_{A\to B}$ and the corresponding tuple $(H_C',s_C',Q_C',V_{CA}')$ is:
\begin{equation}
\begin{aligned}
  H_C' &= [H_C | V_{CB} H_B], \\
  s_C' &= s_C \oplus V_{CB} s_B, \\
  Q_C' &= Q_C + V_{CB} Q_B V_{CB}^T, \\
  V_{CA}' &= V_{CB}V_{BA}.
\end{aligned}
\end{equation}
Once again, $H_C'$ can become degenerate after appending new columns. Stabilizer states can be treated as Clifford channels with zero input qubits.

If there are two channels $\Phi_{A\to B}$ and $\Phi_{A'\to B'}$, then the tuple $(H_{B B'}, s_{B B'}, Q_{B B'}, V_{BB'\, AA'})$ corresponding to their parallel composition
\begin{equation}
  \Phi_{AA' \to BB'} = \Phi_{A\to B}\otimes \Phi_{A'\to B'}
\end{equation}
consists of direct sums of the corresponding matrices:
\begin{equation}
\begin{gathered}
  H_{B B'} = \begin{bmatrix} H_B & 0 \\ 0 & H_{B'} \end{bmatrix}\!, \;
  Q_{B B'} = \begin{bmatrix} Q_B & 0 \\ 0 & Q_{B'} \end{bmatrix}, \\
  s_{B B'} = \begin{bmatrix} s_B \\ s_{B'} \end{bmatrix}, \quad
  V_{BB'\, AA'} = \begin{bmatrix} V_{AB} & 0 \\ 0 & V_{A'B'} \end{bmatrix}.
\end{gathered}
\end{equation}
So, parallel composition of Clifford channels is easy to work with.

\subsubsection{Heisenberg evolution using standard quadratic form expansions} \label{subsubsec:standard_form_Heisenberg}

Let us discuss the action of Clifford channels in the Heisenberg picture. Given a Clifford channel $\Phi$ in standard quadratic form expansion $(H,s,Q,V)$ and a Pauli element $T(u)$, the action of the dual channel $\Phi^*$ on this element is
\begin{align}
  \braD{u}\Phi &= \delta_{u H = 0} (-1)^{u s} i^{u Q u^T} \braD{u V}, \\
  \Phi^*[T(u)] &= \delta_{u H = 0} (-1)^{u s} i^{- u Q u^T} T(u V),
\end{align}
which resembles the definition of quasifree maps between canonical commutation relations algebras \cite{Petz_1990}.

Next, suppose we want to study the Heisenberg evolution of some Pauli projection $\Pi$ (or a scalar multiple of a Pauli projection) defined by a stabilizer group $\StabGroup\subset\PauliGroup^n$:
\begin{equation}
  \Pi = \frac{1}{2^k} \sum_{P\in\StabGroup} P.
\end{equation}
If $k$ equals to the rank of $\mathcal{S}$, then $\Pi$ is indeed a projection, otherwise it is a scalar multiple of a projection. Also, we can set $k=\infty$ so that $\Pi=0$. Let us write $\Pi$ as a quadratic form expansion
\begin{align}
  \Pi &= \frac{1}{2^k} \sum_{w:\, w G = 0} (-1)^{w a} i^{- w M w^T} T(w K), \\
  \braD{\Pi} &= \frac{1}{2^k} \sum_{w:\, w G = 0} (-1)^{w a} i^{w M w^T} \braD{w K}.
\end{align}
Here, $k$ is an integer, $w$ is a Boolean row-vector enumerating stabilizers, $G$ is some parity-check matrix, $a$ is a column-vector of linear form and $M$ is a matrix of quadratic form (pay attention to the selected sign), $K$ is a generating matrix in which each row represent a stabilizer generator. We can store $\Pi$ in a computer as a tuple $(k; G, a, M, K)$. These matrices satifsy the following $1$-cocycle condition: for all $w, w'$ such that $w G = w' G = 0$ it holds that
\begin{equation} \label{eq:postselection_1-cocycle_condition}
  w \mathcal{M} w' = \tau(w K, w' K).
\end{equation}

Using Gaussian method, one can eliminate variables from $w$, making $G$ empty and $K$ an injective matrix, in this case $k$ equals the rank of $K$ if $\Pi$ is a projection. The advantage of storing quadratic form expansions with both $(G,K)$, however, lies in the possibility to postpone unnecessary Gaussian eliminations until needed.

The action of a channel $\Phi$ in standard quadratic form expansion $(H,s,Q,V)$ on this projection provides output
\begin{equation}
  \Pi' = \Phi^*[\Pi], \qquad \braD{\Pi'} = \braD{\Pi}\Phi,
\end{equation}
with the corresponing tuple $(k',G',a',M',K')$:
\begin{equation}
\begin{aligned}
  k' &= k, \\
  G' &= [G|KH], \\
  a' &= a\oplus K s, \\
  M' &= M + K Q K^T, \\
  K' &= K V.
\end{aligned}
\end{equation}

One important instance is when the channel is a stabilizer state $\rho$ preparation, in this case the expectation value of $\Pi$ on this state is given by a quadratic sum:
\begin{equation}
  \Tr(\Pi\rho) = \braD{\Pi}\rho = \frac{1}{2^{k'}}\sum_{w':\, w' G'=0} (-1)^{w' a'}i^{w' M' w'^T}.
\end{equation}
To compute this sum, one should do the Gaussian elimination on $G'$ and find new transform to new variables $w''$, in which the corresponding parity-check matrix $G''$ is empty and quadratic form data $(a'',M'')$ updates accrodingly:
\begin{equation}
  \Tr(\Pi\rho) = \frac{1}{2^{k'}}\sum_{w''} (-1)^{w'' a''}i^{w'' M'' w''^T}.
\end{equation}
Quadratic form $M''$ only has even values on the diagonal and zeroes out of diagonal [because of $1$-cocycle condition \cref{eq:postselection_1-cocycle_condition}], so the sum is in fact over a linear form. If this linear form is zero then $\Tr(\Pi\rho) = 2^{-k'}$, else $\Tr(\Pi\rho) = 0$. It is easy to modify this process to also jointly compute expectation values for all projections in a chosen stabilizer measurement (this corresponds to varying $a$ in the definition of $\Pi$).

The operation $\braD{\Pi} = \Tr(\Pi\,\cdot\,)$ can be interpreted as a map giving the probability that $\Pi$ is satisfied in a post-selection process \cite{Aaronson_2005,Kenbaev_2022}. More generally, one can work with trace-decreasing Clifford channels using quadratic form expansions. Computations with trace-decreasing channels are more sophisticated than working with trace-preserving channels, but they have some advantages, for example in weak simulation (see \cref{subsec:quadratic_form_simulation}). We discuss the theory of trace-decreasing Clifford channels and provide a formula for composing their standard quadratic form expansions in \hyperref[appendix:trace-decreasing_channels]{Appendix~C}.

\subsection{CSS-stabilizer groups, CSS-stabilizer states and CSS-preserving Clifford channels} \label{subsec:CSS_operations}

Suppose $\StabGroup\subset\PauliGroup^n$ is a stabilizer group with the set of phase points $\mathcal{V}$ and corresponding quadratic form of signs $q : \mathcal{V} \to \Int_4$. As mentioned earlier, the stabilizer group is real if and only if $\tau$ is alternating:
\begin{equation} \label{eq:alternating_cocycle}
  \tau(u,u) = 0 \quad \text{ for all } u\in\mathcal{V}.
\end{equation}

Let us call a stabilizer group $\StabGroup$ \emph{CSS-stabilizer group} if $\tau$ is trivial on the whole $\mathcal{V}$:
\begin{equation}
  \tau(u,u') = 0 \quad \text{ for all } u,u'\in\mathcal{V}.
\end{equation}
Obviously, CSS-stabilizer groups are real stabilizer groups. Also, $1$-cocycle condition \cref{eq:1-cocycle_condition} rewrites to
\begin{equation}
  q(u\oplus u') = q(u) + q(u'),
\end{equation}
so that the form is linear: $q(u) = 2 u s$ for some column-vector $s\in\Int_2^{2n}$. If the stabilizer group is maximal $r=n$, then one can choose a basis of $\StabGroup = \langle P_1,\dots, P_n\rangle$ in which each basis element is a Pauli string made entirely either of $Z$ or of $X$, choosing this basis in done by Gaussian elimination on rows of the stabilizer tableau \cite{Nielsen_2010}. For non-maximal stabilizer groups there may not be such a basis, for example consider the two-qubit group with single stabilizer $\langle X_1 Z_2\rangle$.
% Is it possible to choose some nice basis in arbitrary group?

We will call a stabilizer state $\rho$ \emph{CSS-stabilizer state} if the corresponding stabilizer group is CSS. Linear form expansions of CSS-stabilizer states read
\begin{equation}
  \rho = \sum_{u:\, u H = 0} (-1)^{u s} \ketD{u}.
\end{equation}
It suffices to store a pair $(H,s)$ to represent a CSS-state in the computer.

In \cref{sec:rewriting} we used the fact that if the CSS-stabilizer state is pure, then there exist two commuting stabilizer projections $\Pi_Z$ and $\Pi_X$ that are products of $Z$- and $X$-operators correspondingly and such that
\begin{equation}
  \rho = \Pi_Z \Pi_X.
\end{equation}
If $\rho$ is CSS-stabilizer and mixed, then such decomposition may \emph{not} hold: one example is a two-qubit mixed CSS-state given by a single stabilizer $\langle X_1 Z_2\rangle$. It is known that any stabilizer state $\rho$ can be purified to a pure stabilizer state $\ket{\psi}$:
\begin{equation}
  \rho = \Tr_E \proj{\psi}
\end{equation}
and any real stabilizer state purifies to real pure stabilizer state. The example of $\langle X_1 Z_2\rangle$ shows that some CSS-stabilizer states cannot be purified to a pure CSS-stabilizer state, but only to a pure real stabilizer state.

Let us call a channel $\Phi$ \emph{CSS-preserving Clifford channel} if it's Choi state is CSS-stabilizer. Any CSS-preserving Clifford channel $\Phi$ can be represented as some linear form expansion
\begin{equation}
  \Phi = \sum_{u:\, u H = 0} (-1)^{u\, s} \ketbraD{u}{u V}
\end{equation}
and can be stored as a tuple $(H,s,V)$. Parity-check matrix $H$ and transition matrix $V$ are connected by the relation: for all $u,u'\in\Int_2^{2n}$ if $u H = u' H = 0$ then $\tau(u V, u' V) = \tau(u,u')$.

The set of CSS-preserving Clifford unitaries $\CliffordGroup^n_{\text{CSS}}$ was characterized in Ref.~\cite{Delfosse_2015}, where it was shown that $\CliffordGroup^n_{\text{CSS}}$ is generated by unitary gates $\langle Z,X, \CNOT, \WH \rangle$. Analogously to what was discussed for CSS-stabilizer states, not every CSS-preserving Clifford channel can be purified to a Stinespring dilation with CSS-unitary gates. Indeed, there are some noisy stabilizer operations that cannot be written in Stinespring form with unitaries from $\langle Z,X, \CNOT, \WH \rangle$ but are CSS-preserving. For example, the ``half of the Hadamard gate'' noisy single-qubit channel
\begin{equation}
  \Phi = \sum_{z} \ketbraD{0,z}{z,0}
\end{equation}
that measures a qubit in $Z$-basis and prepares a state $\ket{+}$ or $\ket{-}$ as an outcome. This example is connected with the channel of measuring one qubit in $Z$-basis and feed-forwarding the result to $Z$-Pauli gate
\begin{equation}
  \begin{quantikz}[row sep={0.7cm,between origins},column sep={0.9cm,between origins}]
    & \meterD{Z} &\ctrl{0}\wire[d][1]{c}\setwiretype{c} &\setwiretype{n} \\
    &            &\gate{Z}                              &
  \end{quantikz}
  ,
\end{equation}
which also is CSS-preserving and not purifiable to CSS-preserving unitary circuit.
% Is it possible to easily characterise all noisy CSS-preserving gates?

Given two CSS-preserving Clifford channels $\Phi_{A\to B}$ and $\Phi_{B\to C}$ with corresponding tuples $(H_B,s_B,V_{BA})$ and $(H_C,s_C,V_{CB})$, their composition $\Phi_{A\to C} = \Phi_{B\to C}\circ\Phi_{A\to B}$ corresponds to a tuple $(H_C',s_C',V_{CA}')$:
\begin{equation}
\begin{aligned}
  H_C' &= [H_C | V_{CB} H_B], \\
  s_C' &= s_C \oplus V_{CB} s_B, \\
  V_{CA}' &= V_{CB}V_{BA}.
\end{aligned}
\end{equation}

\subsection{Stabilizer simulation using standard quadratic form expansions} \label{subsec:quadratic_form_simulation}

Suppose there is a stabilizer circuit $\mathsf{QC}$ composed of elementary gates and we want to classically simulate it. Let us introduce a simulation algorithm that uses compositions of standard quadratic form expansions and linear algebra over Boolean vectors. This algorithm is in the spirit of stabilizer-tableau methods \cite{Aaronson_2004, Gidney_2021} and is an advancement of the ideas proposed in Ref.~\cite{Yashin_2025_2}.

\subsubsection{Strong simulation} \label{subsubsec:quadratic_form_simulation_strong}

Suppose $\mathsf{QC}$ is a non-adaptive stabilizer circuit \cite{Yashin_2025,Kliuchnikov_2023}, meaning it can be represented as a composition of Clifford channels. Such circuits can be simulated in the strong sense: given an outcome, one can exactly compute it's probability.

First, find a standard quadratic form expansion for each elementary gate in the circuit. Then, represent this circuit as a computational network: to each elementary gate correspond a vertex of a graph, this vertex is labeled by the data about it's standard quadratic form expansion; and to any qubit correspond a wire (edge of the graph) connecting various gates, some wires may have free ends. The overall network represents the computational procedure done by the circuit. There is a freedom in updating this diagram without changing the computational content:
\begin{enumerate}
  \item One can transform between equivalent standard quadratic form expansions on each vertex. That means, one can perform a Gaussian elimination on a given standard quadratic form expansion to simplify it.
  \item One can contract edges in the diagram, which results in the composition of standard quadratic form expansions on the neighbouring vertices.
  \item One can glue together two parallel vertices in the diagram, taking a direct sum of two parallel stabilizer tableaux.
\end{enumerate}

From this perspective, to simulate the circuit means to fully contract the diagram and to simplify the standard quadratic form expansion on the single remaining vertex. Thus, to simulate the entire circuit reduces to a purely linear-algebraic problem of performing some sequence of matrix multiplications and Gaussian eliminations. On the other hand, this problem is complete in the class of problem computable on non-adaptive stabilizer circuits \cite{Aaronson_2004}.

If the circuit is made of trace-preserving Clifford channels, when simplifying a standard quadratic form expansion on a vertex it suffices to do Gaussian eliminations over parity-check matrix $H$ without transforming $(s,Q,V)$. During strong simulation it is often that one wants to find the probability of an outcome represented by a stabilizer projection $\Pi$. In this case it is natural to employ trace-decreasing Clifford channels (see \cref{subsubsec:standard_form_Heisenberg} and \hyperref[appendix:trace-decreasing_channels]{Appendix~C}), simplification of which is a little more sophisticated. The advantage of CSS-preserving circuits is that one does not have to compute the quadratic parts $Q$, which may be demanding in some cases.

There exist different strategies for such sequences of contractions and simplifications. The most straightforward strategy is to choose a single vertex (e.g., corresponding to initial state preparation) and sequentially update it by contracting all incident edges one-by-one (corresponding to online updates of the state under gates and measurements). This strategy is in fact equivalent to the usual textbook algorithm for Gottesman-Knill theorem \cite{Gottesman_1997,Nielsen_2010}.

Since there are different strategies for contracting the occuring networks, there is a great room for employing methods of discrete optimization to parallelize and reduce execution time of the algorithm \cite{Kliuchnikov_2013,Peham_2023}. For more thoughts on this topic consult Ref.~\cite{Yashin_2025_2}.

We note that the same perspective appears when using the language of Lagrangian relations for studying odd prime qudits stabilizer systems \cite{Booth_2024,Booth_2025}. Also, the idea studying strategies of diagrams contraction is used in the symbolic sums approach of Sum-Over-Paths \cite{Amy_2019, Amy_2023, Vilmart_2021} (there, quadratic form expansions encode pure states, unlike in our case), which was recently used for efficiently simulating hidden shift algorithms \cite{Amy_2025}.

\subsubsection{Weak simulation} \label{subsubsec:quadratic_form_simulation_weak}

Now, suppose $\mathsf{QC}$ is an adaptive stabilizer circuit, meaning the operation of gates can depend on the previous measurement outcomes. Such circuits are applicable for simulation in the weak sense, that is we can efficienlty draw samples from them.

The weak simulation of adaptive circuits can be reduced to strong simulation of a sequence of non-adaptive circuits. In fact, there may be different strategies for such reduction, the most common ones being ``bit-by-bit sampling'' and ``gate-by-gate sampling'' \cite{Bravyi_2022}. In bit-by-bit sampling, one enumerates measured bits and samples them one-after-another. Here is a realization of this strategy using strong simulation described above. First, pick first bit and discard all other measured bits, contract the occuring diagram and sample the outcome of this bit. In case $i$ bits were sampled, to sample the next $i+1$ bit, one should post-select on the already known bits and discard all the subsequent bits, then contract the diagram to sample $i+1$-th bit.

Once again, strategies may be different. Any strategy consists of a sequence of diagram contractions to sample an outcome needed to pass to the next decision time. More clever strategies should be able to reuse already contracted diagrams. So, there is lot to research and optimize \cite{Yashin_2025_2}.

For the weak simulating of CSS-preserving circuits, it is probably better to use simulation method described in \cref{sec:rewriting,subsubsec:weak_simulation}. For arbitrary stabilizer circuits, we will discuss the similar method in \cref{subsubsec:frames_simulation}.

\section{Hidden variable theory for stabilizer circuits} \label{sec:hidden_variables}

In this Section, we discuss hidden variable models for stabilizer circuits simulation. In \cref{subsec:CSS_hidden_variables}, we utilize Walsh-Hadamard-Fourier transform to introduce a quasiprobability representation on multiqubit systems. For CSS-preserving operations, the representation is non-negative, which results in a non-contextual hidden variables model. In \cref{subsec:reference_frames}, we introduce a reference frame formalism applicable to multiqubit systems, each reference frame is encoded as a quadratic form. In \cref{subsec:contextual_hidden_variables}, we utilize reference frames to create a contextual hidden variable model in which stabilizer operations update both phase space and reference frames. Finally, in \cref{subsec:magic_simulation} we apply the described simulator to also simulate stabilizer circuits with magic inputs.

\subsection{Hidden variable theory for CSS-preserving stabilizer operations} \label{subsec:CSS_hidden_variables}

In this Section we will use the linearity of CSS-preserving stabilizer operations to construct a probabilistic model for CSS-preserving circuits using Walsh-Hamadard-Fourier transform.

\subsubsection{Walsh-Hadamard-Fourier transform} \label{subsubsec:Walsh-Hadamard-Fourier_transform}

Let us briefly touch on Fourier analysis over Boolean variables \cite{ODonnell_2021}. Given a complex-valued Boolean function $h : \Int_2^{2n} \to \Comp$, its Walsh-Hadamard-Fourier transform $\tilde{h}$ is defined as
\begin{equation}
\begin{aligned}
  &\tilde{h}(v) = \frac{1}{2^{2n}}\sum_u (-1)^{u v} h(u), \\
  &h(u) = \sum_v (-1)^{u v} \tilde{h}(v),
\end{aligned}
\end{equation}
where we treat variable $v \in \Int_2^{2n}$ as a column vector and $u\in\Int_2^{2n}$ as a row vector. (We use the naming ``Walsh-Hadamard-Fourier transform'' for the sake of not interfering with Walsh-Hadamard gate $\WH$.) Suppose $h$ is a linear function with support on a vector subspace defined by a $2n\times k$ parity-check matrix $H$:
\begin{equation}
  h(u) = (-1)^{u s} \delta_{u H = 0}.
\end{equation}
The Walsh-Hadamard-Fourier transform of such function is a uniformly random distribution on an affine subspace of $\Int_2^{2n}$:
\begin{equation}
  \tilde{h}(v) = \frac{1}{2^r} \delta_{v \in s\oplus H\Int_2^k},
\end{equation}
where $r$ is the rank of $H$. Because Walsh-Hadamard-Fourier tranform is invertible, it gives a one-to-one correspondence between linear Boolean functions on vector subspaces and uniform distributions on affine subspaces. We will use this property in \cref{subsubsec:symbols_of_CSS} to construct convenient probabilistic model of CSS-preserving stabilizer circuits.

\subsubsection{Quasiprobability representation using Walsh-Hadamard-Fourier transform} \label{subsubsec:quasiprobability_representation}

Given some $n$-qubit quantum state $\rho$, let us define a function $p_\rho : \Int_2^{2n} \to \Comp$:
\begin{equation}
  p_\rho(v) = \frac{1}{2^{2n}}\sum_u (-1)^{u v} \braD{u}\rho
\end{equation}
Note that this function real-valued for a real state $\rho$, but generally is complex-valued.
Trace condition $\Tr(\rho) = 1$ implies that
\begin{equation}
  \sum_v p_\rho(v) = \braD{0}\rho = 1,
\end{equation}
so that $p_\rho$ is in fact a \emph{(complex) quasiprobability distribution} (probability distribution with possible negative and imaginary signs). We will call $p_\rho$ \emph{the symbol of a state $\rho$}, as in ``the symbol of a differential operator''. We choose not use the term ``Wigner function'' \cite{Wigner_1932, Wootters_1987, Delfosse_2015, Bu_2023, Bu_2023_2} because it uses the symplectic Fourier transform instead of the usual Fourier transform used by the term ``symbol'', but of course these notions are almost identical.

Alternative approach to defining symbols is to introduce \emph{phase point operators} \cite{Gross_2006,Gross_2006_2,Delfosse_2015}
\begin{equation}
  A(v) = \frac{1}{2^n}\sum_{u} (-1)^{u v} T(u).
\end{equation}
Phase point operators are real but not Hermitian. Phase point operators have unit trace $\Tr A(v) = 1$, they sum to identity $\frac{1}{2^n}\sum_v A(v) = I$ and constitute an orthonormal basis with respect to Hilbert-Schmidt inner product:
\begin{equation}
  \frac{1}{2^n}\Tr(A(v)^\dag A(v')) = \delta_{v,v'}.
\end{equation}
The symbol $p_\rho$ corresponds to the decomposition of $\rho$ in this basis:
\begin{align}
  &p_\rho(v) = \frac{1}{2^n}\Tr(A(v)^\dag \rho), \\
  &\rho = \sum_v p_\rho(v) A(v).
\end{align}

Suppose $\Phi$ is a multiqubit quantum channel. Let us define \emph{the symbol of a channel $\Phi$} by a two-sided Walsh-Hadamard-Fourier tranform:
\begin{equation}
  p_\Phi(v|v') = \frac{1}{2^{2n}}\sum_{u,u'} (-1)^{u v}\braD{u}\Phi\ketD{u'} (-1)^{u' v'}.
\end{equation}
Trace-preserving condition implies that $p_\Phi$ is a quasiprobability transition function (Markov map):
\begin{equation}
  \sum_v p_\Phi(v|v') = \sum_{u'} \braD{0}\Phi\ketD{u'} (-1)^{u' v'} = 1.
\end{equation}
The symbol of the composition of two channels $\Phi_{A\to C} = \Phi_{B\to C}\circ \Phi_{A\to B}$ is the composition of two symbols:
\begin{equation}
\begin{aligned}
  &p_{\Phi_{A\to C}}(v_C|v_A) \\
  &= \sum_{v_B}p_{\Phi_{B\to C}}(v_C|v_B) p_{\Phi_{A\to B}}(v_B|v_A).
\end{aligned}
\end{equation}

In terms of phase point operators, the channel $\Phi$ is expressed as
\begin{align}
  & p_\Phi(v|v') = \frac{1}{2^n} \Tr\left(A(v)^\dag \,\Phi[A(v')]\right), \\
  & \Phi[\rho] = \sum_{v,v'} A(v) p_\Phi(v|v') \; \frac{1}{2^n}\Tr(A(v')^\dag\,\rho).
\end{align}

Thus, using Pauli operators and Walsh-Hadamard-Fourier transform, we construct a basis for the space of matrices and a quasiprobability representation of quantum theory \cite{Ferrie_2011, Yashin_2020, Kulikov_2024}. This quasiprobability representation turns out to be well-suited for CSS-preserving subtheory.

\subsubsection{Symbols of CSS-preserving operations} \label{subsubsec:symbols_of_CSS}

If $\rho$ is a CSS-stabilizer state
\begin{equation}
  \rho = \sum_{u:\, u H = 0} (-1)^{u s} \ketD{u},
\end{equation}
then the corresponding symbol is a uniform distribution over affine subspace:
\begin{equation}
  p_\rho(u) = \frac{1}{2^r} \delta_{v \in s\oplus H\Int_2^k}.
\end{equation}
That means, CSS-stabilizer states correspond to affine distributions on $\Int_2^{2n}$. In fact, a version of the discrete Hudson's theorem proven in Ref.~\cite{Delfosse_2015} shows that the pure state $\ket{\psi}$ corresponds to a positive symbol if and only if this state is CSS-stabilizer.

If $\Phi$ is a  CSS-preserving Clifford channel
\begin{equation}
  \Phi = \sum_{u:\, u H = 0} (-1)^{u s}\ketbraD{u}{u V},
\end{equation}
then the corresponding symbol reads
\begin{equation} \label{eq:symbol_of_Clifford_channel}
  p_\Phi(v|v') = \frac{1}{2^r} \delta_{v \in Vv' \oplus s \oplus H\Int_2^k},
\end{equation}
meaning that CSS-preserving Clifford channels act as Markov mappings defined by probabilistic affine Boolean circuits: given an input bit string $v'$, the transition matrix $V$ transforms the bit string to $v'\mapsto V v'$, vector $s$ introduces a shift $V v'\mapsto V v'\oplus s$, and parity-check matrix $H$ introduces additional randomness.

Moreover, symbol representation also works well with trace-decreasing CSS-preserving Clifford channels (see \cref{subsubsec:standard_form_Heisenberg} and \hyperref[appendix:trace-decreasing_channels]{Appendix~C}). If we want to check the post-selection condition for CSS-projection $\Pi$
\begin{equation}
  \braD{\Pi} = \frac{1}{2^k} \sum_{w: w G = 0} (-1)^{w a}\braD{w K},
\end{equation}
where $K$ is a Boolean matrix with linear independent row, $c$ is a column-vector, then the corresponding symbol reads
\begin{equation}
  p_{\braD{\Pi}}(v) = \frac{1}{2^{d}}\delta_{K v \in c\oplus G \Int_2^{b}},
\end{equation}
where $d$ and $b$ are suitable integers. For symbols of general trace-decreasing Clifford channels consult \hyperref[appendix:trace-decreasing_channels]{Appendix~C}.

\subsubsection{Hidden variables simulation of CSS-preserving stabilizer circuits} \label{subsubsec:CSS_hidden_variables_simulation}

By comparing the standard quadratic form expansions of elementary operations (\cref{tab:standard_forms}) with the rewriting rules from \cref{sec:rewriting}, we see that rewriting rules perform exactly the Markov maps described by the symbols of corresponding Clifford channels. Indeed, the elementary operations $X,Z,\CNOT,\Deph_Z,\Deph_X,\Tr$ all work as described by their standard quadratic form expansions $(H,s,V)$ see \cref{tab:standard_forms}. As for the Walsh-Hadamard gate $\WH$ on $n$ qubits, it has an additional quadratic phase factor $\sum_{i=1}^n z_i x_i \; (\mathrm{mod}\,2)$ which equals $\tau(u,u)$ for all $u = (z,x)$ and is trivial for all real inputs [recall \cref{eq:alternating_cocycle}]. That clarifies and gives another proof for why \cref{sec:rewriting} is correct.

The constructed quasiprobability theory realizes a non-contextual hidden variable model for CSS-preserving stabilizer operations also known as Spekkens' toy model \cite{Spekkens_2007, Catani_2017}. So, the simulation methods described in \cref{sec:rewriting} rely on the correctness of this model.

Let us emphasize one more time that the described symbol representation $p_\Phi$ is almost exactly the same thing as discrete Wigner function for rebits described in Ref.~\cite{Delfosse_2015}, except that we do the usual Fourier transform instead of symplectic Fourier transform, which results in different enumeration of the phase space incoherent with rewriting rules in \cref{sec:rewriting}. The novelty of our work is that we reason using Clifford channels, which is arguably more natural and straightforward, and we show a concrete relationship between the structure of the distribution and the information about the stabilizer tableau manifested in standard quadratic form expansion \cref{eq:standard_form}.

\subsection{Reference frames for stabilizer formalism} \label{subsec:reference_frames}

In the previous Section, we constructed a convenient phase space representation for CSS-preserving stabilizer operations, which allows for direct rewriting of a CSS-preserving stabilizer to a classical circuit while providing correct outputs. Is it possible to extend the approach to include arbitrary multiqubit stabilizer circuits? In the strict sense, the answer is negative: Ref.~\cite{Schmid_2022} proves that there do not exist non-contextual hidden variable models for even-dimensional stabilizer theories.

Still, in the broader sense, it turns out that we can construct probability representations that are dependent on the reference frame chosen on the system, in such representation gates should update both quasiprobabilities and reference frames. This idea was proposed in Ref.~\cite{Park_2024} under the naming \emph{framed Wigner function}. We will enhance the method of Ref.~\cite{Park_2024} by using $\Int_4$-valued quadratic forms.

\subsubsection{Stabilizer reference frames as quadratic forms} \label{subsubsec:reference_frames_as_quadratic_forms}

Trying to figure out the correct definition for general quantum reference frames is a prominent problem in quantum foundations and quantum field theory \cite{Loveridge_2018, Giacomini_2019, Giacomini_2019_2, Carette_2025, Fewster_2024, Mangiarotti_1998}. One important fact is that the presence of reference frames on physical systems allows for information transmission between agents \cite{Bartlett_2007, Skotiniotis_2012, Chitambar_2019}. Without going deep into discussion on this fascinating subject, let us propose the definition for one class of reference frames in multiqubit systems that is convenient for studying stabilizer theory.

Firstly, let us consider the example of reference frames on the system of classical bit strings $\Int_2^n$. In this system, the ability to choose different reference frames may come from the ability to shift the origin of a bit string. That is, when going from one reference frame to another, we reinterpret a bit string as $x\mapsto x\oplus f$ for some bit string $f\in\Int_2^n$, so we may identify reference frames with vectors $f\in\Int_2^n$. We will use the analogous identification for quantum bits, but instead of Boolean vectors we will employ $\Int_4$-valued quadratic forms.

Consider a $n$-qubit system with defined Pauli group $\PauliGroup^n$. In \cref{subsubsec:Pauli_group}, we defined Pauli operators by a set of sign-free operators: for all $u = (z,x) \in\Int_2^{2n}$ we write
\begin{equation}
  T(u) = Z(z) X(x).
\end{equation}
Actually, we could have chosen the phases for $T(u)$ differently. Let us identify a reference frame $F$ with a choice of some quadratic form $f : \Int_2^{2n} \to \Int_4$ described by (here we abuse the notation) symmetric matrix $F$:
\begin{equation}
  f(u) = u F u^T.
\end{equation}
Note that we do not distinguish between linear and quadratic parts in $f$. In case of zero qubits $n=0$, the frame $F$ is unique.

Let us define \emph{framed Pauli operators} (interpretation of Pauli matrices inside of reference frame $F$) as
\begin{equation}
 T^F(u) = i^{f(u)} T(u) = i^{f(u)} Z(z) X(x).
\end{equation}
The ``standard'' reference frame used in the previous Sections corresponds to zero quadratic form, we will call it \emph{CSS-reference frame}. We will indicate CSS-reference frame by the lack of upper index ${}^F$. The $2$-cocycle $\tau^F$ which appears when multiplying framed Pauli operators
\begin{equation}
  T^F(u) T^F(u') = (-1)^{\tau^F(u,u')} T^F(u\oplus u')
\end{equation}
is expressed as a sum of $\tau$ with coboundary of $f$:
\begin{equation}
  \tau^F = \tau \oplus df.
\end{equation}
The choice of a frame $F$ does not have impact on the symplectic structure of $\Int_2^{2n}$:
\begin{equation}
\begin{aligned}
  [u,u']
  &= \tau^F(u,u') \oplus \tau^F(u',u) \\
  &= \tau(u,u') \oplus \tau(u',u).
\end{aligned}
\end{equation}
Speaking in terms of group cohomology, $f$ is a $1$-chain corresponding to the transition between transversals $T$ and $T^F$. The fact that $f$ is a quadratic form results in $\tau^F$ being bilinear Boolean form.

In general, one could take $f$ to be an arbitrary function $f:\Int_2^{2n}\to \Int_4$, but that would result in inefficiencies during simulation. In fact, Ref.~\cite{Park_2024} uses degree $3$ Boolean polynomials for $f$, which results in inefficient $\BigO(n^3)$-memory requirements. In case one takes $f(u) = - \sum_i z_i x_i$, one gets a more popular frame in which $T^F(11) = Y$. Such $f$ is not a Boolean bilinear form, so $\tau^F$ is not bilinear and not so pleasing to compute; also, $\tau^F$ becomes non-Boolean and takes values in $\Int_4$ (consider $Z X = i Y$), unlike in reference frames defined by quadratic forms.

\subsubsection{Stabilizer operations in varying reference frames} \label{subsubsec:reference_frames_operations}

Let us introduce operator bra-ket notation in the reference frame $F$ as
\begin{equation}
\begin{aligned}
  &\ketD{u^F} = i^{f(u)}\ketD{u} = \frac{1}{2^n}T^F(u), \\
  &\braD{u^F} = i^{-f(u)}\braD{u} = \Tr(T^F(u)^\dag\,\cdot\,).
\end{aligned}
\end{equation}
The orthogonality condition for operators in different frames $F$ and $F'$ with corresponding $f$ and $f'$ reads
\begin{equation}
  \braketD{u^F}{u'^{F'}} = i^{f'(u')-f(u)} \delta_{u,u'}.
\end{equation}

If a stabilizer state $\rho$ was written as quadratic form expansion in CSS-reference frame
\begin{equation}
  \rho = \sum_{u: u H = 0} i^{q(u)}\ketD{u},
\end{equation}
then in reference frame $F$ it rewrites to
\begin{equation}
    \rho = \sum_{u: u H = 0} i^{q^F(u)}\ketD{u^F},
\end{equation}
where $q^F(u) = q(u)-f(u)$. Setting $f = q$, we see that for any stabilizer state there exists a reference frame in which it's quadratic form is trivial.

Given a stabilizer state $\rho$, let us call a reference frame $F$ \emph{proper} if the state has linear form expansion in this reference frame. For $F$ to be proper, it is necessary and sufficient that $\tau^F = 0$ on $\mathcal{V}\!\times\!\mathcal{V}$, or equivalently
\begin{equation}
  u \mathcal{F} u'^T = \tau(u,u') \quad (\textrm{mod}\,2)
\end{equation}
for all $u H = u' H = 0$.

As an illustrative example, consider a $n$-qubit graph state $\ket{\Gamma}$ with incidence matrix $\Gamma$ \cite{Hein_2006, Anders_2006}:
\begin{equation} \label{eq:graph_state_definition}
  \ket{\Gamma} = \prod_{(i,j)\in \Gamma}\!\CZ_{ij} \cdot \ket{+}^{\otimes n}.
\end{equation}
For this state, it is natural to choose the reference frame $F$ defined by quadratic form
\begin{equation} \label{eq:graph_state_frame}
  f(u) = x \Gamma x^T
\end{equation}
for all $u = (z,x) \in \Int_2^{2n}$. In this reference frame, the state $\ket{\Gamma}$ has expansion with trivial quadratic form:
\begin{equation}
  \proj{\Gamma} = \sum_{u :\, z = \Gamma x} \ketD{u^F}.
\end{equation}
So, graph states $\ket{\Gamma}$ are improper in CSS-reference frame but proper in the reference frame defined by $\Gamma$.

Suppose $\Phi_{A\to B}$ is a Clifford channel with quadratic form expansion
\begin{equation}
  \Phi_{A\to B} = \sum_{u:\, u H_B = 0} i^{q_B(u)}\ketbraD{u}{u V_{BA}}.
\end{equation}
If on the input system we choose a frame $F_A$ and on the output system we choose a frame $F_B$, the quadratic form expansion reads
\begin{equation}
  \Phi_{A\to B}^{F_A\to F_B} = \hspace*{-0.5em} \sum_{u:\, u H_B = 0} \hspace*{-0.5em} i^{q_B^{F_A\to F_B}(u)}\ketbraD{u^{F_B}}{u V_{BA}{}^{F_A}},
\end{equation}
where the quadratic form $q_B^{F_A\to F_B}$ is
\begin{equation}
  q_B^{F_A\to F_B} (u) = q_B(u)-f_B(u)+f_A(u V_{BA}).
\end{equation}
Setting $f_B(u) = f_A(u V_{BA}) + q_B(u)$, we see that for any Clifford channel $\Phi_{A\to B}$ and any input reference frame $F_A$ there exists an output reference frame $F_B$ in which the quadratic form of this channel is trivial.

Given a Clifford channel $\Phi_{A\to B}$, we call a pair of input and output reference frames $F_A$ and $F_B$ \emph{proper} if the channel has linear form expansion with respect to these reference frames.
For the quadratic form $q_B^{F_A\to F_B}$ to contain only linear part, it is necessary and sufficient that
\begin{align}
  d f_B(u,u') &= d q_B(u,u') \oplus d f_A(u V_{BA}, u' V_{BA}), \\
  u \mathcal{F}_B u'^T &= u (\mathcal{Q}_B \oplus V_{BA} \mathcal{F}_A V_{BA}^T) u'^T
\end{align}
for all $u H_B = u' H_B = 0$, both equations are taken modulo $2$.

\subsubsection{Symbols of stabilizer operations in varying reference frames} \label{subsubsec:reference_frames_symbols}

Given a state $\rho$, let us call the \emph{framed symbol} of this state a function $p_\rho^F : \Int_2^{2n}\to\Comp$ defined as
\begin{equation}
  p_\rho^F(v) = \frac{1}{2^{2n}}\sum_u (-1)^{u v} \braD{u^F}\rho.
\end{equation}
The \emph{framed phase point operators} can be interpreted as quadratic Fourier tranforms of Pauli operators:
\begin{equation}
  A^F(v) = \frac{1}{2^n}\sum_u (-1)^{u v} i^{f(u)} T(u).
\end{equation}

Suppose a stabilizer state $\rho$ has linear form expansion in a proper reference frame $F$:
\begin{equation}
  \rho = \sum_{u: u H = 0} (-1)^{u s} \ketD{u^F}.
\end{equation}
The symbol of this state is a uniform distribution over affine subspace
\begin{equation}
  p_\rho^F(v) = \frac{1}{2^r} \delta_{v\in s\oplus H\Int_2^k}.
\end{equation}
Note that in case the reference frame $F$ is not proper, $p_\rho^F$ is not a distribution but generally a complex quasidistribution.

Likewise, if we consider a Clifford channel $\Phi$ and proper reference frames $F$ and $F'$ such that
\begin{equation}
  \Phi^{F'\to F} = \sum_{u:\, u H = 0} (-1)^{u s}\ketbraD{u^F}{u V{}^{F'}},
\end{equation}
then the corresponding symbol reads
\begin{equation} \label{eq:framed_symbol_of_Clifford_channel}
  p_\Phi^{F'\to F}(v|v') = \frac{1}{2^r} \delta_{v \in Vv' \oplus s \oplus H\Int_2^k},
\end{equation}
it acts as a Markov mapping defined by probabilistic affine Boolean circuit.

When considering post-processing tasks, it is not always possible to choose a frame that makes it proper, i.e. having only linear part in the expansion. Indeed, suppose that $\Pi$ is a stabilizer projection and $\braD{\Pi^F}$ has a non-linear quadratic form expansion. Considered as a map with zero output qubits, one cannot choose a reference frame $F'$ on the output that makes it into a linear form expansion. More generally, if $\Phi$ is a trace-decreasing Clifford channel (see \hyperref[appendix:trace-decreasing_channels]{Appendix~C}) with a chosen reference frame $F$ on the input, sometimes one cannot choose a proper reference frame $F'$ on the output that makes $\Phi^{F\to F'}$ into a linear form expansion. Fortunately, it is not crucial to consider post-processing for weak simulation using quasiprobability models.

\subsection{Contextual hidden variable theory for stabilizer operations} \label{subsec:contextual_hidden_variables}

The introduced encoding of reference frames allows us to construct a sort of contextual hidden variable theory describing stabilizer operations, in which the chosen reference frame plays the role of context. For previous explorations in contextual hidden variable models of stabilizer theory and beyond, consult Ref.~\cite{Hindlycke_2022} and a series of works Refs.~\cite{Zurel_2020,Raussendorf_2020,Okay_2021,Zurel_2024,Zurel_2024_2,Zurel_2024_3}. Our approach to encoding context appears to be new with respect to the listed works, but it lacks generality to describe non-stabilizer operations.

\subsubsection{Reference frames as contexts} \label{subsubsec:frames_as_contexts}

Consider a system of $n$ qubits. Let us store the information about a reference frame $F$ as a quadratic form $2n\times 2n$-matrix $F$:
\begin{equation}
  f(u) = u F u^T \qquad (\mathrm{mod}\,4).
\end{equation}
In case of zero qubits $n=0$, the matrix $F$ is $0\times 0$ and unique.

If some channel $\Phi$ with standard quadratic form expansion $(H,s,Q,V)$ acts on this system, let us \emph{choose} the proper reference frame on the output system by the rule
\begin{equation} \label{eq:reference_update_rule}
  F' = Q + V F V^T.
\end{equation}
That is, we update the frame in accordance with transition matrix $V$ and a quadratic part $Q$, ignoring the linear part $s$. Optionally, one can also add a correction to the frame:
\begin{equation} \label{eq:frames_correction}
  F' \mapsto F' + H N^T + N H^T + H M H^T
\end{equation}
for some Boolean matrix $N$ with suitable dimensions and a quadratic form matrix $M$ (recall \cref{subsubsec:standard_form}). This correction does not affect the system, but it might come handy for simplifying the matrix $F'$.

If during the evolution of the system one updates reference frame by the rule \cref{eq:reference_update_rule}, then at any point in time the frame is proper and all stabilizer operations have linear form expansion between the chosen reference frames. In other words, any stabilizer circuit effectively becomes a CSS-preserving circuit when interpreted in terms of evolving reference frames.

This gives a construction of a hidden variables theory generalizing the one described in \cref{subsec:CSS_hidden_variables}, which stores and processes additional information about reference frames. To be more precise, the set of all possible phase points in this model consists of pairs $(v,F)$ where $v\in\Int_2^{2n}$ is a column-vector and $F$ is a quadratic form matrix. Any stabilizer state can be represented as a probability distribution over phase points $(u,F)$ for the proper fixed reference frame $F$. Such representation is not unique because there are many proper $F$. Any stabilizer operation is described as a probabilistic map that transforms $(v,F)$ to $(v',F')$ according to \cref{eq:framed_symbol_of_Clifford_channel,eq:reference_update_rule}.

Let us examine how the reference frame updates work on the counterexample discussed in \cref{subsec:non-CSS_incorrectness}. During initialization and after first $\CNOT$, we live in the CSS-frame. After doing the $\CZ$-gate, we update the frame so that
\begin{equation}
  F =
  \begin{bmatrix}
    0 & 0 & 0 & 0 \\
    0 & 0 & 0 & 1 \\
    0 & 0 & 0 & 0 \\
    0 & 1 & 0 & 0
  \end{bmatrix}
  .
\end{equation}
After doing the second $\CNOT$, the reference frame updates to
\begin{equation}
  F =
  \begin{bmatrix}
    0 & 0 & 0 & 0 \\
    0 & 2 & 0 & 1 \\
    0 & 0 & 0 & 0 \\
    0 & 1 & 0 & 0
  \end{bmatrix}
  .
\end{equation}
The diagonal element $2$ results in a bit flip that appears during the measurement of the first qubit in $X$-basis if we transform back to a CSS-reference frame. This explains the disagreement of the outcomes observed in \cref{subsec:non-CSS_incorrectness}.

How to correctly find classical outcomes of qubit measurements in CSS-reference frame? As discussed in \cref{subsec:preliminaries_circuits}, we can treat classical bits as dephased qubits. Consider a stabilizer state $\rho$ defined by phase points $\mathcal{V}$ and quadratic form $q$. When some qubits are dephased, it is an instance of the situation when there is subspace $\mathcal{V}_c\subseteq\mathcal{V}$ such that $\tau(u_c,u) = 0$ for all $u_c\in\mathcal{V}_c$ and $u\in\mathcal{V}$. Then by $1$-cocycle condition for quadratic form it holds
\begin{equation}
  q(u_c\oplus u) = q(u_c) + q(u),
\end{equation}
implying that $q$ is linear on $\mathcal{V}_c$. That means, CSS-reference frame restricted to classical bits is necessarily proper for this state, the meaningful part of quadratic form $q$ on the classical bits is linear, and there exists a proper correction \cref{eq:frames_correction} of the matrix $F$ to have only even diagonal elements on classical bits. It is not necessary to find the concrete correction but suffices to look at even diagonal elements of the matrix $F$.

To illustrate this behaviour, let us look at a graph state again:
\begin{equation}
  \proj{\Gamma} = \sum_{u:\,z=x\Gamma} \ketD{u^F},\qquad f(u) = x\Gamma x.
\end{equation}
Dephasing it on $X$ over all qubits, the result is
\begin{equation}
  \Deph_X^{\otimes n}[\proj{\Gamma}] = \sum_{x:\,x\Gamma=0}i^{x\Gamma x} \ketD{0,x} = \sum_{x:\,x\Gamma=0} \ketD{0,x},
\end{equation}
because the sum is taken over $x\Gamma = 0$. We conclude that the $X$-dephased graph state is linear in both reference frame $F$ and CSS-reference frame.

As another example, consider the circuit
\begin{equation}
\begin{quantikz}
  \lstick{$\ket{+}$} & \gate{S} & \meterD{X} & \setwiretype{c}
\end{quantikz}
\end{equation}
After preparing the state $\ket{+}$ and using the gate $S$, the proper reference frame can be chosen as
\begin{equation}
  F =
  \begin{bmatrix}
    0 & 0 \\
    0 & -1
  \end{bmatrix}
  .
\end{equation}
Measuring the qubit in $X$-basis results in imaginary reference frame on the classical bit, which looks as nonsense. An explanation to this nonsense is that if diagonal values $\{1,3\}$ appear on a classical bit, then necessary this bit is random and the reference frame $F$ can be corrected by \cref{eq:frames_correction}. In the example under consideration, $H$ is an indentity matrix and one can substitute $M=-F$ into \cref{eq:frames_correction} to correct the diagonal. A more confident way to circumvent this problem is to introduce R-I qubit (see \cref{subsec:preliminaries_circuits}) and work inside of real stabilizer theory.

\subsubsection{Hidden variables simulation of non-CSS-preserving stabilizer circuits} \label{subsubsec:frames_simulation}

The introduction of reference frames allows us to enhance the simulation method of \cref{sec:rewriting} to include non-CSS-preserving stabilizer circuits. The difference between CSS-preserving and non-CSS-preserving case is that additional reference frames processing is required.

Suppose $\mathsf{QC}$ is a non-adaptive stabilizer circuit. We rewrite it to a classical circuit $\mathsf{CC}$ using rewriting rules from \cref{subsec:rewriting_rules} (including non-CSS-preserving rules from \cref{subsubsec:non-CSS_rules}). Arbitrary stabilizer operation in standard quadratic form expansion $(H,s,Q,V)$ also rewrites to a classical gate depending on $(H,s,V)$. Note that if we store the linear part $s$ inside of the quadratic form $Q$ for all operations, then $\mathsf{CC}$ becomes a linear Boolean circuit instead of affine. One can simulate the classical circuit $\mathsf{CC}$ using the methods discussed in \cref{subsec:classical_simulation}.

The classical circuit $\mathsf{CC}$ models the stabilizer circuit $\mathsf{QC}$ inside of evolving reference frames, but the measurement outcomes of $\mathsf{CC}$ and $\mathsf{QC}$ can differ in CSS-reference frame, as shown in \cref{subsec:non-CSS_incorrectness} (note that this explains why finding if the measurement outcome is deterministic or random is easier than finding the value of the deterministic outcome \cite{Aaronson_2004}). To adjust for this difference, one should store and update reference frame information. At each time point the reference frame is stored as a quadratic form matrix $F$, and updated by the rule \cref{eq:reference_update_rule} after the action of stabilizer operations.

If adaptivity is present in the circuit $\mathsf{QC}$, to ensure the feedforward is correct it is necessary to synchronize with CSS-reference frame (to look at even diagonal elements of matrices $F$) during qubit measurements. This can also be seen as simulating the sequence of non-adaptive stabilizer circuits between decision points.

Computing reference frame updates is more demanding than doing weak simulation on $\mathsf{CC}$: drawing a single sample [in $\BigO(L)$ time] is faster than finding a correct reference frame [in $\BigO(n L)$ time]. If the circuit $\mathsf{QC}$ is non-adaptive, then the reference frame updates do not depend on the preceding measurements and one can preliminarily compute the reference frame on the final time point. If the circuit is adaptive, then one has to compute reference frames on all decision points depending on the previously sampled outcomes.

\subsubsection{Refined understanding of non-CSS-ness} \label{subsubsec:non-CSS-ness}

Because CSS-preserving stabilizer operations do not have quadratic part, they cannot take the reference frame ``further away'' from CSS-frame. Thus, non-CSS-preserving Clifford channels (such as $H$ or $S$ or $\CZ$) can be considered in the spirit of resource theories \cite{Chitambar_2019} as carrying the resource of ``non-CSS-ness''. Resource-free processes (CSS-preserving stabilizer circuits) have non-contextual hidden variable theory, and the introduction of non-CSS-ness results in contextuality. It might be useful to explore this resource more deeply.

Non-CSS-ness of a reference frame $F$ is explained by its quadratic form matrix. Let us divide this resource into two parts: diagonal and non-diagonal. By \emph{imaginarity} we will understand a resource described by diagonal elements (modulo $2$) of the reference frame matrix $F$; by \emph{graphness} we will understand the resource described by out-of-diagonal elements of the reference frame matrix $F$.

\begin{figure}[h]
  \centering
  \resizebox{0.45\textwidth}{!}{
  \begin{tikzpicture}[font=\large\sffamily]
    \draw[thick,fill=blue!10!white] (0pt,0pt) ellipse (140pt and 70pt);
    \draw[fill=blue!15!white] (-30pt,-10pt) ellipse (100pt and 50pt);
    \draw[fill=blue!20!white] (-0pt,-20pt) ellipse (60pt and 30pt);

    \node at (0pt,-20pt) {\begin{tabular}{c}CSS-preserving\\ operations\end{tabular}};
    \node[rotate=30] at (-87pt,7pt) {\begin{tabular}{c}real stabilizer\\ operations\end{tabular}};
    \node[rotate=-30] at (90pt,32pt) {\begin{tabular}{c}stabilizer\\ operations\end{tabular}};

    \path (-55pt,-30pt) -- (-65pt,-30pt)
      node[
        draw=black,
        fill=blue!5!white,
        single arrow,
        minimum height=55pt,
        inner sep=2pt,
        single arrow head extend=4pt,
        shape border rotate=180,
        midway
      ]
      {\footnotesize graphness};

    \path (55pt,-30pt) -- (65pt,-30pt)
      node[
        draw=black,
        fill=blue!5!white,
        single arrow,
        minimum height=55pt,
        inner sep=2pt,
        single arrow head extend=4pt,
        midway
      ]
      {\footnotesize imaginarity};

  \end{tikzpicture}
  }
  \caption{
    Venn diagram showing subclasses of stabilizer operations. Adding the resource of graphness to CSS-preserving operations results in the class of real stabilizer operations, adding imaginarity results in the class of general stabilizer operations.
  }
  \label{fig:imaginarity_and_graphness}
\end{figure}
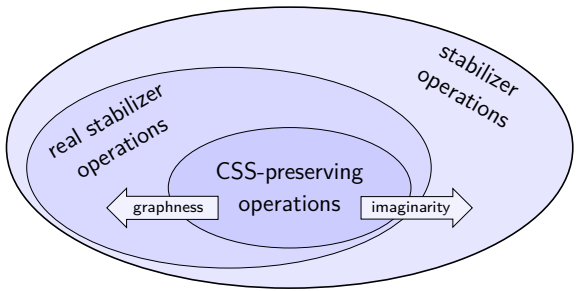

The resource of non-CSS-ness depends on the choice of primary CSS-reference frame, but being resourceful is a property of operations and not frames. When talking about the imaginarity and graphness of stabilizer operations, we will understand the diagonal and non-diagonal parts of the corresponding quadratic form $Q$ (minimizing it over correct possibilities) between CSS-reference frames. We present the refined classification of stabilizer operations on \cref{fig:imaginarity_and_graphness}. Full classification of universality classes of Clifford unitaries was given in Ref.~\cite{Grier_2022}.

The resource of imaginarity is basically the same as described in Ref.~\cite{Hickey_2018}. It can be generated by the gate $S$, the maximal resourceful state of imaginarity is $\ket{+i}$, imaginarity can be introduced into CSS-preserving circuits by a state injection protocol [see \cref{eq:S_injection}]. Indeed, distilling and injecting $\ket{+i}$ states is an important task for surface codes computations \cite{Fowler_2012,Gidney_2024}. Introducing imaginarity can influence the outcome of a stabilizer circuit, at least because $S^2 = Z$.

The resource of graphness is contained in all non-trivial graph states [see \cref{eq:graph_state_definition}]. It can be generated by gates $H$ or $\CZ$, the maximal resourceful state of graphness is $\ket{\CZ}$, graphness can be introduced into CSS-preserving circuits by a state injection protocol [see \cref{eq:H_injection,eq:CZ_injection}]. CSS-preserving stabilizer circuits with introduced graphness are exactly the real stabilizer circuits. As shown in \cref{subsec:non-CSS_incorrectness}, graphness can influence the outcomes of stabilizer circuits. Recent works on measurement-based quantum computations \cite{Raussendorf_2022, Wong_2024} investigate the structure of gauge group on a cluster state that reflects the freedom of formulating the computation in different local reference frames. Let us note that stabilizer cluster states (i.e., graph states) have inherent resource of non-local non-CSS-ness (graphness) in them. Possibly, the resource of graphness has important connections with gauge groups and the computational abilities of cluster states.

Imaginarity is a stronger resource than graphness. One can generate graphness from imaginarity, for example one can prepare a state $\ket{CZ}$ by injecting three states $\ket{+i}$ into a CSS-preserving circuit:
\begin{equation}
  \begin{quantikz}[row sep={0.8cm,between origins},column sep={0.9cm,between origins},align equals at=2]
    \lstick{$\ket{+i}$} &\ctrl{1} &[-0.3cm] &           &[-0.2cm]        &[0.6cm]\rstick[2]{$\ket{\CZ}$} \\
    \lstick{$\ket{-i}$} &\targ{}  &\ctrl{1} &           &\gate{Z}        &                        \\
    \lstick{$\ket{+i}$} &         &\targ{}  &\meterD{Z} &\ctrl{0}\wire[u][1]{c}\setwiretype{c} &\setwiretype{n}
  \end{quantikz}
\end{equation}
where $\ket{-i} = Z\ket{+i}$; but one cannot obtain $\ket{+i}$ by injecting $\ket{CZ}$ into CSS-preserving circuits. On the other hand, imaginarity can be reduced to graphness by introducing R-I rebit (see \cref{subsubsec:state_injection_protocols}). That means, imaginarity can be interpreted as a form of graphness with additional access to ``remote'' entangled qubit. Generally, it is known that imaginary quantum mechanics differs from real quantum mechanics in the setting of many-parties entanglement \cite{Renou_2021, Sarkar_2025}.

\subsection{Simulation of stabilizer circuits with magic} \label{subsec:magic_simulation}

Let us discuss how to simulate Clifford circuits with magic. Because arbitrary non-Clifford operation can be realized through magic state injection protocols \cite{Gottesman_1999,Zhou_2000,Bravyi_2005}, without great loss to generality we can condiser a circuit in which preparations can be non-stabilizer (i.e.\! magical) and all other operations are stabilizer.

In this case, the symbol $p_\rho$ of an initial state $\rho$ is a \emph{quasiprobability distribution} with non-zero negativity. In order to simulate such circuit, we can use a standard procedure of sampling from quasiprobability distributions \cite{Pashayan_2015,Kulikov_2024}. The effectiveness of such procedures is determined by the negativity of the occuring distributions. The time required to estimate a measurement outcome with accuracy $\varepsilon>0$ and probability of failure $p_f>0$ usually scales as
\begin{equation}
  \textrm{TIME} \; \sim \; \mathcal{N}^2\, \frac{1}{\varepsilon^2} \log\frac{1}{p_f},
\end{equation}
where $\mathcal{N}$ is the total negativity ($l_1$-norm of quasiprobabilistic maps) of the circuit elements. Negativity of the composition of operations usually grows as a product of the negativities, so the overall $\mathcal{N}$ grows exponentially with the number of negative gates (or magic states).

Let us consider a magic state $\ket{H}$, which is the eigenvector of Hadamard gate $H$ \cite{Bravyi_2005}:
\begin{equation}
\begin{aligned}
  \proj{H}
  &= \frac{1}{2}\left[ I + \frac{Z+X}{\sqrt{2}}\right] \\
  &= \ketD{00} + \frac{1}{\sqrt{2}}\ketD{10} + \frac{1}{\sqrt{2}}\ketD{01}.
\end{aligned}
\end{equation}
The symbol of this state in CSS-reference frame is
\begin{equation}
\begin{aligned}
  &p_{\ket{H}}(00) = \frac{1+\sqrt{2}}{4}, \quad p_{\ket{H}}(01) = \frac{1}{4}, \\
  &p_{\ket{H}}(10) = \frac{1}{4}, \quad p_{\ket{H}}(11) = \frac{1-\sqrt{2}}{4}.
\end{aligned}
\end{equation}
The negativity of this quasiprobability distribution is
\begin{equation}
  \norm{p_{\ket{H}}}_1 = \frac{1 + \sqrt{2}}{2} \approx 1.207.
\end{equation}

That means, if $\mathsf{QC}$ is a stabilizer circuit with $t$ magic states $\ket{H}$ on the input, one can compute reference frames information to construct a quasiprobability model for the circuit, which can be weakly simulated in time $\BigO(1.207^{2t})$ using standard quasiprobability simulation methods \cite{Veitch_2012}. Surprisingly, this value is a lower bound on the regularized robustness of magic for such state \cite{Howard_2017,Heinrich_2019}, so quasiprobability simulation allows for better time estimates than methods based on robustness of magic.

Now, suppose the circuit $\mathsf{QC}$ is adaptive and feedforward influences reference frame updates. In this case, for the simulation to be correct, it is necessary that all outcomes of intermediate measurements (or at least measurements on the decision points) are correctly estimated, which cannot be done inside of hidden variable model because they have to be averaged over many samples. So, for adaptive circuits the direct quasiprobability simulation may break. In this case it is better to implement sum-over-Cliffords methods, the upper bounds on which are given by robustness of magic or related quantities \cite{Bravyi_2016,Bravyi_2019,Pashayan_2022,Bennink_2017,Seddon_2019,Seddon_2021}. Note that the difference between simulation complexity of adaptive and non-adative stabilizer circuits with magic inputs was first investigated in Ref.~\cite{Yoganathan_2019}, where the authors raised a question whether non-adaptive stabilizer circtuis with magic are easier to simulate and implement than non-adaptive stabilizer circuits with magic. Also, the protocols of injecting magic into CSS-preserving circuits were investigated in Ref.~\cite{Alexander_2023}.

\section{Final remarks} \label{sec:conclusion}

Let us summarise the results of this work and outline several promising directions for future research.

\subsection{Conclusion} \label{subsec:conclusion}

We have examined a natural set of rewriting rules that transform any stabilizer circuit into a classical circuit. The result of this rewriting procedure was shown to exactly reproduce measurement outcomes for the class of CSS-preserving stabilizer circuits. The correctness of the rewriting was established using elementary circuit transformations. The resulting classical circuit can thus be executed efficiently on a standard computer, providing a valid simulation algorithm.

We then developed a framework based on $\Int_4$-valued quadratic forms over Boolean vector spaces and demonstrated that such forms naturally arise in the study of stabilizer groups. In this setting, every stabilizer operation (Clifford channel) admits a representation as a quadratic form expansion over Pauli operators, while CSS-preserving operations correspond to the linear form expansions. We proposed a standard representation of these expansions, which may be practical for constructing optimized stabilizer tableau simulators.

Applying the Walsh–Fourier–Hadamard transform over Pauli group leads to a non-contextual hidden variable model for CSS-preserving stabilizer operations. To extend this framework to general stabilizer operations, we introduced the concept of quantum reference frames encoded by quadratic forms. Within this theory, any stabilizer operation can be described by a probabilistic model defined in appropriately chosen input and output reference frames, yielding a context-dependent hidden variable description of stabilizer circuits. This approach enables efficient simulation of stabilizer and near-stabilizer dynamics using classical circuits supplemented with reference frame processing.

\subsection{Outlook} \label{subsec:outlook}

The present results naturally invite several extensions.

First, the framework should generalize to qudit systems. For odd prime qudit dimensions $d$, this generalization is straightforward, and it has already been well studied from different perspectives \cite{Gross_2006,Gross_2006_2,Booth_2022,Booth_2024,Comfort_2023,Bauer_2026}. However, composite dimensions introduce additional technical subtleties -- we plan to address these challenges in future work \cite{Yashin_TBA}.

Second, our results reveal a deep connection between stabilizer operations and quadratic forms. It would be highly interesting to explore the role of higher-degree forms in quantum theory. We conjecture that operations belonging to the $k$-th level of the Clifford hierarchy \cite{Gottesman_1999} correspond to non-linear affine varieties over $\Int_2$ and $k$-degree forms defined on them. Consequently, studying higher-level Clifford operations may benefit from methods of algebraic geometry (see also Refs.~\cite{Chen_2024, de_Silva_2025, Bu_2025}).

Finally and most importantly, the proposed framework offers promising computational applications. Implementing the described methods in highly optimized code may lead to significant improvements over existing stabilizer simulators \cite{Gidney_2021, de_Beaudrap_2022, Cirq_2024, Qiskit_2024, Aaronson_2004, Garner_2025}, and can also benefit the performance of modern near-Clifford simulators \cite{Bravyi_2019,Pashayan_2022}.

%----------------------------------------------------------------------------------------
%  ACKNOWLEDGEMENTS
%----------------------------------------------------------------------------------------
\section*{Acknowledgements}
The authors thank A.V.~Antipov for useful comments and help with working on surface codes.
The work of V.I.~Yashin was performed at the Steklov International Mathematical Center and supported by the Ministry of Science and Higher Education of the Russian Federation (agreement no. 075-15-2025-303).
The work of E.O.K. and A.K.F. was supported by the Priority 2030 program at the NUST ``MISIS'' under the project K1-2022-027.

%----------------------------------------------------------------------------------------
%  BIBLIOGRAPHY
%----------------------------------------------------------------------------------------
\nocite{Kay_2023}
\bibliographystyle{quantum}
\bibliography{bibliography}

@article{Aaronson_2004,
  title = {Improved simulation of stabilizer circuits},
  author = {Scott Aaronson and Daniel Gottesman},
  year = 2004,
  month = nov,
  journal = {Physical Review A},
  publisher = {American Physical Society ({APS})},
  volume = 70,
  number = 5,
  doi = {10.1103/physreva.70.052328},
  url = {https://doi.org/10.1103/physreva.70.052328}
}

@article{Aaronson_2005,
  title = {Quantum computing, postselection, and probabilistic polynomial-time},
  author = {Aaronson, Scott},
  year = 2005,
  journal = {Proceedings of the Royal Society A: Mathematical, Physical and Engineering Sciences},
  volume = 461,
  number = 2063,
  pages = {3473--3482},
  doi = {10.1098/rspa.2005.1546},
  url = {https://royalsocietypublishing.org/doi/abs/10.1098/rspa.2005.1546}
}

@article{Alexander_2023,
  title = {General Entropic Constraints on {C}alderbank-{S}hor-{S}teane Codes within Magic Distillation Protocols},
  author = {Rhea Alexander and Si Gvirtz-Chen and Nikolaos Koukoulekidis and David Jennings},
  year = 2023,
  month = jun,
  journal = {{PRX} Quantum},
  publisher = {American Physical Society ({APS})},
  volume = 4,
  number = 2,
  doi = {10.1103/prxquantum.4.020359},
  url = {https://doi.org/10.1103/prxquantum.4.020359}
}

@article{Amy_2019,
  title = {Towards Large-scale Functional Verification of Universal Quantum Circuits},
  volume = {287},
  ISSN = {2075-2180},
  url = {http://dx.doi.org/10.4204/EPTCS.287.1},
  DOI = {10.4204/eptcs.287.1},
  journal = {Electronic Proceedings in Theoretical Computer Science},
  publisher = {Open Publishing Association},
  author = {Amy, Matthew},
  year = {2019},
  month = jan,
  pages = {1–21}
}

@article{Amy_2023,
  title = {Symbolic Synthesis of {C}lifford Circuits and Beyond},
  volume = {394},
  ISSN = {2075-2180},
  url = {http://dx.doi.org/10.4204/EPTCS.394.17},
  DOI = {10.4204/eptcs.394.17},
  journal = {Electronic Proceedings in Theoretical Computer Science},
  publisher = {Open Publishing Association},
  author = {Amy, Matthew and Bennett-Gibbs, Owen and Ross, Neil J.},
  year = {2023},
  month = nov,
  pages = {343–362}
}

@misc{Amy_2025,
  title = {Polynomial-Time Classical Simulation of Hidden Shift Circuits via Confluent Rewriting of Symbolic Sums},
  author = {Matthew Amy and Lucas Shigeru Stinchcombe},
  year = {2025},
  eprint = {2408.02778},
  archivePrefix = {arXiv},
  primaryClass = {quant-ph},
  url = {https://arxiv.org/abs/2408.02778},
}

@article{Anders_2006,
  title = {Fast simulation of stabilizer circuits using a graph-state representation},
  author = {Anders, Simon and Briegel, Hans J.},
  year = 2006,
  month = feb,
  journal = {Physical Review A},
  publisher = {American Physical Society (APS)},
  volume = 73,
  number = 2,
  doi = {10.1103/physreva.73.022334},
  issn = {1094-1622},
  url = {http://dx.doi.org/10.1103/PhysRevA.73.022334}
}

@article{Audenaert_2005,
  title = {Entanglement on mixed stabilizer states: normal forms and reduction procedures},
  author = {Koenraad M R Audenaert and Martin B Plenio},
  year = 2005,
  month = aug,
  journal = {New Journal of Physics},
  volume = 7,
  number = 1,
  pages = 170,
  doi = {10.1088/1367-2630/7/1/170},
  url = {https://dx.doi.org/10.1088/1367-2630/7/1/170}
}

@article{Backens_2014,
  title = {The ZX-calculus is complete for stabilizer quantum mechanics},
  author = {Backens, Miriam},
  year = 2014,
  month = sep,
  journal = {New Journal of Physics},
  publisher = {IOP Publishing},
  volume = 16,
  number = 9,
  pages = 093021,
  issn = {1367-2630},
  doi = {10.1088/1367-2630/16/9/093021},
  url = {http://dx.doi.org/10.1088/1367-2630/16/9/093021}
}

@article{Bartlett_2007,
  title = {Reference frames, superselection rules, and quantum information},
  author = {Bartlett, Stephen D. and Rudolph, Terry and Spekkens, Robert W.},
  year = 2007,
  month = apr,
  journal = {Rev. Mod. Phys.},
  publisher = {American Physical Society},
  volume = 79,
  pages = {555--609},
  doi = {10.1103/RevModPhys.79.555},
  url = {https://link.aps.org/doi/10.1103/RevModPhys.79.555},
  issue = 2,
  numpages = {0}
}

@article{Batista_2023,
  title = {Comments on ``Efficient classical simulation of the {D}eutsch--{J}ozsa and {S}imon’s algorithms''},
  author = {Batista, Carlos A and de Veras, Tiago ML and da Silva, Leon D and da Silva, Adenilton J},
  year = 2023,
  journal = {Quantum Information Processing},
  publisher = {Springer},
  volume = 22,
  number = 11,
  pages = 399,
  doi = {10.1007/s11128-023-04134-7}
}

@misc{Bauer_2026,
  title = {Quadratic tensors as a unification of {C}lifford, {G}aussian, and free-fermion physics},
  author = {Andreas Bauer and Seth Lloyd},
  year = {2026},
  eprint = {2601.15396},
  archivePrefix = {arXiv},
  primaryClass = {quant-ph},
  url = {https://arxiv.org/abs/2601.15396},
}

@article{Bennett_1991,
  title = {Communication via one- and two-particle operators on {E}instein-{P}odolsky-{R}osen states},
  author = {Bennett, Charles H. and Wiesner, Stephen J.},
  year = 1992,
  month = nov,
  journal = {Phys. Rev. Lett.},
  publisher = {American Physical Society},
  volume = 69,
  pages = {2881--2884},
  doi = {10.1103/PhysRevLett.69.2881},
  url = {https://link.aps.org/doi/10.1103/PhysRevLett.69.2881},
  issue = 20,
  numpages = {0}
}

@article{Bennett_1993,
  title = {Teleporting an unknown quantum state via dual classical and {E}instein-{P}odolsky-{R}osen channels},
  author = {Bennett, Charles H. and Brassard, Gilles and Cr\'epeau, Claude and Jozsa, Richard and Peres, Asher and Wootters, William K.},
  year = 1993,
  month = mar,
  journal = {Phys. Rev. Lett.},
  publisher = {American Physical Society},
  volume = 70,
  pages = {1895--1899},
  doi = {10.1103/PhysRevLett.70.1895},
  url = {https://link.aps.org/doi/10.1103/PhysRevLett.70.1895},
  issue = 13,
  numpages = {0}
}

@article{Bennink_2017,
  title = {Unbiased simulation of near-{C}lifford quantum circuits},
  author = {Ryan S. Bennink and Erik M. Ferragut and Travis S. Humble and Jason A. Laska and James J. Nutaro and Mark G. Pleszkoch and Raphael C. Pooser},
  year = 2017,
  month = jun,
  journal = {Physical Review A},
  publisher = {American Physical Society ({APS})},
  volume = 95,
  number = 6,
  doi = {10.1103/physreva.95.062337},
  url = {https://doi.org/10.1103/physreva.95.062337}
}

@article{Bernstein_1997,
  title = {Quantum Complexity Theory},
  author = {Bernstein, Ethan and Vazirani, Umesh},
  year = 1997,
  journal = {SIAM Journal on Computing},
  volume = 26,
  number = 5,
  pages = {1411--1473},
  doi = {10.1137/S0097539796300921}
}

@misc{Booth_2022,
  title = {Complete {ZX}-calculi for the stabiliser fragment in odd prime dimensions},
  author = {Robert I. Booth and Titouan Carette},
  year = {2022},
  eprint = {2204.12531},
  archivePrefix = {arXiv},
  primaryClass = {quant-ph}
}

@misc{Booth_2024,
  title = {Graphical Symplectic Algebra},
  author = {Robert I. Booth and Titouan Carette and Cole Comfort},
  year = 2024,
  eprint = {2401.07914},
  archivePrefix = {arXiv},
  primaryClass = {cs.LO},
  url = {https://arxiv.org/abs/2401.07914}
}

@misc{Booth_2025,
  title = {Denotational semantics for stabiliser quantum programs},
  author = {Robert I. Booth and Cole Comfort},
  year = 2025,
  url = {https://robertbooth.fr/pdfs/Denotational_semantics_for_stabiliser_quantum_programs.pdf}
}

@article{Bravyi_2005,
  title = {Universal quantum computation with ideal {C}lifford gates and noisy ancillas},
  author = {Bravyi, Sergey and Kitaev, Alexei},
  year = 2005,
  month = feb,
  journal = {Physical Review A},
  publisher = {American Physical Society (APS)},
  volume = 71,
  number = 2,
  doi = {10.1103/physreva.71.022316},
  issn = {1094-1622},
  url = {http://dx.doi.org/10.1103/PhysRevA.71.022316}
}

@article{Bravyi_2016,
  title = {Improved Classical Simulation of Quantum Circuits Dominated by {C}lifford Gates},
  author = {Sergey Bravyi and David Gosset},
  year = 2016,
  month = jun,
  journal = {Physical Review Letters},
  publisher = {American Physical Society ({APS})},
  volume = 116,
  number = 25,
  doi = {10.1103/physrevlett.116.250501},
  url = {https://doi.org/10.1103/physrevlett.116.250501}
}

@article{Bravyi_2019,
  title = {Simulation of quantum circuits by low-rank stabilizer decompositions},
  author = {Bravyi, Sergey and Browne, Dan and Calpin, Padraic and Campbell, Earl and Gosset, David and Howard, Mark},
  year = 2019,
  month = sep,
  journal = {Quantum},
  publisher = {Verein zur Forderung des Open Access Publizierens in den Quantenwissenschaften},
  volume = 3,
  pages = 181,
  doi = {10.22331/q-2019-09-02-181},
  issn = {2521-327X},
  url = {http://dx.doi.org/10.22331/q-2019-09-02-181}
}

@article{Bravyi_2022,
  title = {How to Simulate Quantum Measurement without Computing Marginals},
  author = {Bravyi, Sergey and Gosset, David and Liu, Yinchen},
  year = 2022,
  month = jun,
  journal = {Physical Review Letters},
  publisher = {American Physical Society (APS)},
  volume = 128,
  number = 22,
  doi = {10.1103/physrevlett.128.220503},
  issn = {1079-7114},
  url = {http://dx.doi.org/10.1103/PhysRevLett.128.220503}
}

@article{Brown_1972,
  title = {Generalizations of the {K}ervaire Invariant},
  author = {Edgar H. Brown},
  year = 1972,
  journal = {Annals of Mathematics},
  publisher = {[Annals of Mathematics, Trustees of Princeton University on Behalf of the Annals of Mathematics, Mathematics Department, Princeton University]},
  volume = 95,
  number = 2,
  pages = {368--383},
  issn = {0003486X, 19398980},
  url = {http://www.jstor.org/stable/1970804},
  urldate = {2025-08-22}
}

@article{Bu_2022,
  title = {Classical Simulation of Quantum Circuits by Half Gauss Sums},
  author = {Bu, Kaifeng and Koh, Dax Enshan},
  year = 2022,
  month = jan,
  journal = {Communications in Mathematical Physics},
  publisher = {Springer Science and Business Media LLC},
  volume = 390,
  number = 2,
  pages = {471–500},
  doi = {10.1007/s00220-022-04320-1},
  issn = {1432-0916},
  url = {http://dx.doi.org/10.1007/s00220-022-04320-1}
}

@article{Bu_2023,
  title = {Quantum entropy and central limit theorem},
  author = {Bu, Kaifeng and Gu, Weichen and Jaffe, Arthur},
  year = 2023,
  month = jun,
  journal = {Proceedings of the National Academy of Sciences},
  publisher = {Proceedings of the National Academy of Sciences},
  volume = 120,
  number = 25,
  doi = {10.1073/pnas.2304589120},
  issn = {1091-6490},
  url = {http://dx.doi.org/10.1073/pnas.2304589120}
}

@misc{Bu_2023_2,
  title = {Discrete Quantum Gaussians and Central Limit Theorem},
  author = {Kaifeng Bu and Weichen Gu and Arthur Jaffe},
  year = 2023,
  url = {https://arxiv.org/abs/2302.08423},
  eprint = {2302.08423},
  archiveprefix = {arXiv},
  primaryclass = {quant-ph}
}

@misc{Bu_2025,
  title = {Quantum Higher Order {F}ourier Analysis and the {C}lifford Hierarchy},
  author = {Kaifeng Bu and Weichen Gu and Arthur Jaffe},
  year = 2025,
  url = {https://arxiv.org/abs/2508.15908},
  eprint = {2508.15908},
  archiveprefix = {arXiv},
  primaryclass = {quant-ph}
}

@article{Buca_2012,
  title = {A note on symmetry reductions of the {L}indblad equation: transport in constrained open spin chains},
  author = {Buča, Berislav and Prosen, Tomaž},
  year = 2012,
  month = jul,
  journal = {New Journal of Physics},
  publisher = {IOP Publishing},
  volume = 14,
  number = 7,
  pages = {073007},
  doi = {10.1088/1367-2630/14/7/073007},
  issn = {1367-2630},
  url = {http://dx.doi.org/10.1088/1367-2630/14/7/073007}
}

@inproceedings{Buhman_2006,
  title = {New Limits on Fault-Tolerant Quantum Computation},
  author = {Buhrman, Harry and Cleve, Richard and Laurent, Monique and Linden, Noah and Schrijver, Alexander and Unger, Falk},
  year = 2006,
  booktitle = {2006 47th Annual IEEE Symposium on Foundations of Computer Science (FOCS'06)},
  pages = {411--419},
  doi = {10.1109/FOCS.2006.50}
}

@article{Calderbank_1996,
  title = {Good quantum error-correcting codes exist},
  author = {Calderbank, A. R. and Shor, Peter W.},
  year = 1996,
  month = aug,
  journal = {Physical Review A},
  publisher = {American Physical Society (APS)},
  volume = 54,
  number = 2,
  pages = {1098–1105},
  doi = {10.1103/physreva.54.1098},
  issn = {1094-1622},
  url = {http://dx.doi.org/10.1103/PhysRevA.54.1098}
}

@article{Cappellini_2007,
  title = {Subnormalized states and trace-nonincreasing maps},
  author = {Cappellini, Valerio and Sommers, Hans-Jürgen and Życzkowski, Karol},
  year = 2007,
  month = may,
  journal = {Journal of Mathematical Physics},
  publisher = {AIP Publishing},
  volume = 48,
  number = 5,
  doi = {10.1063/1.2738359},
  issn = {1089-7658},
  url = {http://dx.doi.org/10.1063/1.2738359}
}

@article{Carette_2025,
  title = {Operational Quantum Reference Frame Transformations},
  author = {Carette, Titouan and Glowacki, Jan and Loveridge, Leon},
  year = 2025,
  month = mar,
  journal = {Quantum},
  publisher = {Verein zur Forderung des Open Access Publizierens in den Quantenwissenschaften},
  volume = 9,
  pages = 1680,
  doi = {10.22331/q-2025-03-27-1680},
  issn = {2521-327X},
  url = {http://dx.doi.org/10.22331/q-2025-03-27-1680}
}

@article{Catani_2017,
  title = {{S}pekkens' toy model in all dimensions and its relationship with stabiliser quantum mechanics},
  author = {Catani, Lorenzo and Browne, Dan E},
  year = 2017,
  month = jul,
  journal = {New Journal of Physics},
  publisher = {IOP Publishing},
  volume = 19,
  number = 7,
  pages = {073035},
  doi = {10.1088/1367-2630/aa781c},
  issn = {1367-2630},
  url = {http://dx.doi.org/10.1088/1367-2630/aa781c}
}

@book{Ceccherini-Silberstein_2022,
  title = {Representation Theory of Finite Group Extensions: {C}lifford Theory, Mackey Obstruction, and the Orbit Method},
  author = {Ceccherini-Silberstein, Tullio and Scarabotti, Fabio and Tolli, Filippo},
  year = 2022,
  publisher = {Springer Cham},
  series = {Springer Monographs in Mathematics},
  doi = {https://doi.org/10.1007/978-3-031-13873-7},
  isbn = {978-3-031-13872-0},
  issn = {1439-7382}
}

@misc{Chen_2024,
  title = {Characterising semi-{C}lifford gates using algebraic sets},
  author = {Imin Chen and Nadish de Silva},
  year = 2024,
  url = {https://arxiv.org/abs/2309.15184},
  eprint = {2309.15184},
  archiveprefix = {arXiv},
  primaryclass = {quant-ph}
}

@article{Chitambar_2019,
  title = {Quantum resource theories},
  author = {Chitambar, Eric and Gour, Gilad},
  year = 2019,
  month = apr,
  journal = {Reviews of Modern Physics},
  publisher = {American Physical Society (APS)},
  volume = 91,
  number = 2,
  doi = {10.1103/revmodphys.91.025001},
  issn = {1539-0756},
  url = {http://dx.doi.org/10.1103/RevModPhys.91.025001}
}

@misc{Cirq_2024,
  title = {{Cirq}},
  author = {Cirq Developers},
  year = 2024,
  month = may,
  publisher = {Zenodo},
  doi = {10.5281/zenodo.4586899},
  url = {https://doi.org/10.5281/zenodo.4586899},
  note = {{See full list of authors on Github: \url{https://github.com/quantumlib/Cirq/graphs/contributors}}},
  version = {v0.10.0}
}

@article{Comfort_2022,
  title = {A Graphical Calculus for {L}agrangian Relations},
  author = {Comfort, Cole and Kissinger, Aleks},
  year = 2022,
  month = nov,
  journal = {Electronic Proceedings in Theoretical Computer Science},
  publisher = {Open Publishing Association},
  volume = 372,
  pages = {338--351},
  issn = {2075-2180},
  url = {http://dx.doi.org/10.4204/EPTCS.372.24},
  doi = {10.4204/eptcs.372.24}
}

@misc{Comfort_2023,
  title = {The Algebra for Stabilizer Codes},
  author = {Cole Comfort},
  year = {2023},
  eprint = {2304.10584},
  archivePrefix = {arXiv},
  primaryClass = {quant-ph},
  url = {https://arxiv.org/abs/2304.10584},
}

@article{Damm_1990,
  title = {Problems complete for {$\mathtt{\oplus L}$}},
  author = {Damm, Carsten},
  year = 1990,
  month = dec,
  journal = {Inf. Process. Lett.},
  publisher = {Elsevier North-Holland, Inc.},
  address = {USA},
  volume = 36,
  number = 5,
  pages = {247–250},
  doi = {10.1016/0020-0190(90)90150-V},
  issn = {0020-0190},
  url = {https://doi.org/10.1016/0020-0190(90)90150-V},
  issue_date = {Dec. 1, 1990},
  numpages = 4
}

@article{de_Beaudrap_2022,
  title = {Fast Stabiliser Simulation with Quadratic Form Expansions},
  author = {de Beaudrap, Niel and Herbert, Steven},
  year = 2022,
  month = sep,
  journal = {Quantum},
  publisher = {Verein zur Forderung des Open Access Publizierens in den Quantenwissenschaften},
  volume = 6,
  pages = 803,
  doi = {10.22331/q-2022-09-15-803},
  url = {https://doi.org/10.22331/q-2022-09-15-803}
}

@misc{de_Silva_2025,
  title = {The {C}lifford hierarchy for one qubit or qudit},
  author = {Nadish de Silva and Oscar Lautsch},
  year = 2025,
  url = {https://arxiv.org/abs/2501.07939},
  eprint = {2501.07939},
  archiveprefix = {arXiv},
  primaryclass = {quant-ph}
}

@article{Dehaene_2003,
  title = {{C}lifford group, stabilizer states, and linear and quadratic operations over {$GF(2)$}},
  author = {Jeroen Dehaene and Bart De Moor},
  year = 2003,
  month = oct,
  journal = {Physical Review A},
  publisher = {American Physical Society ({APS})},
  volume = 68,
  number = 4,
  doi = {10.1103/physreva.68.042318},
  url = {https://doi.org/10.1103/physreva.68.042318}
}

@article{Delfosse_2015,
  title = {{W}igner Function Negativity and Contextuality in Quantum Computation on Rebits},
  author = {Nicolas Delfosse and Philippe Allard Guerin and Jacob Bian and Robert Raussendorf},
  year = 2015,
  month = apr,
  journal = {Physical Review X},
  publisher = {American Physical Society ({APS})},
  volume = 5,
  number = 2,
  doi = {10.1103/physrevx.5.021003},
  url = {https://doi.org/10.1103/physrevx.5.021003}
}

@misc{Ermakov_2025,
  title = {Operator growth in many-body systems of higher spins},
  author = {Igor Ermakov},
  year = 2025,
  url = {https://arxiv.org/abs/2504.07833},
  eprint = {2504.07833},
  archiveprefix = {arXiv},
  primaryclass = {quant-ph}
}

@misc{Fattal_2004,
  title = {Entanglement in the stabilizer formalism},
  author = {David Fattal and Toby S. Cubitt and Yoshihisa Yamamoto and Sergey Bravyi and Isaac L. Chuang},
  year = 2004,
  eprint = {quant-ph/0406168},
  archiveprefix = {arXiv},
  primaryclass = {quant-ph}
}

@article{Ferrie_2011,
  title = {Quasi-probability representations of quantum theory with applications to quantum information science},
  author = {Ferrie, Christopher},
  year = 2011,
  month = oct,
  journal = {Reports on Progress in Physics},
  publisher = {IOP Publishing},
  volume = 74,
  number = 11,
  pages = 116001,
  doi = {10.1088/0034-4885/74/11/116001},
  issn = {1361-6633},
  url = {http://dx.doi.org/10.1088/0034-4885/74/11/116001}
}

@misc{Fewster_2024,
  title = {Quantum reference frames, measurement schemes and the type of local algebras in quantum field theory},
  author = {Christopher J. Fewster and Daan W. Janssen and Leon Deryck Loveridge and Kasia Rejzner and James Waldron},
  year = 2024,
  doi = {https://doi.org/10.1007/s00220-024-05180-7},
  url = {https://arxiv.org/abs/2403.11973},
  eprint = {2403.11973},
  archiveprefix = {arXiv},
  primaryclass = {math-ph}
}

@article{Filippov_2021,
  title = {Capacity of trace decreasing quantum operations and superadditivity of coherent information for a generalized erasure channel},
  author = {Filippov, Sergey N},
  year = 2021,
  month = may,
  journal = {Journal of Physics A: Mathematical and Theoretical},
  publisher = {IOP Publishing},
  volume = 54,
  number = 25,
  pages = 255301,
  doi = {10.1088/1751-8121/abfd61},
  url = {https://dx.doi.org/10.1088/1751-8121/abfd61}
}

@article{Fowler_2012,
  title = {Surface codes: Towards practical large-scale quantum computation},
  author = {Fowler, Austin G. and Mariantoni, Matteo and Martinis, John M. and Cleland, Andrew N.},
  year = 2012,
  month = sep,
  journal = {Phys. Rev. A},
  publisher = {American Physical Society},
  volume = 86,
  pages = {032324},
  doi = {10.1103/PhysRevA.86.032324},
  url = {https://link.aps.org/doi/10.1103/PhysRevA.86.032324},
  issue = 3,
  numpages = 48
}

@book{Fujii_2015,
  title = {Quantum Computation with Topological Codes: from qubit to topological fault-tolerance},
  author = {Fujii, Keisuke},
  year = 2015,
  volume = 8,
  publisher = {Springer Singapore},
  doi = {https://doi.org/10.1007/978-981-287-996-7},
  url = {https://link.springer.com/book/10.1007/978-981-287-996-7}
}

@misc{Garner_2025,
  title = {{STABSim}: A Parallelized {C}lifford Simulator with Features Beyond Direct Simulation},
  author = {Sean Garner and Chenxu Liu and Meng Wang and Samuel Stein and Ang Li},
  year = 2025,
  url = {https://arxiv.org/abs/2507.03092},
  eprint = {2507.03092},
  archiveprefix = {arXiv},
  primaryclass = {quant-ph}
}

@article{Giacomini_2019,
  title = {Quantum mechanics and the covariance of physical laws in quantum reference frames},
  author = {Giacomini, Flaminia and Castro-Ruiz, Esteban and Brukner, Časlav},
  year = 2019,
  month = jan,
  journal = {Nature Communications},
  publisher = {Springer Science and Business Media LLC},
  volume = 10,
  number = 1,
  doi = {10.1038/s41467-018-08155-0},
  issn = {2041-1723},
  url = {http://dx.doi.org/10.1038/s41467-018-08155-0}
}

@article{Giacomini_2019_2,
  title = {Relativistic Quantum Reference Frames: The Operational Meaning of Spin},
  author = {Giacomini, Flaminia and Castro-Ruiz, Esteban and Brukner, Caslav},
  year = 2019,
  month = aug,
  journal = {Phys. Rev. Lett.},
  publisher = {American Physical Society},
  volume = 123,
  pages = {090404},
  doi = {10.1103/PhysRevLett.123.090404},
  url = {https://link.aps.org/doi/10.1103/PhysRevLett.123.090404},
  issue = 9,
  numpages = 6
}

@article{Gidney_2021,
  title = {Stim: a fast stabilizer circuit simulator},
  author = {Craig Gidney},
  year = 2021,
  month = jul,
  journal = {Quantum},
  publisher = {Verein zur Forderung des Open Access Publizierens in den Quantenwissenschaften},
  volume = 5,
  pages = 497,
  doi = {10.22331/q-2021-07-06-497},
  url = {https://doi.org/10.22331/q-2021-07-06-497}
}

@article{Gidney_2024,
  title = {Inplace Access to the Surface Code Y Basis},
  author = {Gidney, Craig},
  year = 2024,
  month = apr,
  journal = {Quantum},
  publisher = {Verein zur Forderung des Open Access Publizierens in den Quantenwissenschaften},
  volume = 8,
  pages = 1310,
  doi = {10.22331/q-2024-04-08-1310},
  issn = {2521-327X},
  url = {http://dx.doi.org/10.22331/q-2024-04-08-1310}
}

@misc{Gottesman_1997,
  title = {Stabilizer Codes and Quantum Error Correction},
  author = {Daniel Gottesman},
  year = 1997,
  eprint = {quant-ph/9705052},
  archiveprefix = {arXiv},
  primaryclass = {quant-ph}
}

@misc{Gottesman_1998,
  title = {The {H}eisenberg Representation of Quantum Computers},
  author = {Daniel Gottesman},
  year = 1998,
  eprint = {quant-ph/9807006},
  archiveprefix = {arXiv},
  primaryclass = {quant-ph}
}

@article{Gottesman_1999,
  title = {Demonstrating the viability of universal quantum computation using teleportation and single-qubit operations},
  author = {Daniel Gottesman and Isaac L. Chuang},
  year = 1999,
  month = nov,
  journal = {Nature},
  publisher = {Springer Science and Business Media {LLC}},
  volume = 402,
  number = 6760,
  pages = {390--393},
  doi = {10.1038/46503},
  url = {https://doi.org/10.1038/46503}
}

@article{Grier_2022,
  title = {The Classification of {C}lifford Gates over Qubits},
  author = {Grier, Daniel and Schaeffer, Luke},
  year = 2022,
  month = jun,
  journal = {Quantum},
  publisher = {Verein zur Forderung des Open Access Publizierens in den Quantenwissenschaften},
  volume = 6,
  pages = 734,
  doi = {10.22331/q-2022-06-13-734},
  issn = {2521-327X},
  url = {http://dx.doi.org/10.22331/q-2022-06-13-734}
}

@article{Gross_2006,
  title = {{H}udson's theorem for finite-dimensional quantum systems},
  author = {D. Gross},
  year = 2006,
  month = dec,
  journal = {Journal of Mathematical Physics},
  publisher = {{AIP} Publishing},
  volume = 47,
  number = 12,
  doi = {10.1063/1.2393152},
  url = {https://doi.org/10.1063/1.2393152}
}

@article{Gross_2006_2,
  title = {Non-negative {W}igner functions in prime dimensions},
  author = {D. Gross},
  year = 2006,
  month = dec,
  journal = {Applied Physics B},
  publisher = {Springer Science and Business Media {LLC}},
  volume = 86,
  number = 3,
  pages = {367--370},
  doi = {10.1007/s00340-006-2510-9},
  url = {https://doi.org/10.1007s00340-006-2510-9}
}

@article{Hashagen_2018,
  title = {Real Randomized Benchmarking},
  author = {A. K. Hashagen and S. T. Flammia and D. Gross and J. J. Wallman},
  year = 2018,
  month = aug,
  journal = {Quantum},
  publisher = {Verein zur Forderung des Open Access Publizierens in den Quantenwissenschaften},
  volume = 2,
  pages = 85,
  doi = {10.22331/q-2018-08-22-85},
  url = {https://doi.org/10.22331/q-2018-08-22-85}
}

@article{Heimendahl_2022,
  title = {The axiomatic and the operational approaches to resource theories of magic do not coincide},
  author = {Heimendahl, Arne and Heinrich, Markus and Gross, David},
  year = 2022,
  month = nov,
  journal = {Journal of Mathematical Physics},
  publisher = {AIP Publishing},
  volume = 63,
  number = 11,
  doi = {10.1063/5.0085774},
  issn = {1089-7658},
  url = {http://dx.doi.org/10.1063/5.0085774}
}

@misc{Hein_2006,
  title = {Entanglement in Graph States and its Applications},
  author = {M. Hein and W. Dür and J. Eisert and R. Raussendorf and M. Van den Nest and H. -J. Briegel},
  year = 2006,
  url = {https://arxiv.org/abs/quant-ph/0602096},
  eprint = {quant-ph/0602096},
  archiveprefix = {arXiv},
  primaryclass = {quant-ph}
}

@article{Heinrich_2019,
  title = {Robustness of Magic and Symmetries of the Stabiliser Polytope},
  author = {Heinrich, Markus and Gross, David},
  year = 2019,
  month = apr,
  journal = {Quantum},
  publisher = {Verein zur Forderung des Open Access Publizierens in den Quantenwissenschaften},
  volume = 3,
  pages = 132,
  doi = {10.22331/q-2019-04-08-132},
  issn = {2521-327X},
  url = {http://dx.doi.org/10.22331/q-2019-04-08-132}
}

@phdthesis{Heinrich_2021,
  title = {On stabiliser techniques and their application to simulation  and certification of quantum devices},
  author = {Markus Heinrich},
  year = 2021,
  url = {https://kups.ub.uni-koeln.de/50465/},
  school = {Universit{\"a}t zu K{\"o}ln}
}

@article{Hickey_2018,
  title = {Quantifying the imaginarity of quantum mechanics},
  author = {Hickey, Alexander and Gour, Gilad},
  year = 2018,
  month = sep,
  journal = {Journal of Physics A: Mathematical and Theoretical},
  publisher = {IOP Publishing},
  volume = 51,
  number = 41,
  pages = 414009,
  doi = {10.1088/1751-8121/aabe9c},
  issn = {1751-8121},
  url = {http://dx.doi.org/10.1088/1751-8121/aabe9c}
}

@article{Hindlycke_2022,
  title = {Efficient Contextual Ontological Model of {$n$}-Qubit Stabilizer Quantum Mechanics},
  author = {Hindlycke, Christoffer and Larsson, Jan-\AA{}ke},
  year = 2022,
  month = sep,
  journal = {Phys. Rev. Lett.},
  publisher = {American Physical Society},
  volume = 129,
  pages = 130401,
  doi = {10.1103/PhysRevLett.129.130401},
  url = {https://link.aps.org/doi/10.1103/PhysRevLett.129.130401},
  issue = 13,
  numpages = 6
}

@article{Howard_2017,
  title = {Application of a Resource Theory for Magic States to Fault-Tolerant Quantum Computing},
  author = {Howard, Mark and Campbell, Earl},
  year = 2017,
  month = mar,
  journal = {Physical Review Letters},
  publisher = {American Physical Society (APS)},
  volume = 118,
  number = 9,
  doi = {10.1103/physrevlett.118.090501},
  issn = {1079-7114},
  url = {http://dx.doi.org/10.1103/PhysRevLett.118.090501}
}

@article{Hu_2022,
  title = {Improved graph formalism for quantum circuit simulation},
  author = {Hu, Alexander Tianlin and Khesin, Andrey Boris},
  year = 2022,
  month = feb,
  journal = {Physical Review A},
  publisher = {American Physical Society (APS)},
  volume = 105,
  number = 2,
  doi = {10.1103/physreva.105.022432},
  issn = {2469-9934},
  url = {http://dx.doi.org/10.1103/PhysRevA.105.022432}
}

@article{Johansson_2017,
  title = {Efficient classical simulation of the {D}eutsch–{J}ozsa and {S}imon's algorithms},
  author = {Johansson, Niklas and Larsson, Jan-Ake},
  year = 2017,
  month = aug,
  journal = {Quantum Information Processing},
  publisher = {Springer Science and Business Media LLC},
  volume = 16,
  number = 9,
  doi = {10.1007/s11128-017-1679-7},
  issn = {1573-1332},
  url = {http://dx.doi.org/10.1007/s11128-017-1679-7}
}

@article{Johansson_2019,
  title = {Quantum Simulation Logic, Oracles, and the Quantum Advantage},
  author = {Johansson, Niklas and Larsson, Jan-Åke},
  year = 2019,
  month = aug,
  journal = {Entropy},
  publisher = {MDPI AG},
  volume = 21,
  number = 8,
  pages = 800,
  doi = {10.3390/e21080800},
  issn = {1099-4300},
  url = {http://dx.doi.org/10.3390/e21080800}
}

@article{Jones_2024,
  title = {The {H}adamard gate cannot be replaced by a resource state in universal quantum computation},
  author = {Jones, Benjamin D. M. and Linden, Noah and Skrzypczyk, Paul},
  year = 2024,
  month = sep,
  journal = {Quantum},
  publisher = {Verein zur Forderung des Open Access Publizierens in den Quantenwissenschaften},
  volume = 8,
  pages = 1470,
  doi = {10.22331/q-2024-09-11-1470},
  issn = {2521-327X},
  url = {http://dx.doi.org/10.22331/q-2024-09-11-1470}
}

@misc{Kay_2023,
  title = {Tutorial on the Quantikz Package},
  author = {Alastair Kay},
  year = 2023,
  eprint = {1809.03842},
  archiveprefix = {arXiv},
  primaryclass = {quant-ph}
}

@article{Kenbaev_2022,
  title = {Quantum postselective measurements: Sufficient condition for overcoming the Holevo bound and the role of max-relative entropy},
  author = {Kenbaev, N. R. and Kronberg, D. A.},
  year = 2022,
  month = jan,
  journal = {Phys. Rev. A},
  publisher = {American Physical Society},
  volume = 105,
  pages = {012609},
  doi = {10.1103/PhysRevA.105.012609},
  url = {https://link.aps.org/doi/10.1103/PhysRevA.105.012609},
  issue = 1,
  numpages = 6
}

@misc{Kissinger_2022,
  title = {Phase-free {ZX} diagrams are {CSS} codes (...or how to graphically grok the surface code)},
  author = {Aleks Kissinger},
  year = 2022,
  eprint = {2204.14038},
  archiveprefix = {arXiv},
  primaryclass = {quant-ph}
}

@article{Kliuchnikov_2013,
  title = {Optimization of {C}lifford circuits},
  volume = {88},
  ISSN = {1094-1622},
  url = {http://dx.doi.org/10.1103/PhysRevA.88.052307},
  DOI = {10.1103/physreva.88.052307},
  number = {5},
  journal = {Physical Review A},
  publisher = {American Physical Society (APS)},
  author = {Kliuchnikov, Vadym and Maslov, Dmitri},
  year = {2013},
  month = nov
}

@misc{Kliuchnikov_2023,
  title = {Stabilizer circuit verification},
  author = {Vadym Kliuchnikov and Michael Beverland and Adam Paetznick},
  year = 2023,
  eprint = {2309.08676},
  archiveprefix = {arXiv},
  primaryclass = {quant-ph}
}

@article{Kulikov_2024,
  title = {Minimizing the negativity of quantum circuits in overcomplete quasiprobability representations},
  author = {Kulikov, Denis A. and Yashin, Vsevolod I. and Fedorov, Aleksey K. and Kiktenko, Evgeniy O.},
  year = 2024,
  month = jan,
  journal = {Physical Review A},
  publisher = {American Physical Society (APS)},
  volume = 109,
  number = 1,
  doi = {10.1103/physreva.109.012219},
  issn = {2469-9934},
  url = {http://dx.doi.org/10.1103/PhysRevA.109.012219}
}

@article{Lessa_2025,
  title = {Strong-to-Weak Spontaneous Symmetry Breaking in Mixed Quantum States},
  author = {Lessa, Leonardo A. and Ma, Ruochen and Zhang, Jian-Hao and Bi, Zhen and Cheng, Meng and Wang, Chong},
  year = 2025,
  month = mar,
  journal = {PRX Quantum},
  publisher = {American Physical Society (APS)},
  volume = 6,
  number = 1,
  doi = {10.1103/prxquantum.6.010344},
  issn = {2691-3399},
  url = {http://dx.doi.org/10.1103/PRXQuantum.6.010344}
}

@article{Loveridge_2018,
  title = {Symmetry, Reference Frames, and Relational Quantities in Quantum Mechanics},
  author = {Loveridge, Leon and Miyadera, Takayuki and Busch, Paul},
  year = 2018,
  month = feb,
  journal = {Foundations of Physics},
  publisher = {Springer Science and Business Media LLC},
  volume = 48,
  number = 2,
  pages = {135–198},
  doi = {10.1007/s10701-018-0138-3},
  issn = {1572-9516},
  url = {http://dx.doi.org/10.1007/s10701-018-0138-3}
}

@misc{Lunt_2025,
  title = {Emergent random matrix universality in quantum operator dynamics},
  author = {Oliver Lunt and Thomas Kriecherbauer and Kenneth T-R McLaughlin and Curt von Keyserlingk},
  year = 2025,
  url = {https://arxiv.org/abs/2504.18311},
  eprint = {2504.18311},
  archiveprefix = {arXiv},
  primaryclass = {quant-ph}
}

@misc{Mangiarotti_1998,
  title = {Dynamic Connections in Analytical Mechanics},
  author = {L. Mangiarotti and G. Sardanashvily},
  year = 1998,
  url = {https://arxiv.org/abs/math-ph/9805024},
  eprint = {math-ph/9805024},
  archiveprefix = {arXiv},
  primaryclass = {math-ph}
}

@article{McKague_2009,
  title = {Simulating Quantum Systems Using Real {H}ilbert Spaces},
  author = {McKague, Matthew and Mosca, Michele and Gisin, Nicolas},
  year = 2009,
  month = jan,
  journal = {Phys. Rev. Lett.},
  publisher = {American Physical Society},
  volume = 102,
  pages = {020505},
  doi = {10.1103/PhysRevLett.102.020505},
  url = {https://link.aps.org/doi/10.1103/PhysRevLett.102.020505},
  issue = 2,
  numpages = 4
}

@article{Nandy_2025,
  title = {Quantum dynamics in {K}rylov space: Methods and applications},
  author = {Nandy, Pratik and Matsoukas-Roubeas, Apollonas S. and Martínez-Azcona, Pablo and Dymarsky, Anatoly and del Campo, Adolfo},
  year = 2025,
  month = jun,
  journal = {Physics Reports},
  publisher = {Elsevier BV},
  volume = {1125–1128},
  pages = {1–82},
  doi = {10.1016/j.physrep.2025.05.001},
  issn = {0370-1573},
  url = {http://dx.doi.org/10.1016/j.physrep.2025.05.001}
}

@book{Nielsen_2010,
  title = {Quantum Computation and Quantum Information: 10th Anniversary Edition},
  author = {Nielsen, Michael A. and Chuang, Isaac L.},
  year = 2010,
  publisher = {Cambridge University Press},
  address = {Cambridge},
  doi = {10.1017/CBO9780511976667},
  isbn = 9781107002173
}

@misc{ODonnell_2021,
  title = {Analysis of {B}oolean Functions},
  author = {Ryan O'Donnell},
  year = 2021,
  url = {https://arxiv.org/abs/2105.10386},
  eprint = {2105.10386},
  archiveprefix = {arXiv},
  primaryclass = {cs.DM}
}

@article{Okay_2021,
  title = {On the extremal points of the Lambda polytopes and classical simulation of quantum computation with magic states},
  author = {Okay, Cihan and Zurel, Michael and Raussendorf, Robert},
  year = 2021,
  month = sep,
  journal = {Quantum Information and Computation},
  publisher = {Rinton Press},
  volume = 21,
  number = {13 & 14},
  pages = {1091–1110},
  doi = {10.26421/qic21.13-14-2},
  issn = {1533-7146},
  url = {http://dx.doi.org/10.26421/QIC21.13-14-2}
}

@misc{Pang_2025,
  title = {Simulating {C}lifford Circuits with {G}aussian Elimination},
  author = {Yuchen Pang and Edgar Solomonik},
  year = 2025,
  eprint = {2511.06127},
  archivePrefix = {arXiv},
  primaryClass = {quant-ph},
  url = {https://arxiv.org/abs/2511.06127},
}

@article{Park_2024,
  title = {Extending Classically Simulatable Bounds of {C}lifford Circuits with Nonstabilizer States via Framed {W}igner Functions},
  author = {Park, Guedong and Kwon, Hyukjoon and Jeong, Hyunseok},
  year = 2024,
  month = nov,
  journal = {Physical Review Letters},
  publisher = {American Physical Society (APS)},
  volume = 133,
  number = 22,
  doi = {10.1103/physrevlett.133.220601},
  issn = {1079-7114},
  url = {http://dx.doi.org/10.1103/PhysRevLett.133.220601}
}

@article{Parker_2019,
  title = {A Universal Operator Growth Hypothesis},
  author = {Parker, Daniel E. and Cao, Xiangyu and Avdoshkin, Alexander and Scaffidi, Thomas and Altman, Ehud},
  year = 2019,
  month = oct,
  journal = {Physical Review X},
  publisher = {American Physical Society (APS)},
  volume = 9,
  number = 4,
  doi = {10.1103/physrevx.9.041017},
  issn = {2160-3308},
  url = {http://dx.doi.org/10.1103/PhysRevX.9.041017}
}

@article{Pashayan_2015,
  title = {Estimating Outcome Probabilities of Quantum Circuits Using Quasiprobabilities},
  author = {Hakop Pashayan and Joel J. Wallman and Stephen D. Bartlett},
  year = 2015,
  month = aug,
  journal = {Physical Review Letters},
  publisher = {American Physical Society ({APS})},
  volume = 115,
  number = 7,
  doi = {10.1103/physrevlett.115.070501},
  url = {https://doi.org/10.1103/physrevlett.115.070501}
}

@article{Pashayan_2022,
  title = {Fast Estimation of Outcome Probabilities for Quantum Circuits},
  author = {Pashayan, Hakop and Reardon-Smith, Oliver and Korzekwa, Kamil and Bartlett, Stephen D.},
  year = 2022,
  month = jun,
  journal = {PRX Quantum},
  publisher = {American Physical Society (APS)},
  volume = 3,
  number = 2,
  doi = {10.1103/prxquantum.3.020361},
  issn = {2691-3399},
  url = {http://dx.doi.org/10.1103/PRXQuantum.3.020361}
}

@misc{Peham_2023,
  title = {Depth-Optimal Synthesis of {C}lifford Circuits with SAT Solvers},
  author = {Tom Peham and Nina Brandl and Richard Kueng and Robert Wille and Lukas Burgholzer},
  year = {2023},
  eprint = {2305.01674},
  archivePrefix = {arXiv},
  primaryClass = {quant-ph},
  url = {https://arxiv.org/abs/2305.01674},
}

@book{Petz_1990,
  title = {An Invitation to the algebra of canonical commutation relations},
  author = {Petz, D.},
  year = 1990,
  publisher = {Leuven University Press},
  address = {Leuven, Belgium},
  series = {Leuven notes in mathematical and theoretical physics},
  isbn = 9789061863601,
  nolink = {}
}

@misc{Qiskit_2024,
  title = {Quantum computing with {Qiskit}},
  author = {Ali Javadi-Abhari and Matthew Treinish and Kevin Krsulich and Christopher J. Wood and Jake Lishman and Julien Gacon and Simon Martiel and Paul D. Nation and Lev S. Bishop and Andrew W. Cross and Blake R. Johnson and Jay M. Gambetta},
  year = 2024,
  url = {https://arxiv.org/abs/2405.08810},
  eprint = {2405.08810},
  archiveprefix = {arXiv},
  primaryclass = {quant-ph}
}

@article{Raussendorf_2007,
  title = {Fault-Tolerant Quantum Computation with High Threshold in Two Dimensions},
  author = {Raussendorf, Robert and Harrington, Jim},
  year = 2007,
  month = may,
  journal = {Phys. Rev. Lett.},
  publisher = {American Physical Society},
  volume = 98,
  pages = 190504,
  doi = {10.1103/PhysRevLett.98.190504},
  url = {https://link.aps.org/doi/10.1103/PhysRevLett.98.190504},
  issue = 19,
  numpages = 4
}

@article{Raussendorf_2017,
  title = {Contextuality and {W}igner-function negativity in qubit quantum computation},
  author = {Robert Raussendorf and Dan E. Browne and Nicolas Delfosse and Cihan Okay and Juan Bermejo-Vega},
  year = 2017,
  month = may,
  journal = {Physical Review A},
  publisher = {American Physical Society ({APS})},
  volume = 95,
  number = 5,
  doi = {10.1103/physreva.95.052334},
  url = {https://doi.org/10.1103/physreva.95.052334}
}

@article{Raussendorf_2020,
  title = {Phase-space-simulation method for quantum computation with magic states on qubits},
  author = {Robert Raussendorf and Juani Bermejo-Vega and Emily Tyhurst and Cihan Okay and Michael Zurel},
  year = 2020,
  month = jan,
  journal = {Physical Review A},
  publisher = {American Physical Society ({APS})},
  volume = 101,
  number = 1,
  doi = {10.1103/physreva.101.012350},
  url = {https://doi.org/10.1103/physreva.101.012350}
}

@misc{Raussendorf_2022,
  title = {Putting paradoxes to work: contextuality in measurement-based quantum computation},
  author = {Robert Raussendorf},
  year = 2022,
  url = {https://arxiv.org/abs/2208.06624},
  eprint = {2208.06624},
  archiveprefix = {arXiv},
  primaryclass = {quant-ph}
}

@article{Renou_2021,
  title = {Quantum theory based on real numbers can be experimentally falsified},
  author = {Renou, Marc-Olivier and Trillo, David and Weilenmann, Mirjam and Le, Thinh P and Tavakoli, Armin and Gisin, Nicolas and Ac{\'\i}n, Antonio and Navascu{\'e}s, Miguel},
  year = 2021,
  journal = {Nature},
  publisher = {Nature Publishing Group UK London},
  volume = 600,
  number = 7890,
  pages = {625--629},
  doi = {https://doi.org/10.1038/s41586-021-04160-4},
  url = {https://www.nature.com/articles/s41586-021-04160-4}
}

@misc{Rijlaarsdam_2020,
  title = {Improvements of the classical simulation of quantum circuits: Using graph states with local {C}liffords},
  author = {Rijlaarsdam, Matthijs},
  year = 2020,
  url = {http://resolver.tudelft.nl/uuid:d5143594-80d1-465f-8dce-8cae5432bf6b}
}

@misc{Rudolph_2002,
  title = {A {$2$} rebit gate universal for quantum computing},
  author = {Terry Rudolph and Lov Grover},
  year = 2002,
  eprint = {quant-ph/0210187},
  archiveprefix = {arXiv},
  primaryclass = {quant-ph}
}

@misc{Sarkar_2025,
  title = {Gap between quantum theory based on real and complex numbers is arbitrarily large},
  author = {Sarkar, Shubhayan and Trillo, David and Renou, Marc Olivier and Augusiak, Remigiusz},
  year = 2025,
  url = {https://arxiv.org/abs/2503.09724},
  eprint = {2503.09724},
  archiveprefix = {arXiv},
  primaryclass = {quant-ph}
}

@article{Schmid_2022,
  title = {Uniqueness of Noncontextual Models for Stabilizer Subtheories},
  author = {Schmid, David and Du, Haoxing and Selby, John H. and Pusey, Matthew F.},
  year = 2022,
  month = sep,
  journal = {Physical Review Letters},
  publisher = {American Physical Society (APS)},
  volume = 129,
  number = 12,
  doi = {10.1103/physrevlett.129.120403},
  issn = {1079-7114},
  url = {http://dx.doi.org/10.1103/PhysRevLett.129.120403}
}

@article{Schmidt_2009,
  title = {{$\Int_4$} -Valued Quadratic Forms and Quaternary Sequence Families},
  author = {Schmidt, Kai-Uwe},
  year = 2009,
  journal = {IEEE Transactions on Information Theory},
  volume = 55,
  number = 12,
  pages = {5803--5810},
  doi = {10.1109/TIT.2009.2032818}
}

@article{Seddon_2019,
  title = {Quantifying magic for multi-qubit operations},
  author = {James R. Seddon and Earl T. Campbell},
  year = 2019,
  month = jul,
  journal = {Proceedings of the Royal Society A: Mathematical, Physical and Engineering Sciences},
  publisher = {The Royal Society},
  volume = 475,
  number = 2227,
  pages = 20190251,
  doi = {10.1098/rspa.2019.0251},
  url = {https://doi.org/10.1098/rspa.2019.0251}
}

@article{Seddon_2021,
  title = {Quantifying Quantum Speedups: Improved Classical Simulation From Tighter Magic Monotones},
  author = {Seddon, James R. and Regula, Bartosz and Pashayan, Hakop and Ouyang, Yingkai and Campbell, Earl T.},
  year = 2021,
  month = mar,
  journal = {PRX Quantum},
  publisher = {American Physical Society (APS)},
  volume = 2,
  number = 1,
  doi = {10.1103/prxquantum.2.010345},
  issn = {2691-3399},
  url = {http://dx.doi.org/10.1103/PRXQuantum.2.010345}
}

@article{Shi_2003,
  title = {Both {T}offoli and {C}ontrolled-{NOT} Need Little Help to Do Universal Quantum Computing},
  author = {Shi, Yaoyun},
  year = 2003,
  month = jan,
  journal = {Quantum Info. Comput.},
  publisher = {Rinton Press, Incorporated},
  address = {Paramus, NJ},
  volume = 3,
  number = 1,
  pages = {84–92},
  issn = {1533-7146},
  issue_date = {January 2003},
  numpages = 9,
  nolink = {}
}

@misc{Shirokov_2007,
  title = {On approximation of quantum channels},
  author = {M. E. Shirokov and A. S. Holevo},
  year = 2007,
  url = {https://arxiv.org/abs/0711.2245},
  eprint = {0711.2245},
  archiveprefix = {arXiv},
  primaryclass = {quant-ph}
}

@article{Skotiniotis_2012,
  title = {Alignment of reference frames and an operational interpretation for the {$G$}-asymmetry},
  author = {Skotiniotis, Michael and Gour, Gilad},
  year = 2012,
  month = jul,
  journal = {New Journal of Physics},
  publisher = {IOP Publishing},
  volume = 14,
  number = 7,
  pages = {073022},
  doi = {10.1088/1367-2630/14/7/073022},
  issn = {1367-2630},
  url = {http://dx.doi.org/10.1088/1367-2630/14/7/073022}
}

@article{Spekkens_2007,
  title = {Evidence for the epistemic view of quantum states: A toy theory},
  author = {Spekkens, Robert W.},
  year = 2007,
  month = mar,
  journal = {Physical Review A},
  publisher = {American Physical Society (APS)},
  volume = 75,
  number = 3,
  doi = {10.1103/physreva.75.032110},
  issn = {1094-1622},
  url = {http://dx.doi.org/10.1103/PhysRevA.75.032110}
}

@article{Steane_1996,
  title = {Multiple-particle interference and quantum error correction},
  author = {Steane, Andrew},
  year = 1996,
  journal = {Proceedings of the Royal Society of London. Series A: Mathematical, Physical and Engineering Sciences},
  volume = 452,
  number = 1954,
  pages = {2551--2577},
  doi = {10.1098/rspa.1996.0136},
  url = {https://royalsocietypublishing.org/doi/abs/10.1098/rspa.1996.0136}
}

@article{Van_den_Nest_2010,
  title = {Classical simulation of quantum computation, the {G}ottesman-{K}nill theorem, and slightly beyond},
  author = {Van den Nest, M.},
  year = 2010,
  month = mar,
  journal = {Quantum Information and Computation},
  publisher = {Rinton Press},
  volume = 10,
  number = {3 \& 4},
  pages = {258–271},
  doi = {10.26421/qic10.3-4-6},
  issn = {1533-7146},
  url = {http://dx.doi.org/10.26421/QIC10.3-4-6}
}

@article{Veitch_2012,
  title = {Negative quasi-probability as a resource for quantum computation},
  author = {Victor Veitch and Christopher Ferrie and David Gross and Joseph Emerson},
  year = 2012,
  month = nov,
  journal = {New Journal of Physics},
  publisher = {{IOP} Publishing},
  volume = 14,
  number = 11,
  pages = 113011,
  doi = {10.1088/1367-2630/14/11/113011},
  url = {https://doi.org/10.1088/1367-2630/14/11/113011}
}

@inproceedings{Vilmart_2021,
  title = {The Structure of {Sum-Over-Paths}, its Consequences, and Completeness for {C}lifford},
  author = {Vilmart, Renaud},
  url = {https://hal.science/hal-02651473},
  booktitle = {Foundations of Software Science and Computation Structures},
  address = {Luxembourg, Luxembourg},
  editor = {Stefan Kiefer, Christine Tasson},
  series = {Foundations of Software Science and Computation Structures},
  volume = {12650},
  pages = {531-550},
  year = {2021},
  month = Mar,
  doi = {10.1007/978-3-030-71995-1\_27},
  HAL_ID = {hal-02651473},
  HAL_VERSION = {v1}
}

@article{Weedbrook_2012,
  title = {Gaussian quantum information},
  author = {Weedbrook, Christian and Pirandola, Stefano and García-Patrón, Raúl and Cerf, Nicolas J. and Ralph, Timothy C. and Shapiro, Jeffrey H. and Lloyd, Seth},
  year = 2012,
  month = may,
  journal = {Reviews of Modern Physics},
  publisher = {American Physical Society (APS)},
  volume = 84,
  number = 2,
  pages = {621–669},
  doi = {10.1103/revmodphys.84.621},
  issn = {1539-0756},
  url = {http://dx.doi.org/10.1103/RevModPhys.84.621}
}

@article{Wigner_1932,
  title = {On the Quantum Correction For Thermodynamic Equilibrium},
  author = {Wigner, Eugene},
  year = 1932,
  month = jun,
  journal = {Phys. Rev.},
  publisher = {American Physical Society},
  volume = 40,
  pages = {749--759},
  doi = {10.1103/PhysRev.40.749},
  url = {https://link.aps.org/doi/10.1103/PhysRev.40.749},
  issue = 5,
  numpages = {0}
}

@article{Wong_2024,
  title = {The Gauge Theory of Measurement-Based Quantum Computation},
  author = {Wong, Gabriel and Raussendorf, Robert and Czech, Bartlomiej},
  year = 2024,
  month = jul,
  journal = {Quantum},
  publisher = {Verein zur Forderung des Open Access Publizierens in den Quantenwissenschaften},
  volume = 8,
  pages = 1397,
  doi = {10.22331/q-2024-07-04-1397},
  issn = {2521-327X},
  url = {http://dx.doi.org/10.22331/q-2024-07-04-1397}
}

@article{Wootters_1987,
  title = {A {W}igner-function formulation of finite-state quantum mechanics},
  author = {William K Wootters},
  year = 1987,
  journal = {Annals of Physics},
  volume = 176,
  number = 1,
  pages = {1--21},
  doi = {https://doi.org/10.1016/0003-4916(87)90176-X},
  issn = {0003-4916},
  url = {https://www.sciencedirect.com/science/article/pii/000349168790176X}
}

@article{Yashin_2020,
  title = {Minimal informationally complete measurements for probability representation of quantum dynamics},
  author = {Yashin, V I and Kiktenko, E O and Mastiukova, A S and Fedorov, A K},
  year = 2020,
  month = oct,
  journal = {New Journal of Physics},
  publisher = {IOP Publishing},
  volume = 22,
  number = 10,
  pages = 103026,
  doi = {10.1088/1367-2630/abb963},
  issn = {1367-2630},
  url = {http://dx.doi.org/10.1088/1367-2630/abb963}
}

@article{Yashin_2025,
  title = {Characterization of non-adaptive {C}lifford channels},
  author = {Yashin, Vsevolod I. and Elovenkova, Maria A.},
  year = 2025,
  month = mar,
  journal = {Quantum Information Processing},
  publisher = {Springer Science and Business Media LLC},
  volume = 24,
  number = 3,
  doi = {10.1007/s11128-025-04682-0},
  issn = {1573-1332},
  url = {http://dx.doi.org/10.1007/s11128-025-04682-0}
}

@misc{Yashin_2025_2,
  title = {A streamlined demonstration that stabilizer circuits simulation reduces to {B}oolean linear algebra},
  author = {Vsevolod I. Yashin},
  year = 2025,
  url = {https://arxiv.org/abs/2504.14101},
  eprint = {2504.14101},
  archiveprefix = {arXiv},
  primaryclass = {quant-ph}
}

@misc{Yashin_TBA,
  title = {Properties of {C}lifford channels in stabilizer formalisms over arbitrary finite abelian groups},
  author = {Yashin, Vsevolod I.},
  year = {In Preparation}
}

@article{Yoganathan_2019,
  title = {Quantum advantage of unitary {C}lifford circuits with magic state inputs},
  author = {Yoganathan, Mithuna  and Jozsa, Richard  and Strelchuk, Sergii},
  year = 2019,
  journal = {Proceedings of the Royal Society A: Mathematical, Physical and Engineering Sciences},
  volume = 475,
  number = 2225,
  pages = 20180427,
  doi = {10.1098/rspa.2018.0427},
  url = {https://royalsocietypublishing.org/doi/abs/10.1098/rspa.2018.0427}
}

@article{Zhou_2000,
  title = {Methodology for quantum logic gate construction},
  author = {Zhou, Xinlan and Leung, Debbie W. and Chuang, Isaac L.},
  year = 2000,
  month = oct,
  journal = {Phys. Rev. A},
  publisher = {American Physical Society},
  volume = 62,
  pages = {052316},
  doi = {10.1103/PhysRevA.62.052316},
  url = {https://link.aps.org/doi/10.1103/PhysRevA.62.052316},
  issue = 5,
  numpages = 12
}

@article{Zurel_2020,
  title = {Hidden Variable Model for Universal Quantum Computation with Magic States on Qubits},
  author = {Michael Zurel and Cihan Okay and Robert Raussendorf},
  year = 2020,
  month = dec,
  journal = {Physical Review Letters},
  publisher = {American Physical Society ({APS})},
  volume = 125,
  number = 26,
  doi = {10.1103/physrevlett.125.260404},
  url = {https://doi.org/10.1103/physrevlett.125.260404}
}

@article{Zurel_2024,
  title = {Hidden variable model for quantum computation with magic states on qudits of any dimension},
  author = {Zurel, Michael and Okay, Cihan and Raussendorf, Robert and Heimendahl, Arne},
  year = 2024,
  month = apr,
  journal = {Quantum},
  publisher = {Verein zur Forderung des Open Access Publizierens in den Quantenwissenschaften},
  volume = 8,
  pages = 1323,
  doi = {10.22331/q-2024-04-30-1323},
  issn = {2521-327X},
  url = {http://dx.doi.org/10.22331/q-2024-04-30-1323}
}

@article{Zurel_2024_2,
  title = {Simulating Quantum Computation: How Many `Bits' for `It'?},
  author = {Zurel, Michael and Okay, Cihan and Raussendorf, Robert},
  year = 2024,
  month = sep,
  journal = {PRX Quantum},
  publisher = {American Physical Society (APS)},
  volume = 5,
  number = 3,
  doi = {10.1103/prxquantum.5.030343},
  issn = {2691-3399},
  url = {http://dx.doi.org/10.1103/PRXQuantum.5.030343}
}

@misc{Zurel_2024_3,
  title = {Efficient classical simulation of quantum computation beyond {W}igner positivity},
  author = {Michael Zurel and Arne Heimendahl},
  year = 2024,
  url = {https://arxiv.org/abs/2407.10349},
  eprint = {2407.10349},
  archiveprefix = {arXiv},
  primaryclass = {quant-ph}
}

%----------------------------------------------------------------------------------------
%  APPENDICES
%----------------------------------------------------------------------------------------
% \onecolumn
\appendix
\section{Classical rewriting of one-cell defect movement in surface code} \label{appendix:surface_code}

In this Appendix, we illustrate how the rewriting method of \cref{sec:rewriting} can be applied to describe the operation of surface code computations with defect braidings \cite{Raussendorf_2007, Fowler_2012}. We will consider the simplest yet essential step in this approach: a one-cell movement of a defect. Using classical bit updates, we will demonstrate why the procedure described in Ref.~\cite{Fowler_2012} is correct.

Surface codes is an important and widely studied family of quantum error correcting code that has a number of good properties: they are CSS (therefore, allow for transversal $\CNOT$ gates), they have beautiful geometrical description, they have relatively low overhead and relatively high error threshold.

A surface code consists of a two-dimensional lattice of \emph{data qubits}, fixed by local $Z$- and $X$-stabilizers of form $Z_1Z_2Z_3Z_4$ and $X_1X_2X_3X_4$ respectively. In order to measure the stabilizer syndromes, there is an additional lattice of \emph{ancilla qubits}. Surface codes are illustrated as a carpet of data qubits and ancilla qubits joined by $Z$- and $X$-stabilizers:
\begin{equation} \label{eq:surface_code_depiction}
\adjustbox{width=0.4\textwidth,valign=m}{
  \begin{tikzpicture}
    % size of the lattice
    \def\Nx{8};
    \def\Ny{4};
    % primal lattice
    \draw[step=2cm, color=blue!70!black!30!white, line width=2.5pt]
      (-0.5cm,-0.5cm) grid (\Nx cm + 0.5cm,\Ny cm + 0.5cm);
    % dual lattice
    \draw[step=2cm, xshift=1cm, yshift=1cm, color=red!70!black!30!white, line width=2.5pt]
      (-1.5cm,-1.5cm) grid (\Nx cm - 0.5cm, \Ny cm - 0.5cm);
    % qubits
    \foreach \x in {0,1,...,\Nx}
    \foreach \y in {0,1,...,\Ny}{
      \pgfmathparse{\x+\y}
      \ifodd\pgfmathresult
        % data qubits
        \shade[ball color=white,draw=black,thick]
          (\x,\y) circle[radius=3pt];
      \else
        % ancilla qubits
        \fill[black,draw=black,thick]
          (\x,\y) circle[radius=3pt];
      \fi
    };
  \end{tikzpicture}
}
\end{equation}
Here and thereafter, we indicate data qubits as shaded balls $\tikz{ \shade[ball color=white,draw=black,thick] circle[radius=3pt]; }$, ancilla qubits as black balls $\tikz{ \fill[black,draw=black,thick] circle[radius=3pt]; }$; they are joined by a grid of blue lines \adjustbox{height=1.2em,raise=-0.3em}{$\tikz{  \draw[step=2cm, color=blue!70!black!30!white, line width=2.5pt] (-0.2cm,-0.2cm) grid (0.2cm,0.2cm); }$} (called \emph{primary lattice}), and a grid of red lines \adjustbox{height=1.2em,raise=-0.3em}{$\tikz{  \draw[step=2cm, color=red!70!black!30!white, line width=2.5pt] (-0.2cm,-0.2cm) grid (0.2cm,0.2cm); }$} (called \emph{dual lattice}). These grids indicate the occurence of a stabilizer measurement: $Z$-stabilizer measurements are performed by introducing ancilla qubit in state $\ket{0}$, entangling it using $\CNOT$s and measuring ancilla:
\begin{equation} \label{eq:surface_Z_stabilizer}
  \adjustbox{valign=m,scale=0.9}{
  \begin{tikzpicture}
    \foreach \n [count=\ni] in {(0cm,1cm),(1cm,0cm),(0cm,-1cm),(-1cm,0cm)} {
      \draw[color=blue!70!black!30!white, line width=2.5pt] (0cm,0cm) -- \n ;
      \shade[ball color=white,draw=black,thick]
        \n circle[radius=3pt];
      \node[above left,color=blue!70!black] at \n {$Z_{\ni}$};
    };
    \fill[black,draw=black,thick] (0cm,0cm) circle[radius=3pt];
  \end{tikzpicture}
  }
  ~
  \adjustbox{valign=m,scale=0.9}{
  \begin{quantikz}[wire types={n,q,q,q,q}, row sep={0.5cm,between origins}, column sep={0.6cm,between origins}]
    &\lstick{$\ket{0}$} &[-0.2cm]\targ{}\setwiretype{q} &\targ{} &\targ{} &\targ{} &\meterD{Z} &\setwiretype{c} \\[0.3cm]
    &&\ctrl{-1} &&&&& \\
    &&&\ctrl{-2} &&&& \\
    &&&&\ctrl{-3} &&& \\
    &&&&&\ctrl{-4} &&
  \end{quantikz}
  }
\end{equation}
and similarly to measure $X$-stabilizer one performs a circuit:
\begin{equation} \label{eq:surface_X_stabilizer}
  \adjustbox{valign=m,scale=0.9}{
  \begin{tikzpicture}
    \foreach \n [count=\ni] in {(0cm,1cm),(1cm,0cm),(0cm,-1cm),(-1cm,0cm)} {
      \draw[color=red!70!black!30!white, line width=2.5pt] (0cm,0cm) -- \n ;
      \shade[ball color=white,draw=black,thick]
        \n circle[radius=3pt];
      \node[above right,color=red!70!black] at \n {$X_{\ni}$};
    };
    \fill[black,draw=black,thick] (0cm,0cm) circle[radius=3pt];
  \end{tikzpicture}
  }
  ~
  \adjustbox{valign=m,scale=0.9}{
  \begin{quantikz}[wire types={n,q,q,q,q}, row sep={0.5cm,between origins}, column sep={0.6cm,between origins}]
    &\lstick{$\ket{+}$} &[-0.2cm]\ctrl{1}\setwiretype{q} &\ctrl{2} &\ctrl{3} &\ctrl{4} &\meterD{X} &\setwiretype{c} \\[0.3cm]
    &&\targ{} &&&&& \\
    &&&\targ{} &&&& \\
    &&&&\targ{} &&& \\
    &&&&&\targ{} &&
  \end{quantikz}
  }
\end{equation}
Rounds of error correction consist of measuring all local stabilizers according to  \cref{eq:surface_Z_stabilizer,eq:surface_X_stabilizer} and decoding the errors from the obtained syndromes.

Logical Pauli gates in a surface code depend on the topology of the surface: they correspond to non-contractible paths on the primal or dual lattice, which appear if the surface has non-zero genus or has boundaries. Note that in \cref{eq:surface_code_depiction}, we illustrated the bulk of the surface and did not draw boundaries.

One possible technique to process quantum data in surface codes is called \emph{computations using defect braiding}. It is based on the idea of puncturing holes inside of the carpet and manipulating these holes by moving them around the surface. There are two types of holes: $Z$-holes and $X$-holes. Creating a hole is done by deleting a ($Z$- or $X$-) stabilizer out of the code, therefore more holes means more logical qubits in the system. Pauli gates are processed on a classical computer, and one can manipulate hole positions by movement protocols while correcting emerging byproduct Pauli gates. Clifford gates are implemented by braiding the holes, e.g. a $\CNOT$ gate is done via a detour of $Z$-hole over $X$-hole. To measure a logical qubit in computational basis is to measure the stabilizer closing the hole. By braiding the trajectories of holes, one can fault-tolerantly implement logical CSS-preserving stabilizer operations; to produce arbitrary operations, one can can inject imaginary states $\ket{+i}$ and magic states $\ket{H}$. For deep and concrete explanation of this framework, consult \cite{Fowler_2012,Fujii_2015}.

The simplest elementary operation employed in surface code computing using defect braiding is moving a hole to the neighbouring position. The procedure of moving a $Z$-hole one position below is as follows:
\begin{equation} \label{eq:one-cell_defect_movement}
\adjustbox{valign=m,width=0.4\textwidth}{
  \begin{tikzpicture}[font=\large]
    \node[font=\sffamily\Huge] at (2cm,13.2cm) {Step~I};
    % size of the lattice
    \def\Nx{4};
    \def\Ny{12};
    % primal lattice
    \draw[step=2cm, color=blue!70!black!30!white, line width=2.5pt]
      (-0.5cm,-0.5cm) grid (\Nx cm + 0.5cm,\Ny cm + 0.5cm);
    % dual lattice
    \draw[step=2cm, xshift=1cm, yshift=1cm, color=red!70!black!30!white, line width=2.5pt]
      (-1.5cm,-1.5cm) grid (\Nx cm - 0.5cm, \Ny cm - 0.5cm);
    % holes
    \fill[color=white]
      (1cm+1.25pt,9cm+1.25pt) rectangle (3cm-1.25pt,11cm-1.25pt);
    \fill[color=white]
      (1cm+1.25pt,3cm+1.25pt) rectangle (3cm-1.25pt,5cm-1.25pt);
    % qubits
    \foreach \x in {0,1,...,\Nx}
    \foreach \y in {0,1,...,\Ny}{
      \pgfmathparse{\x+\y}
      \ifodd\pgfmathresult
        % data qubits
        \shade[ball color=white,draw=black,thick]
          (\x,\y) circle[radius=3pt];
      \else
        % ancilla qubits
        \fill[black,draw=black,thick]
          (\x,\y) circle[radius=3pt];
      \fi
    };
    % logical X
    \draw[line width=3.5pt,color=red] (2cm,9cm) -- (2cm,5cm);
    \node[above left,color=red!70!black] at (2cm,9cm) {$X_9$};
    \node[above left,color=red!70!black] at (2cm,7cm) {$X_8$};
    \node[above left,color=red!70!black] at (2cm,5cm) {$X_1$};
    \node[font=\huge,color=red!70!black,fill=white] at (2.8cm,7cm) {$\boldsymbol{X_L}$};
    %logical Z
    \draw[line width=3.5pt,color=blue] (1cm,5cm) -- (3cm,5cm) -- (3cm,3cm) -- (1cm,3cm) -- cycle;
    \draw[dashed,line width=3.5pt,color=blue] (3cm,3cm) -- (3cm,1cm) -- (1cm,1cm) -- (1cm,3cm);
    \node[below,color=blue!70!black] at (2cm,5cm) {$Z_1$};
    \node[right,color=blue!70!black] at (1cm,4cm) {$Z_2$};
    \node[left,color=blue!70!black] at (3cm,4cm) {$Z_3$};
    \node[above,color=blue!70!black] at (2cm,3cm) {$Z_4$};
    \node[font=\huge,color=blue!70!black,fill=white] at (3.8cm,4cm) {$\boldsymbol{Z_L}$};
  \end{tikzpicture}
  \quad
  \begin{tikzpicture}[font=\large]
    \node[font=\sffamily\Huge] at (2cm,13.2cm) {Step~II};
    % size of the lattice
    \def\Nx{4};
    \def\Ny{12};
    % primal lattice
    \draw[step=2cm, color=blue!70!black!30!white, line width=2.5pt]
      (-0.5cm,-0.5cm) grid (\Nx cm + 0.5cm,\Ny cm + 0.5cm);
    % dual lattice
    \draw[step=2cm, xshift=1cm, yshift=1cm, color=red!70!black!30!white, line width=2.5pt]
      (-1.5cm,-1.5cm) grid (\Nx cm - 0.5cm, \Ny cm - 0.5cm);
    % holes
    \fill[color=white]
      (1cm+1.25pt,9cm+1.25pt) rectangle (3cm-1.25pt,11cm-1.25pt);
    \fill[color=white]
      (1cm+1.25pt,1cm+1.25pt) rectangle (3cm-1.25pt,5cm-1.25pt);
    % qubits
    \foreach \x in {0,1,...,\Nx}
    \foreach \y in {0,1,...,\Ny}{
      \pgfmathparse{\x+\y}
      \ifodd\pgfmathresult
        % data qubits
        \shade[ball color=white,draw=black,thick]
          (\x,\y) circle[radius=3pt];
      \else
        % ancilla qubits
        \fill[black,draw=black,thick]
          (\x,\y) circle[radius=3pt];
      \fi
    };
    % logical X
    \draw[line width=3.5pt,color=red] (2cm,9cm) -- (2cm,5cm);
    \node[above left,color=red!70!black] at (2cm,9cm) {$X_9$};
    \node[above left,color=red!70!black] at (2cm,7cm) {$X_8$};
    \node[above left,color=red!70!black] at (2cm,5cm) {$X_1$};
    \node[font=\huge,color=red!70!black,fill=white] at (2.8cm,7cm) {$\boldsymbol{X_L}$};
    \node[right,color=red!70!black] at (2cm,3cm) {$X_4$};
    %logical Z
    \draw[line width=3.5pt,color=blue] (1cm,5cm) -- (3cm,5cm) -- (3cm,1cm) -- (1cm,1cm) -- cycle;
    \node[below,color=blue!70!black] at (2cm,5cm) {$Z_1$};
    \node[right,color=blue!70!black] at (1cm,4cm) {$Z_2$};
    \node[left,color=blue!70!black] at (3cm,4cm) {$Z_3$};
    \node[right,color=blue!70!black] at (1cm,2cm) {$Z_5$};
    \node[left,color=blue!70!black] at (3cm,2cm) {$Z_6$};
    \node[above,color=blue!70!black] at (2cm,1cm) {$Z_7$};
    \node[font=\huge,color=blue!70!black,fill=white] at (3.8cm,3cm) {$\boldsymbol{Z_L^e}$};
  \end{tikzpicture}
  \quad
  \begin{tikzpicture}[font=\large]
    \node[font=\sffamily\Huge] at (2cm,13.2cm) {Step~III};
    % size of the lattice
    \def\Nx{4};
    \def\Ny{12};
    % primal lattice
    \draw[step=2cm, color=blue!70!black!30!white, line width=2.5pt]
      (-0.5cm,-0.5cm) grid (\Nx cm + 0.5cm,\Ny cm + 0.5cm);
    % dual lattice
    \draw[step=2cm, xshift=1cm, yshift=1cm, color=red!70!black!30!white, line width=2.5pt]
      (-1.5cm,-1.5cm) grid (\Nx cm - 0.5cm, \Ny cm - 0.5cm);
    % holes
    \fill[color=white]
      (1cm+1.25pt,9cm+1.25pt) rectangle (3cm-1.25pt,11cm-1.25pt);
    \fill[color=white]
      (1cm+1.25pt,1cm+1.25pt) rectangle (3cm-1.25pt,3cm-1.25pt);
    % qubits
    \foreach \x in {0,1,...,\Nx}
    \foreach \y in {0,1,...,\Ny}{
      \pgfmathparse{\x+\y}
      \ifodd\pgfmathresult
        % data qubits
        \shade[ball color=white,draw=black,thick]
          (\x,\y) circle[radius=3pt];
      \else
        % ancilla qubits
        \fill[black,draw=black,thick]
          (\x,\y) circle[radius=3pt];
      \fi
    };
    % logical X
    \draw[line width=3.5pt,color=red] (2cm,9cm) -- (2cm,3cm);
    \node[above left,color=red!70!black] at (2cm,9cm) {$X_9$};
    \node[above left,color=red!70!black] at (2cm,7cm) {$X_8$};
    \node[above left,color=red!70!black] at (2cm,5cm) {$X_1$};
    \node[above left,color=red!70!black] at (2cm,3cm) {$X_4$};
    \node[font=\huge,color=red!70!black,fill=white] at (2.8cm,6cm) {$\boldsymbol{X_L'}$};
    %logical Z
    \draw[line width=3.5pt,color=blue] (1cm,3cm) -- (3cm,3cm) -- (3cm,1cm) -- (1cm,1cm) -- cycle;
    \node[below,color=blue!70!black] at (2cm,3cm) {$Z_4$};
    \node[right,color=blue!70!black] at (1cm,2cm) {$Z_5$};
    \node[left,color=blue!70!black] at (3cm,2cm) {$Z_6$};
    \node[above,color=blue!70!black] at (2cm,1cm) {$Z_7$};
    \node[font=\huge,color=blue!70!black,fill=white] at (3.8cm,2cm) {$\boldsymbol{Z_L'}$};
  \end{tikzpicture}
}
\end{equation}
Let us discuss what happens on this figure. On \textsf{Step~I}, the surface contains two $Z$-holes (two $Z$-stabilizers were disabled), which makes it possible to encode logical Pauli operators as non-contractible paths $\boldsymbol{X_L} = X_9X_8X_1$ and $\boldsymbol{Z_L} = Z_1Z_2Z_3Z_4$. All active stabilizers are measured and remain in zero eigenvalue. On \textsf{Step~II}, the $Z$-stabilizer neighbouring to a hole is disabled and the qubit $X_4$ is measured to widen the hole, resulting in updated logical Pauli operator $\boldsymbol{Z_L^e} = Z_1Z_2Z_3Z_5Z_6Z_7$. On \textsf{Step~III}, the upper part of the hole is shrinked as a result of measuring $Z$-stabilizer $Z_1Z_2Z_3Z_4$, the final logical Pauli operators are $\boldsymbol{Z_L'} = Z_4Z_5Z_6Z_7$ and $\boldsymbol{X_L'} = X_9X_8X_1X_4$.

As a result of this code update, the logical operators $\boldsymbol{Z_L'}$ and $\boldsymbol{X_L'}$ may differ from $\boldsymbol{Z_L}$ and $\boldsymbol{X_L}$ up to signs, that is, a logical state can be transformed to
\begin{equation}
  \ket{\psi}\mapsto \boldsymbol{X_L'}^{d_Z}\boldsymbol{Z_L'}^{d_X}\ket{\psi'},
\end{equation}
where $d_Z$ and $d_X$ can be calculated from the measurement outcomes: $d_X$ equals the result of $X_4$-measurement and $d_Z$ equals the result of $Z_1Z_2Z_3Z_4$ measurement.

Let us examine how this procedure works if we rewrite this circuit to a classical circuit. To each data qubit we correspond two classical bits $z$ and $x$, we will depict them as
\begin{equation}
\scalebox{1.2}{
  \begin{tikzpicture}
    \shade[ball color=white,draw=black,thick] (0,0) circle[radius=3pt];
    \node[above left,color=blue!70!black] at (0,0) {$z$};
    \node[above right,color=red!70!black] at (0,0) {$x$};
  \end{tikzpicture}
}
\end{equation}

When reduced to a classical circuit, the operation \cref{eq:surface_Z_stabilizer} of measuring $Z$-stabilizer $Z_1Z_2Z_3Z_4$ corresponds to obtaining a classical bit $z_1\oplus z_2\oplus z_3\oplus z_4$ and introducting a correlated error $x_a$ to each $X$-bit:
\begin{equation}
  \adjustbox{valign=m,width=0.39\textwidth}{
  \begin{quantikz}[wire types={cz,cx,cz,cx,cz,cx,cz,cx,cz,cx}, row sep={0.25cm,between origins}, column sep={0.5cm,between origins}]
    \lstick{$0$}   & &\targ{}\wire[d][2]{c} & &\targ{}\wire[d][4]{c} & &\targ{}\wire[d][6]{c} & &\targ{}\wire[d][8]{c} &\rstick{$z_1\oplus z_2\oplus z_3\oplus z_4$} \\
    \lstick{$x_a$} &\ctrl{0}\wire[d][2]{c} & &\ctrl{0}\wire[d][4]{c} & &\ctrl{0}\wire[d][6]{c} & &\ctrl{0}\wire[d][8]{c} & &\ground{} \\[0.8cm]
    \lstick{$z_1$} & &\ctrl{0} & & & & & & &\rstick{$z_1$} \\
    \lstick{$x_1$} &\targ{} & & & & & & & &\rstick{$x_1\oplus x_a$} \\[0.4cm]
    \lstick{$z_2$} & & & &\ctrl{0} & & & & &\rstick{$z_2$} \\
    \lstick{$x_2$} & & &\targ{} & & & & & &\rstick{$x_2\oplus x_a$} \\[0.4cm]
    \lstick{$z_3$} & & & & & &\ctrl{0} & & &\rstick{$z_3$} \\
    \lstick{$x_3$} & & & & &\targ{} & & & &\rstick{$x_3\oplus x_a$} \\[0.4cm]
    \lstick{$z_4$} & & & & & & & &\ctrl{0} &\rstick{$z_4$} \\
    \lstick{$x_4$} & & & & & & &\targ{} & &\rstick{$x_4\oplus x_a$}
  \end{quantikz}
  }
\end{equation}
Analogously, the operation \cref{eq:surface_X_stabilizer} of measuring $X$-stabilizer rewrites to a obtaining bit value $x_1\oplus x_2\oplus x_3\oplus x_4$ and introducing a correlated error $z_a$ to each $Z$-bit:
\begin{equation}
  \adjustbox{valign=m,width=0.39\textwidth}{
  \begin{quantikz}[wire types={cz,cx,cz,cx,cz,cx,cz,cx,cz,cx}, row sep={0.25cm,between origins}, column sep={0.5cm,between origins}]
    \lstick{$z_a$}   & &\ctrl{0}\wire[d][2]{c} & &\ctrl{0}\wire[d][4]{c} & &\ctrl{0}\wire[d][6]{c} & &\ctrl{0}\wire[d][8]{c} &\ground{} \\
    \lstick{$0$} &\targ{}\wire[d][2]{c} & &\targ{}\wire[d][4]{c} & &\targ{}\wire[d][6]{c} & &\targ{}\wire[d][8]{c} & &\rstick{$x_1\oplus x_2\oplus x_3\oplus x_4$}  \\[0.8cm]
    \lstick{$z_1$} & &\targ{} & & & & & & &\rstick{$z_1\oplus z_a$} \\
    \lstick{$x_1$} &\ctrl{0} & & & & & & & &\rstick{$x_1$} \\[0.4cm]
    \lstick{$z_2$} & & & &\targ{} & & & & &\rstick{$z_2\oplus z_a$} \\
    \lstick{$x_2$} & & &\ctrl{0} & & & & & &\rstick{$x_2$} \\[0.4cm]
    \lstick{$z_3$} & & & & & &\targ{} & & &\rstick{$z_3\oplus z_a$} \\
    \lstick{$x_3$} & & & & &\ctrl{0} & & & &\rstick{$x_3$} \\[0.4cm]
    \lstick{$z_4$} & & & & & & & &\targ{} &\rstick{$z_4\oplus z_a$} \\
    \lstick{$x_4$} & & & & & & &\ctrl{0} & &\rstick{$x_4$}
  \end{quantikz}
  }
\end{equation}
That means, in the case that all error syndromes are zero, for all stabilizers in the carpet it holds that $z_a\oplus z_b\oplus z_c\oplus z_d=0$ or $x_a\oplus x_b\oplus x_c\oplus x_d=0$, where $a,b,c,d$ are data qubits connected by a plaquette.

Let us describe the procedure of one-cell defect movement in terms of rewriting the circuit to a classical form. Here we depict only the lower part of \cref{eq:one-cell_defect_movement}, and after each step we display only the updated classical bits.
\begin{equation}
  \adjustbox{valign=m,width=0.4\textwidth}{
  \begin{tikzpicture}
    \node[font=\sffamily\huge] at (1cm,5.2cm) {Step~I};
    % size of the lattice
    \def\Nx{2};
    \def\Ny{4};
    % primal lattice
    \draw[step=2cm, color=red!70!black!30!white, line width=2.5pt]
      (-0.5cm,-0.5cm) grid (\Nx cm + 0.5cm,\Ny cm + 0.5cm);
    % dual lattice
    \draw[step=2cm, xshift=1cm, yshift=1cm, color=blue!70!black!30!white, line width=2.5pt]
      (-1.5cm,-1.5cm) grid (\Nx cm - 0.5cm, \Ny cm - 0.5cm);
    % holes
    \fill[color=white]
      (0cm+1.25pt,2cm+1.25pt) rectangle (2cm-1.25pt,4cm-1.25pt);
    % qubits
    \foreach \x in {0,1,...,\Nx}
    \foreach \y in {0,1,...,\Ny}{
      \pgfmathparse{\x+\y}
      \ifodd\pgfmathresult
        % data qubits
        \shade[ball color=white,draw=black,thick]
          (\x,\y) circle[radius=3pt];
      \else
        % ancilla qubits
        \fill[black,draw=black,thick]
          (\x,\y) circle[radius=3pt];
      \fi
    };
    % text
    \node[above left,color=blue!70!black] at (1 cm, 4 cm) {$z_1$};
    \node[above right,color=red!70!black] at (1 cm, 4 cm) {$x_1$};
    \node[above left,color=blue!70!black] at (0 cm, 3 cm) {$z_2$};
    \node[above right,color=red!70!black] at (0 cm, 3 cm) {$x_2$};
    \node[above left,color=blue!70!black] at (2 cm, 3 cm) {$z_3$};
    \node[above right,color=red!70!black] at (2 cm, 3 cm) {$x_3$};
    \node[above left,color=blue!70!black] at (1 cm, 2 cm) {$z_4$};
    \node[above right,color=red!70!black] at (1 cm, 2 cm) {$x_4$};
    \node[above left,color=blue!70!black] at (0 cm, 1 cm) {$z_5$};
    \node[above right,color=red!70!black] at (0 cm, 1 cm) {$x_5$};
    \node[above left,color=blue!70!black] at (2 cm, 1 cm) {$z_6$};
    \node[above right,color=red!70!black] at (2 cm, 1 cm) {$x_6$};
    \node[above left,color=blue!70!black] at (1 cm, 0 cm) {$z_7$};
    \node[above right,color=red!70!black] at (1 cm, 0 cm) {$x_7$};
  \end{tikzpicture}
  \quad
  \begin{tikzpicture}
    \node[font=\sffamily\huge] at (1cm,5.2cm) {Step~II};
    % size of the lattice
    \def\Nx{2};
    \def\Ny{4};
    % primal lattice
    \draw[step=2cm, color=red!70!black!30!white, line width=2.5pt]
      (-0.5cm,-0.5cm) grid (\Nx cm + 0.5cm,\Ny cm + 0.5cm);
    % dual lattice
    \draw[step=2cm, xshift=1cm, yshift=1cm, color=blue!70!black!30!white, line width=2.5pt]
      (-1.5cm,-1.5cm) grid (\Nx cm - 0.5cm, \Ny cm - 0.5cm);
    % holes
    \fill[color=white]
      (0cm+1.25pt,0cm+1.25pt) rectangle (2cm-1.25pt,4cm-1.25pt);
    % qubits
    \foreach \x in {0,1,...,\Nx}
    \foreach \y in {0,1,...,\Ny}{
      \pgfmathparse{\x+\y}
      \ifodd\pgfmathresult
        % data qubits
        \shade[ball color=white,draw=black,thick]
          (\x cm,\y cm) circle[radius=3pt];
      \else
        % ancilla qubits
        \fill[black,draw=black,thick]
          (\x cm,\y cm) circle[radius=3pt];
      \fi
    };
    % text
    \node[above left,color=blue!70!black] at (1 cm, 2 cm) {$z_4^m$};
    \node[above right,color=red!70!black] at (1 cm, 2 cm) {$x_4^m$};
  \end{tikzpicture}
  \quad
  \begin{tikzpicture}
    \node[font=\sffamily\huge] at (1cm,5.2cm) {Step~III};
    % size of the lattice
    \def\Nx{2};
    \def\Ny{4};
    % primal lattice
    \draw[step=2cm, color=red!70!black!30!white, line width=2.5pt]
      (-0.5cm,-0.5cm) grid (\Nx cm + 0.5cm,\Ny cm + 0.5cm);
    % dual lattice
    \draw[step=2cm, xshift=1cm, yshift=1cm, color=blue!70!black!30!white, line width=2.5pt]
      (-1.5cm,-1.5cm) grid (\Nx cm - 0.5cm, \Ny cm - 0.5cm);
    % holes
    \fill[color=white]
      (0cm+1.25pt,0cm+1.25pt) rectangle (2cm-1.25pt,2cm-1.25pt);
    % qubits
    \foreach \x in {0,1,...,\Nx}
    \foreach \y in {0,1,...,\Ny}{
      \pgfmathparse{\x+\y}
      \ifodd\pgfmathresult
        % data qubits
        \shade[ball color=white,draw=black,thick]
          (\x cm,\y cm) circle[radius=3pt];
      \else
        % ancilla qubits
        \fill[black,draw=black,thick]
          (\x cm,\y cm) circle[radius=3pt];
      \fi
    };
    % text
    \node[below left] at (1 cm, 3 cm) {$z_a$};
    \node[above right,color=red!70!black,font=\tiny] at (1 cm, 4 cm) {$x_1\!\oplus\!x_a$};
    \node[above right,color=red!70!black,font=\tiny] at (0 cm, 3 cm) {$x_2\!\oplus\!x_a$};
    \node[above right,color=red!70!black,font=\tiny] at (2 cm, 3 cm) {$x_3\!\oplus\!x_a$};
    \node[above right,color=red!70!black,font=\tiny] at (1 cm, 2 cm) {$x_4^m\!\oplus\!x_a$};
  \end{tikzpicture}
  }
\end{equation}
On \textsf{Step~I}, there are seven data qubits with corresponding bits $(z_1,x_1,\dots,z_7,x_7)$, the upper $Z$-stabilizer is absent, and the lower stabilizer ensures $z_4\oplus z_5\oplus z_6\oplus z_7 = 0$. On \textsf{Step~II}, we disable the lower stabilizer and measure data qubit $4$, obtaining determinate value $x_4^m$ (a sample from $x_4$) and uniformly random bit $z_4^m$. On \textsf{Step~III}, the upper $Z$-stabilizer is measured, giving determinate value $z_a=z_1\oplus z_2\oplus z_3\oplus z_4^m$ and introducing uniformly random bit $x_a$ to qubits $1,2,3,4$.

Let us make sure that logical operations do not break because of defect movement. Before the movement, measuring the logical qubit $\boldsymbol{Z_L}=Z_1Z_2Z_3Z_4$ gives a bit value
\begin{equation}
  m_Z = z_1\oplus z_2\oplus z_3\oplus z_4.
\end{equation}
After the movement, measuring the $\boldsymbol{Z_L'}=Z_4Z_5Z_6Z_7$ gives
\begin{equation}
  m_Z' = z_4^m\oplus z_5\oplus z_6\oplus z_7  = m_Z \oplus d_Z,
\end{equation}
where dy $d_Z$ we denote the difference between such measurements. Then, by definition
\begin{equation}
\begin{aligned}
  d_Z
  &= m_Z \oplus m_Z' \\
  &= (z_1\!\oplus\! z_2\!\oplus\! z_3\!\oplus\! z_4^m) \oplus (z_4\!\oplus\! z_5\!\oplus\! z_6\!\oplus\! z_7) \\
  &= z_a.
\end{aligned}
\end{equation}
Thus, $d_Z$ is a determinate value obtained during the process of movement, and it can be accounted for during the computation. Analogously, when measuring logical $\boldsymbol{X_L}=X_9X_8X_1$ before the movements, we get
\begin{equation}
  m_X = x_9\oplus x_8\oplus x_1,
\end{equation}
and measuring $\boldsymbol{X_L}=X_9X_8X_1X_4$ after the movement gives
\begin{equation}
\begin{aligned}
  m_X'
    &= x_9\oplus x_8\oplus (x_1\oplus x_a)\oplus (x_4^m\oplus x_a) \\
    &= m_X\oplus d_X,
\end{aligned}
\end{equation}
where $d_X = x_4^m$. The values $d_Z$ and $d_X$ are consitent with byproduct bits discussed in Ref.~\cite{Fowler_2012}.

\section{Performance of a simple simulator} \label{appendix:simulator}

\begin{figure}[h]
  \centering
  \includegraphics[width=0.5\textwidth]{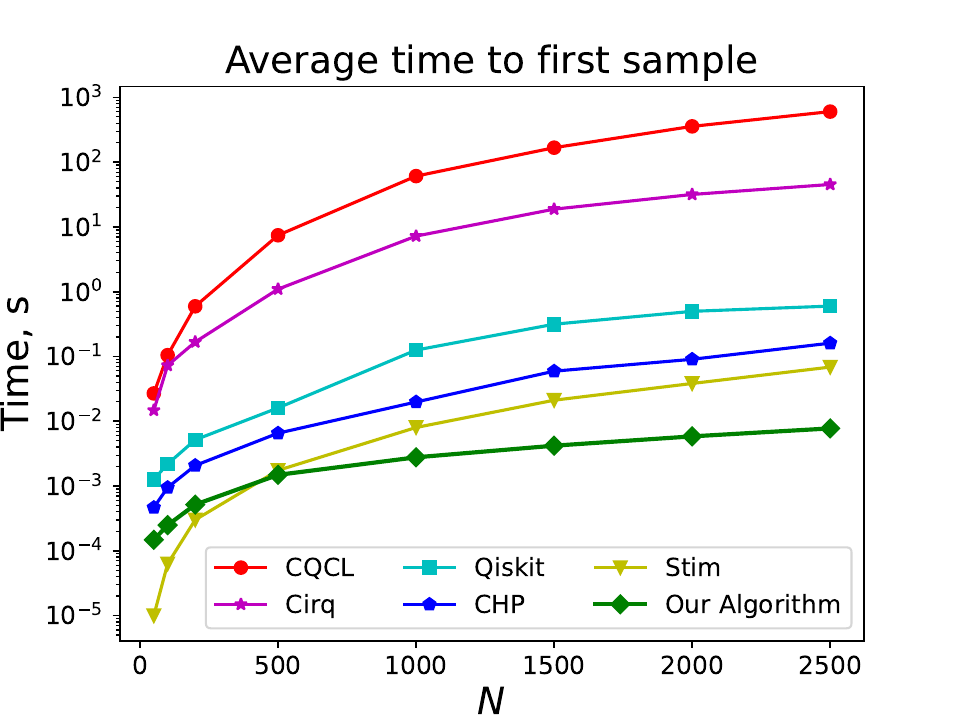}
  \caption{Running times to taking a single sample from random CSS-preserving circuits for various stabilizer circuit simulators. The data is generated by taking average running times over circuits constructed of $N$ qubits with $N$ randomly chosen CSS-preserving gates and measurements. The included stabilizer circuit simulators are: CQCL/Simplex simulator \cite{de_Beaudrap_2022}, Cirq \cite{Cirq_2024}, Qiskit \cite{Qiskit_2024}, CHP \cite{Aaronson_2004}, Stim \cite{Gidney_2021} and \textsf{Method~I} from \cref{subsubsec:weak_simulation}.}
  \label{fig:performance}
\end{figure}

To test the correctness and performance of the proposed procedures, we developed a simple simulator written in \texttt{C++} that realizes \textsf{Methods I-IV} of \cref{subsec:classical_simulation}. The program was used to prepare \cref{fig:sampling_times} and \cref{fig:performance}. The implementation is far from optimal but shows quite impressive results. On \cref{fig:performance} we see that for the task of taking a single sample from a CSS-preserving stabilizer circuit our algorithm asymptotically outperforms other known simulators. The source of this advantage comes from taking into account the CSS structure. Thus, modern stabilizer simulators might benefit from learning to recognize CSS-ness.

\section{Standard quadratic form expansions of trace-decreasing Clifford channels} \label{appendix:trace-decreasing_channels}

Trace-decreasing quantum operations \cite{Cappellini_2007,Filippov_2021,Shirokov_2007} naturally appear when studying post-selective processes or erasure noise \cite{Aaronson_2005,Kenbaev_2022}. In this Appendix, we discuss the notions of subnormal stabilizer states and trace-decreasing Clifford channels, propose standard quadratic form expansion expansions for such operations and show how to compute the standard quadratic form expansion of composition of two trace-decreasing Clifford channels.

\emph{Subnormal quantum state} $\rho$ is a positive Hermitian operator with trace less than unit $\Tr(\rho)\leq 1$. Such states naturally appear when studying post-selection processes and can be interpreted operationally as an ability to prepare a state $\rho/\Tr(\rho)$ with probability $\Tr(\rho)$, otherwise error. Let us call a \emph{subnormal stabilizer state} $\rho$ a state that can be written as
\begin{equation}
  \rho = \frac{1}{2^k} \sum_{u: uH=0} (-1)^{u s}i^{u Q u^T} \ketD{u},
\end{equation}
where $(H,s,Q)$ is a usual standard quadratic form expansion data and $k \in \{0,1,2,\dots,\infty\}$ is an integer appearing in scalar factor $2^k$, in case $k=\infty$ the state is zero $\rho=0$. Subnormal quantum states can be stored in the computer as a tuple $(k;H,s,Q)$.

As discussed in \cref{subsubsec:standard_form_Heisenberg}, during post-processing tasks there often appear operations of checking the condition $\Pi$. Let us now write it as
\begin{equation}
  \braD{\Pi} = \frac{1}{2^k} \sum_{w:\, w G = 0} (-1)^{w a} i^{w M w^T} \braD{w K}.
\end{equation}
Note that the integer $k$ should be big enough to guarantee that $\Pi\leq I$, lower bound on $k$ can be computed using Gaussian elimination (first make $G=0$, then make $K$ injective). Such checks $\braD{\Pi}$ can be stored in computer memory as tuples $(k; G,a,M,K)$.

A superoperator $\Phi$ is called \emph{trace-decreasing channel} if it is completely positive and satisfies the condition $\Tr\circ\Phi \leq \Tr$. Let us call a \emph{trace-decreasing Clifford channel} the trace-decreasing channel $\Phi$ that has subnormal stabilizer Choi state. Using the reasoning from \cref{subsubsec:standard_form}, one can show that any trace-decreasing Clifford channel $\Phi$ can be written as
\begin{center}
  \sffamily
  Standard quadratic form expansion of \\
  trace-decreasing Clifford channel $\Phi$
\end{center}
\begin{equation}
\begin{multlined}
  \Phi = \frac{1}{2^k} \sum_{u,w:\, u H = w G} (-1)^{u s} i^{u Q u^T} \ketD{u} \, \times \\
    \qquad \times (-1)^{w (a\oplus J u^T)} i^{w M w^T}\braD{u V \oplus w K},
\end{multlined}
\end{equation}
where $k \in \{0,1,2,\dots,\infty\}$ is an integer in scalar factor $2^{-k}$, $u$ and $t$ are row-vectors, $H$ and $G$ are parity-check matrices, $(s,Q)$ and $(c,M)$ are datum of quadratic forms, $J$ is a matrix of Boolean bilinear form $(u,w)\mapsto w J u^T$, $V$ and $K$ are transition matrices and the rows of $K$ are linearly independent from the rows of $V$. Thus, trace-decreasing channels are encoded by tuples $(k;H,s,Q,V;G,a,J,M,K)$. Intuitively, $k$ represents overall sub-normalisation of the channel, the part $(H,s,Q,V)$ represents the transformation of the input, and the part $(G,a,J,M,K)$ represents post-selection check on the input.

Discarding the channel's outputs gives a post-selection check
\begin{equation}
  \braD{0}\Phi = \frac{1}{2^k} \sum_{w :\, w G = 0} (-1)^{w a} i^{w M w^T} \braD{w K},
\end{equation}
using Gaussian elimination on this form (first make $G=0$, then make $K$ injective) gives a lower bound on integer $k$.

The matrices $Q$, $J$, $M$ satisfy $1$-cocycle conditions: for all $u, w$ and $u', w'$ such that $u H = w G$ and $u' H = w' G$ it holds that
\begin{align}
  u \mathcal{Q} u'^T &= \tau(u,u')\oplus\tau(u V,u' V), \\
  w J u^T  &= \tau(w K, u V), \\
  w \mathcal{M} w'^T &= \tau(w K, w' K),
\end{align}
these equations are taken modulo $2$.

Suppose we have two trace-decreasing channels $\Phi$ and $\Phi'$ in standard quadratic form expansion $(k;H,s,Q,V;G,a,J,M,K)$ and $(k';H',s',Q',V';G',a',J',M',K')$ respectively. Let us describe the standard quadratic form expansion of their composition $\Phi'' = \Phi \circ \Phi'$. The sub-normalisation integer $k''$ is
\begin{equation}
  k'' = k+k',
\end{equation}
the part $(H'',s'',Q'',V'')$ corresponding to input transformation follows the laws of trace-preserving case
\begin{equation}
\begin{aligned}
  H'' &= [H|V H'],\\
  s'' &= s\oplus V s',\\
  Q'' &= Q + V Q' V^T,\\
  V'' &= V V',\\
\end{aligned}
\end{equation}
while the second part $(G'',a'',J'',M'',K'')$ corresponding to post-selection check satisfies
\begin{equation}
\begin{aligned}
  G'' &= \left[\begin{array}{c|c} G & K H' \\ \hline 0 & G' \end{array} \right],\\
  a'' &= \left[\begin{array}{c}   a\oplus K s'     \\ \hline c'     \end{array} \right],\\
  J'' &= \left[\begin{array}{c}   J\oplus K Q' V^T \\ \hline J' V^T \end{array} \right],\\
  M'' &= \left[\begin{array}{c|c} M + K Q K^T & K J'^T \\ \hline J'K^T & M' \end{array} \right],\\
  K'' &= \left[\begin{array}{c}   K V' \\ \hline K' \end{array} \right].
\end{aligned}
\end{equation}
These relations were derived using the convention $u' = u V + w K$, $u'' = u$, $w''= [w|w']$. The obtained standard quadratic form expansion can be simplified by producing suitable Gaussian elimination on matrices $H$, $G$ and $K$.

Suppose $\Phi$ is a CSS-preserving trace-decreasing stabilizer operation, by which we mean that it has a linear form expansion
\begin{equation}
    \Phi = \frac{1}{2^k} \sum_{u,w:\, u H = w G} \hspace{-1em} (-1)^{u s}\ketD{u} \cdot (-1)^{w a} \braD{u V \oplus w K}.
\end{equation}
The symbol of this superoperator (see \cref{subsubsec:quasiprobability_representation}) is
\begin{equation}
  p_\Phi(v|v') = \frac{1}{2^{d}} \delta_{v\in V v' \oplus s \oplus H\Int_2^k} \delta_{K v' = a\oplus G\Int^b},
\end{equation}
where $d,b$ are suitable integers. In terms of hidden variable model discussed in \cref{subsubsec:CSS_hidden_variables_simulation}, given a point $v'$, the channel $\Phi$ first tosses some coins depending on $a$, then checks the condition $K v' = a \oplus G \Int^b$, and then transforms $v'$ to random $v \in V v'\oplus s \oplus H \Int^k$ (if some checks failed, the output is rejected).

\end{document}